# Predicting molecular phenotypes from histopathology images: a transcriptome-wide expression-morphology analysis in breast cancer

Yinxi Wang[1,§], Kimmo Kartasalo[1,2,§], Masi Valkonen[2], Christer Larsson[3], Pekka Ruusuvuori[2,4,‡], Johan Hartman[5,6 ‡], Mattias Rantalainen[1‡*]

§ These authors contributed equally to the study

‡ These authors contributed equally to the study

* Corresponding author

**Affiliations:**
[1] Department of Medical Epidemiology and Biostatistics, Karolinska Institutet, Stockholm, Sweden
[2] Faculty of Medicine and Health Technology, Tampere University, Tampere, Finland
[3] Division of Translational Cancer Research, Department of Laboratory Medicine, Lund University, Lund, Sweden
[4] Institute of Biomedicine, University of Turku, Turku, Finland
[5] Department of Oncology-Pathology, Karolinska Institutet, Stockholm, Sweden
[6] Department of Clinical Pathology and Cytology, Karolinska University Laboratory, Stockholm, Sweden

**Abstract**

Molecular phenotyping is central in cancer precision medicine, but remains costly and standard methods only provide a tumour average profile. Microscopic morphological patterns observable in histopathology sections from tumours are determined by the underlying molecular phenotype and associated with clinical factors. The relationship between morphology and molecular phenotype has a potential to be exploited for prediction of the molecular phenotype from the morphology visible in histopathology images.

We report the first transcriptome-wide Expression-MOrphology (EMO) analysis in breast cancer, where gene-specific models were optimised and validated for prediction of mRNA expression both as a tumour average and in spatially resolved manner. Individual deep convolutional neural networks (CNNs) were optimised to predict the expression of 17,695 genes from hematoxylin and eosin (HE) stained whole slide images (WSIs). Predictions for 9,334 (52.75%) genes were significantly associated with RNA-sequencing estimates (FDR adjusted p-value < 0.05). 1,011 of the genes were brought forward for validation, with 876 (87%) and 908 (90%) successfully replicated in internal and external test data, respectively. Predicted spatial intra-tumour variabilities in expression were validated in 76 genes, out of which 59 (77.6%) had a significant association (FDR adjusted p-value < 0.05) with spatial transcriptomics estimates. These results suggest that the proposed methodology can be applied to predict both tumour average gene expression and intra-tumour spatial expression directly from morphology, thus providing a scalable approach to characterise intra-tumour heterogeneity.

**Introduction**

Microscopic morphological patterns observable in stained tumour tissue are routinely characterised by pathologists to classify and diagnose cancers. General morphology is assessed using HE staining, while immunohistochemistry enables semi-quantitative assessment of specific markers. However, cancer is a genetic disease where somatic alterations and their interactions with other phenotypic factors and the tumour microenvironment give rise to a complex and dynamic molecular phenotype. Profiling of e.g. somatic DNA alterations, RNA expression or protein abundances provide a comprehensive characterisation of tumours. In breast cancer, the molecular phenotype defined by the mRNA expression profile contains prognostic information[1–4] and also defines the intrinsic molecular subtypes[5]. Compared with routine pathology, molecular profiling represents a more comprehensive characterisation of the individual tumour[6], providing information relevant for precision medicine[7], and information that can contribute to the discovery of novel therapeutic targets and diagnostic markers.

Intra-tumour heterogeneity is a key contributing factor to emerging treatment resistance, or reduced efficacy of treatment, which is caused by either subclonality or as a consequence of plasticity in the dynamic molecular phenotype of a tumour[8,9]. Tumour evolution and subclonality can be inferred from genetic data. However, the more comprehensive phenotype defined by the mRNA expression profile, and other dynamic molecular phenotypes, is generally acquired from a bulk average mRNA pool where intra-tumour variability is lost. Single-cell RNA-sequencing[10,11] enables profiling of thousands of individual cells, providing unique information to characterise intra-tumour heterogeneity[12]. Although techniques for single-cell sequencing now are mainstream, it remains challenging on primary human samples as fresh samples typically are required. Spatial transcriptomic profiling[13,14] is another emerging technology enabling characterisation of intra-tumour heterogeneity, but it is still technically demanding, expensive, and offers low resolution both spatially and in terms of the number of genes that can be detected.

Computational pathology, driven by deep learning based artificial intelligence applied on digital WSIs, has recently emerged and demonstrated human pathologist level performance in cancer detection and classification[15,16]. Deep CNNs have also been applied for prediction of molecular phenotypes from routine formalin-fixed paraffin-embedded (FFPE) HE stained sections[16–20]. More importantly, this approach also enables inference of spatial heterogeneity. To date, three studies with the objective of predicting gene expression phenotypes from histopathology images have been reported[18–20]. However, these studies have substantial limitations in either: (a) the number of genes analysed (250 genes) and sample size (23 patients)[18]; (b) extensive use of transfer-learning, i.e. a single-global CNN model for prediction of all phenotypes rather than optimisation of gene specific models[19,20]; (c) the use of a pan-cancer approach[19,20], where a single model is used across a range of cancer diseases, which by design will lead to models optimised to capture morphologies shared across the majority of diseases included; (d) lack of independent external validation cohorts[19,20], or validation in very small data sets[18] (two tumours); or (e) lack of validation by orthogonal experimental techniques and spatial expression predictions[19,20].

Here we report the first EMO analysis in breast cancer using large-scale deep learning directly from routine HE WSIs. The study is comprehensive in that individual models were optimised for each gene across the mRNA transcriptome. The results were validated in a fully independent external patient cohort at the gene level. Furthermore, we demonstrate that our CNN models enable prediction of spatial expression patterns, which were validated in independent tumours using spatial transcriptomic (ST) profiling.

**Results**

We performed a transcriptome-wide EMO analysis, where individual deep CNN models were optimised separately for each mRNA transcript. RNA-sequencing was used to quantify the expression of 20,477 individual genes (see Supplementary Methods for details). In total, 991 patients (7.28 million HE tiles) from two studies (TCGA breast cancer[6] and Clinseq-breast[21]), each with one WSI, were included and split into training (N=697, 70.3%), validation (N=122, 12.3%) and internal test sets (N=172, 17.4%) prior to model optimisation and validation. The pre-processing of WSIs included segmentation of tissue and invasive cancer, tiling of WSIs into tiles of 598×598 pixels (271 μm×271 μm), quality control for image sharpness, and colour normalisation to adjust for variations in stains and scanners (see Supplementary Methods). In the training set, 17,695 genes remained after excluding transcripts with low variance (see Supplementary Methods). For each of the transcripts, a deep CNN model (InceptionV3[22]) was optimised to predict normalised gene expression using image tiles as predictors. The models were trained in parallel on a high-performance compute cluster (CSC, Kajaani, Finland), with the transcriptome-wide analysis requiring approximately 300,000 GPU hours.

The tile level predictions of each slide were averaged to obtain slide level predicted expression (EMO-average), which was compared with gene expression measured by RNA-seq. The optimised models were subsequently applied and evaluated in validation and test sets (Figure 1a). In the validation set, out of 17,695 genes, the predicted expression of 9,334 (52.75%) genes was significantly correlated with expression levels measured by RNA-seq (Spearman correlation, FDR adjusted p-value < 0.05; Figure 1c-d). We also assessed the proportion of variance predicted: 1,026 (5.80%) genes showed a coefficient of determination ($R^2_{pred}$) higher than 0.2, and 196 (1.11%) and 26 (0.15%) genes had $R^2_{pred}$ higher than 0.3 and 0.4, respectively (Figure 1b). Taken together, these results indicate that morphological patterns in histopathology images can be learned by deep CNN models and exploited to predict gene expression for a substantial proportion of genes across the transcriptome.

To assess the generalisability of the approach, 1,011 genes with $R^2_{pred}$ >0.2 and FDR adjusted p-value <0.001 in the validation set (Supplementary Table 3) were brought forward for validation in the internal (N=172) and external test sets (ABiM study[23], N=350). 876 (86.6%) genes had a significant association between predicted (EMO-average) and observed (RNA-seq) expression (Bonferroni adjusted p-value <0.05, Spearman correlation, Figure 2b). 479 of these genes had an $R^2_{pred}$ >0.2 (Figure 2a) in the internal test set. 908 (91.3%) genes were successfully validated in the external test set (Bonferroni adjusted p-value (Spearman correlation) <0.05, Figure 2c). The estimated correlation coefficients (Spearman's $\rho$) between EMO-average prediction and RNA-seq across the 1,011 genes had a high concordance between the validation, internal and external test sets (Supplementary Figure 1a-c), indicating similar levels of prediction performance across datasets. Concordance between EMO-average predicted and RNA-seq estimated gene expression for the 25 genes with the best prediction performances in the internal test set, ranked by p-value (Spearman correlation), are visualised in Figure 2d, with the corresponding results in the external test set in Figure 2e. To determine if genes involved in particular molecular mechanisms or processes were enriched in the set of transcripts that could be predicted from histopathology images, we conducted a gene set enrichment analysis (GSEA[24]) across all 17,695 genes[25]. 16 pathways (Figure 1e, Supplementary Table S2) were significantly enriched; a majority of these have previously been found to be associated with breast cancer. The functional classes of the gene sets included angiogenesis, cell proliferation, cell cycle, apoptosis, signal transduction, metabolism and immune system (see Supplementary Results and Table S2).

Finally, we validated EMO predictions of intra-tumour expression variability (EMO-spatial) by performing ST analysis. Expressions of 76 genes (Supplementary Table 4) across 12 regions of interest (ROIs) in 22 tumours (FFPE sections from independent sets of tumours, 264 ROIs in total) were measured using the Nanostring GeoMX DSP platform (Figure 3a) and compared with EMO-spatial predictions. To ascertain if intra-tumour heterogeneity in expression could be predicted, we assessed the association between EMO-spatial predictions and ST measurements, using linear mixed-effects models fitted for each gene across all ROIs and slides, with the ST expression as response, EMO-spatial prediction as a fixed effect and the slide ID included as random effect to account for slide level systematic variability. Spatial predictions of 59 genes (79%) were significantly associated with ST estimated expression levels (FDR adjusted p-value <0.05, likelihood ratio test, Figure 3b-c; see also Supplementary Figure 1d for gene-level within-slide estimates of Spearman correlations between EMO-spatial and ST expression; Supplementary Figure 2 and 3 for examples of prediction results across the 22 WSIs). Among the ten genes with the most significant association between ST estimates and EMO-spatial predictions, three genes could be found in the T cell receptor pathway (CD3E, CD8A and CD27) and three genes in the cytokine and chemokine signaling pathway (CXCL9, CXCL10 and CMKLR1), other genes were found in the total immune (PTPRC), B cells (MS4A1), proliferation (MKI67) and cytotoxicity (NKG7) pathways. Taken together, these results indicate that EMO spatial prediction offers a methodology that can enable exploration of intra-tumour gene expression heterogeneity based on routine HE stained sections.

**Discussion**

We have performed the first reported transcriptome-wide expression - morphology study in breast cancer based on individually optimised gene-level models. Tumour-level prediction results were validated in a completely independent external cohort, and spatial expression predictions were validated in independent tumours by ST profiling. 17,695 gene-specific CNN models were optimised for prediction of gene expression, out of these 9,334 had a significant association between EMO-average prediction and RNA-seq estimates. Out of 1,011 genes brought forward for final validation, prediction performance could be confirmed for 86.6% and 91.3% in the internal and external test data, respectively. We also demonstrated that our approach predicted spatial variability in gene expression with significant association with ST profiling in 59 out of 76 genes validated using this technique, suggesting that deep CNN models enable characterisation of intra-tumour heterogeneity in RNA expression.

By developing models for a single cancer disease (breast cancer) and by optimising individual deep CNN models for each gene, we avoid several strong assumptions made in previously reported studies[18–20] that are unlikely to hold. Pan-cancer models assume shared morphologies across cancer diseases, which provides a fundamental limitation given the broad range of morphological characteristics observable in different cancer types. Strong reliance on transfer learning across genes represents another fundamental limitation that is likely to constrain the ability to develop models that are effective for modelling more specific relationships between morphology and gene expression.

This study is limited with respect to the size of the training dataset, and it is expected that with more training data, the prediction performance could improve further. Spatially resolved data for model optimisation also has the potential to improve model performance in the future. One previously reported study has applied that approach, however, their training dataset was limited to only 23 tumours and 250 genes[18]. In our study, the ST validation was limited to a panel of 76 genes, which was dictated by the availability of FFPE compatible ST profiling gene panel at the time of the study.

Prediction of gene expression from routine HE WSIs has the potential to impact both clinical diagnostics as well as cancer research. Prediction of molecular phenotypes can enable cost-effective precision medicine, either by direct predictions of key markers, or as a way to prioritise which patients are likely to benefit from comprehensive but costly molecular profiling. In the research domain, cost-effective predictions of expression will enable large-scale epidemiological studies that include gene expression phenotypes as exposures. Spatial prediction of gene expression provides a complement to single-cell sequencing and ST profiling for studies of intra-tumour heterogeneity and tumour microenvironment, and enables studies at a substantially larger scale compared to what is possible by direct molecular profiling. The results from this study are promising and we expect that our approach will also work well for application in other cancer diseases, and for prediction of other types of molecular phenotypes, such as somatic mutations, copy number alterations, epigenetic factors, metabolite or protein abundances.

Our findings suggest that deep learning-based image analysis for prediction of the tumour average expression of a substantial number of transcripts is possible and feasible. However, more importantly, we demonstrated and experimentally validated that spatial gene expression predictions can be used to characterise intra-tumour heterogeneity.

**Acknowledgements**

The authors would like to acknowledge patients, clinicians, and hospital staff participating in the SCAN-B study, the staff at the central SCAN-B laboratory at Division of Oncology, Lund University, the Swedish national breast cancer quality registry (NKBC), Regional Cancer Center South, and the South Swedish Breast Cancer Group (SSBCG). We would also like to thank Dr. Johan Vallon-Christersson (Lund University) for helping in preparing these data. This project was supported by funding from the Swedish Research Council, Swedish Cancer Society, Karolinska Institutet (Cancer Research KI; StratCan), ERA PerMed (ERAPERMED2019-224 - ABCAP), MedTechLabs, Swedish e-science Research Centre (SeRC) - eCPC, Stockholm Region, Stockholm Cancer Society and Swedish Breast Cancer Association, Academy of Finland (313921, 326463, 314558 & 326364), CSC – IT Center for Science (Finland) (Grand Challenge pilot project AI-EMO, 2001568).

# Figures

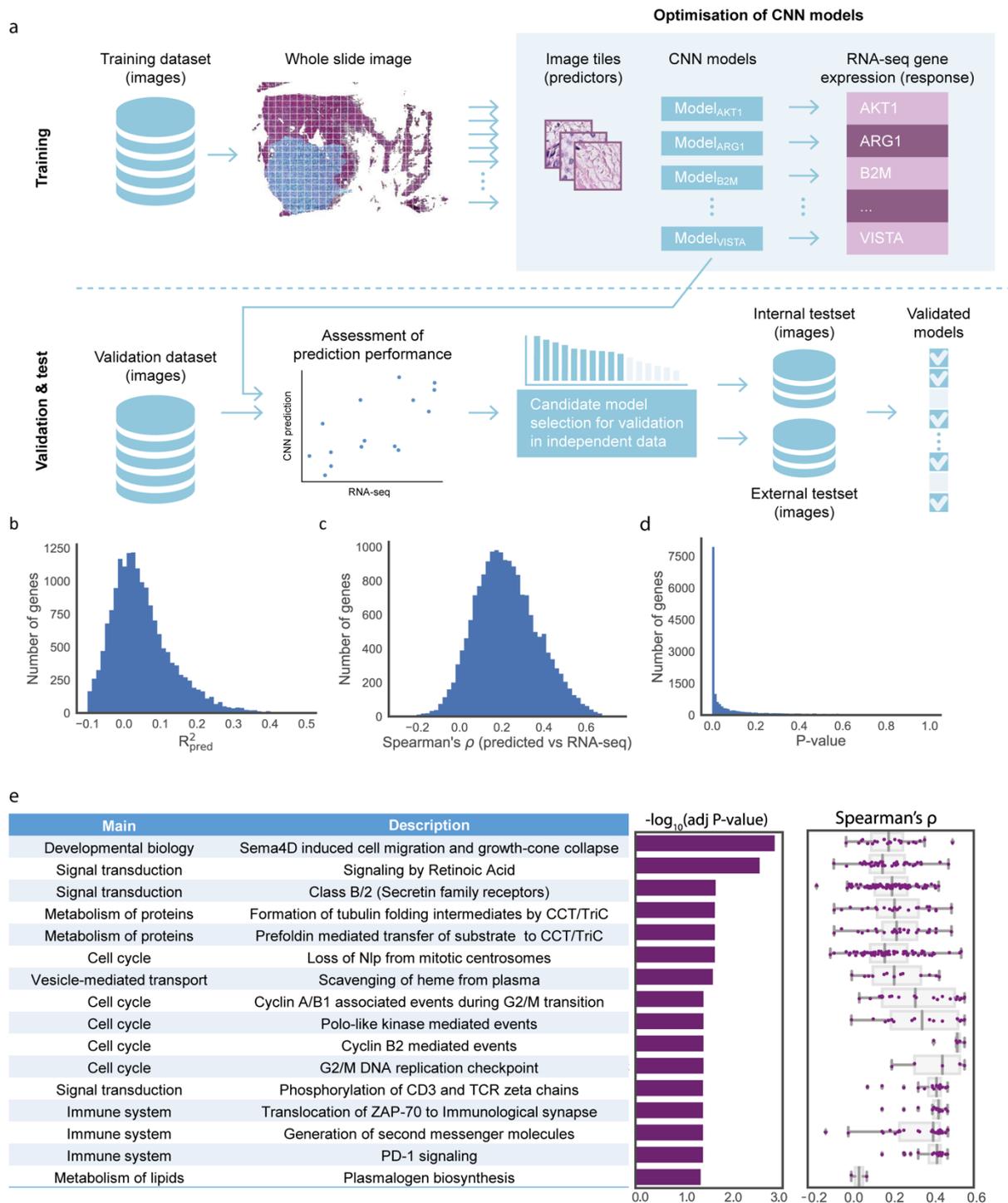

**Figure 1.** Study design and summary statistics for transcriptome-wide predictions. **a**, Overview of the EMO process. In the training phase, training WSIs (N=697) were split into image tiles. The tiles (predictors) together with expression levels (response) across the protein coding transcriptome were used to optimise individual deep CNN models (Inception V3) for each gene. All optimised models were then applied to predict expression in WSIs in the validation set (N=122), association analysis between true gene expression value and predicted value was performed, and candidate models were selected for further validation. The validation was performed on the internal (N=172) and external (N=350) test sets. **b**, Histogram describing the empirical distribution of predicted $R^2$ in validation set (458 genes with a predicted $R^2 < -0.1$ were excluded from the figure for clarity). **c**, Histogram of the empirical distribution of Spearman's **ρ** between EMO predictions and RNA-seq in the validation set. **d**, Histogram of the p-values related to panel **c. e**. Pathway analysis of EMO predictions by GSEA in the Reactome database, revealing 16 significant pathways.

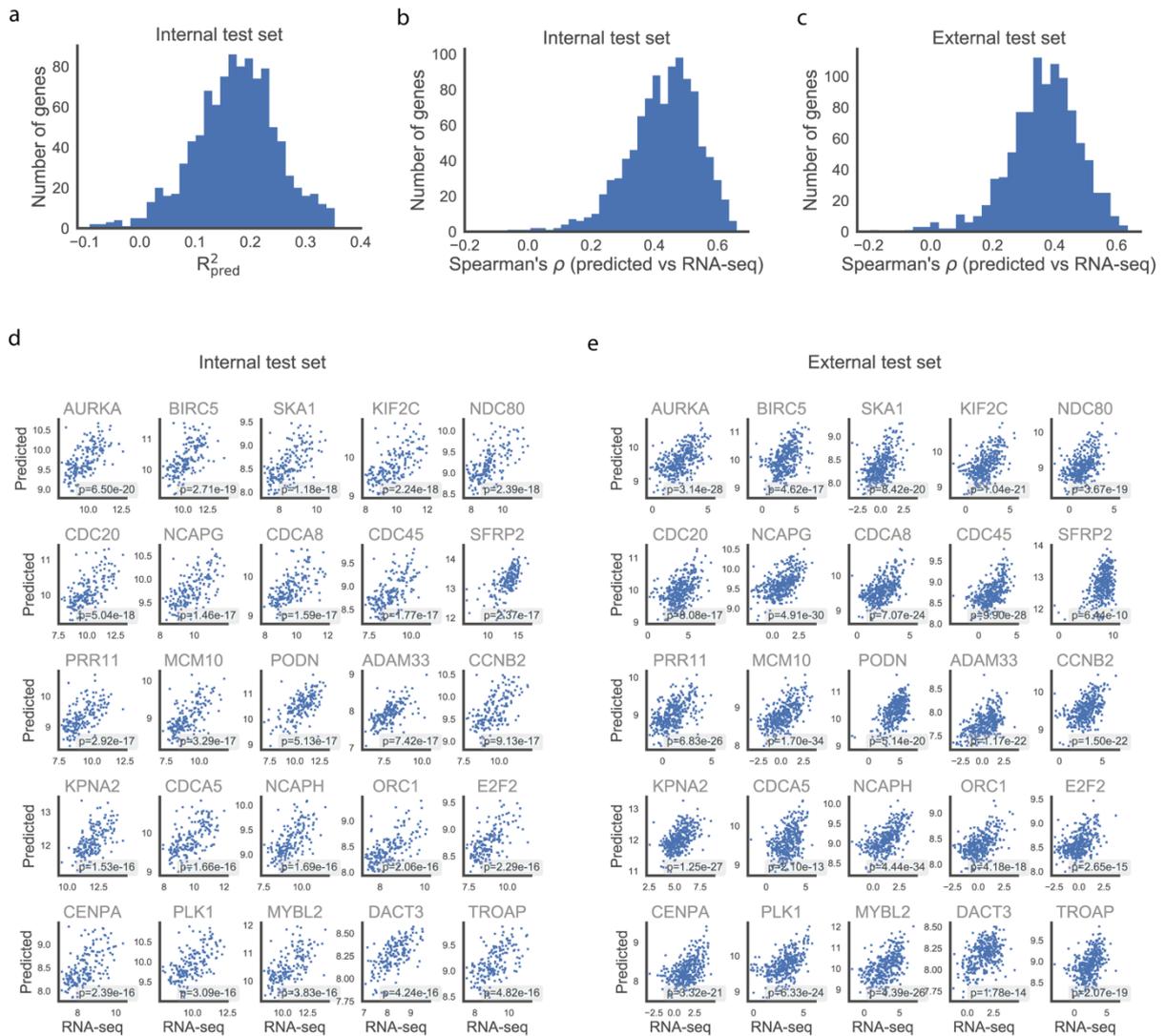

**Figure 2.** Summary of model performance on test sets. **a**, Distribution of $R^2_{pred}$ in the internal test set ($N_{genes}$=1,011; one gene with a predicted $R^2 < -0.1$ was excluded from the figure for clarity). **b,** Distribution of Spearman's $\rho$ in the internal test set. **c**, Distribution of Spearman's $\rho$ in the external test set ($N_{genes}$=995). **d**, Scatter plot of EMO predicted and RNA-seq estimated gene expression values for the 25 top performing genes in the internal test set. **e**, Scatter plot of EMO predicted and RNA-seq estimated gene expression values for the same 25 genes in the external test set.

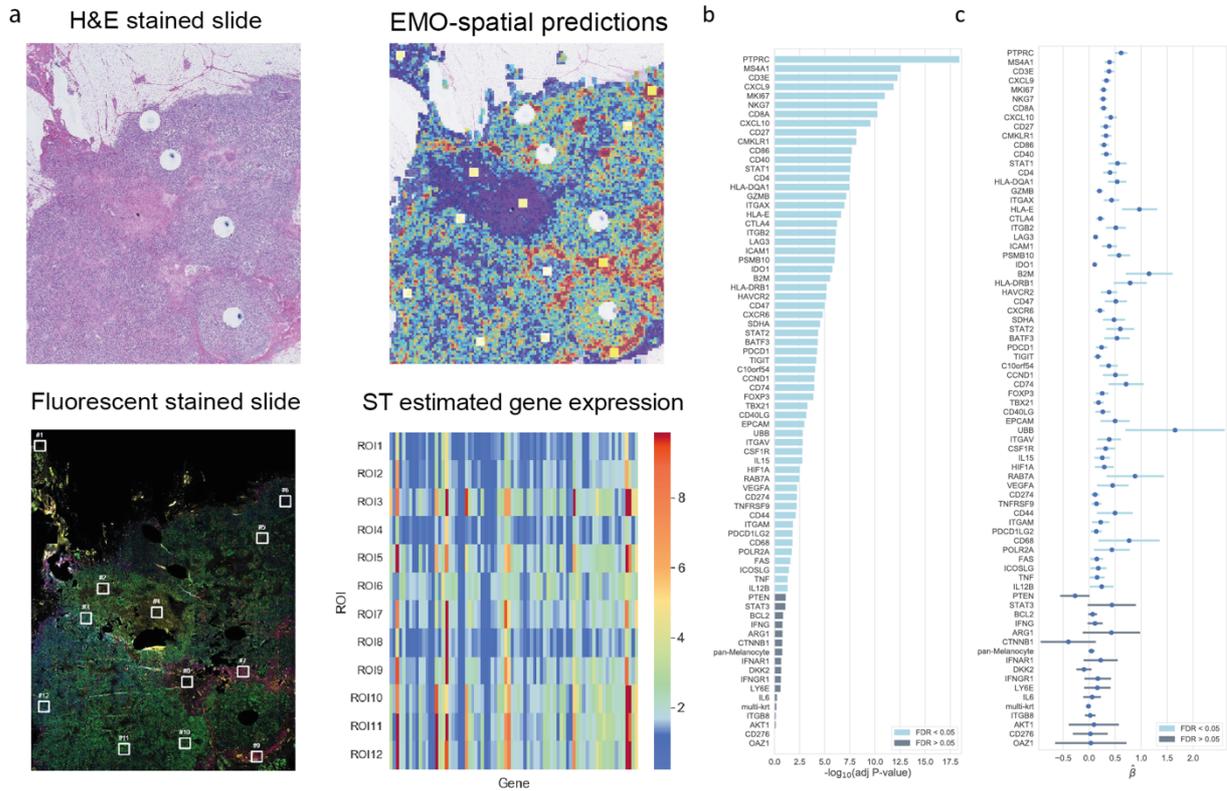

**Figure 3.** ST validation of spatial expression predictions. **a**, Overview of the ST profiling process: For each WSI (top-left), trained models for the genes in the ST gene panel were used to predict spatial (tile-level) expression, visualised as heatmaps. 12 ROI (yellow squares) were subsequently manually selected to obtain a representative set of regions including low, medium and high predicted expression across a range of genes (top-right). The ROI from each slide were then manually registered against fluorescently labeled slides from consecutive FFPE sections (lower-left). ST profiling of the ROIs was performed and subsequently used to validate spatial EMO prediction results (lower-right). **b**, Bar plot for the ranked -$\log_{10}$(FDR adjusted p-value) for genes from each linear mixed effects model. Light blue indicates FDR adjusted p-value <0.05. **c.** Corresponding fixed effect coefficients and 95% CI related to the EMO prediction for each gene (linear mixed effects model).

# Supplementary Online Methods and Results

## 1. Data collection

The study consists of female breast cancer patients from three data sources: Clinseq-BC (N=270), The Cancer Genome Atlas (TCGA-BC) (N=721), and ABiM (N=350) as an external validation cohort. For Clinseq-BC and ABiM, hematoxylin and eosin (HE)-stained formalin-fixed paraffin-embedded (FFPE) histopathology slides were scanned in-house with a Hamamatsu Nanozoomer XR (Hamamatsu Photonics, Hamamatsu, Japan) at 40X magnification (0.226 μm/pixel). Whole slide images (WSIs) from TCGA-BC were downloaded from https://portal.gdc.cancer.gov/. WSIs in TCGA-BC that were scanned at 20X were excluded to ensure image quality. One WSI image was included from each individual. All included patients have corresponding RNA-seq data available for analysis.

Images from Clinseq-BC and TCGA-BC were randomly split into training (N=558, 56.3%; 4.08 million HE tiles), tuning (N=139, 14.0%; 0.97 million HE tiles), validation (N=122, 12.3%; 0.90 million HE tiles) and test sets (N=172, 17.4%; 1.33 million HE tiles). The internal and external test sets remained untouched during the model development and training phase and were used only once for final evaluation of model performance at the end of the project.

## 2. Data pre-processing

### 2.1. Tissue segmentation

Tissue was segmented based on downsampled images that were extracted from the resolution pyramids within the WSI using OpenSlide[1]. After HSV transformation of the image, Otsu's thresholding[2] was applied to the saturation component, and a fixed threshold of 0.75 was applied to the hue component to generate two binary masks. The logical intersection of these masks was then computed. The resulting mask was post-processed by applying binary morphological closing followed by opening to fill holes and to remove small debris and noise. A disk-shaped structuring element with a radius of 10 pixels was used.

### 2.2. Tumour annotation

The invasive cancer regions in the Clinseq-BC slides were collaboratively annotated by two pathologists using QuPath[3]. The resulting coordinates representing the contours of the invasive regions were converted into annotation polygons to generate tumour annotation masks. To produce a final label mask for each WSI, we computed the logical intersection of the tissue mask and the tumour annotation mask. The regions forming the intersection were labeled as 'invasive cancer', whereas all other tissue regions were labeled as 'non-cancerous'.

### 2.3. Tiling

The WSIs were split into small image tiles extracted from the regions representing invasive cancer. For images scanned in-house on the Hamamatsu device, we initially extracted tiles of 1,196 × 1,196 pixels (271 × 271 μm) at full resolution. We then downsampled the tiles by a factor of two using Lanczos interpolation, resulting in dimensions of 598 × 598 pixels. For WSIs scanned on other devices, we adapted the initial tile size and downsampling factor in accordance with the scanner pixel size to obtain 598 × 598 pixel tiles representing the same physical dimensions as in the case of the Hamamatsu images. This resampling ensures there is no scanner-dependent bias in the apparent size of histological structures in the images. We performed the tiling with 50% overlap (that is, a stride of 135.5 μm) between neighboring tiles. Only tiles containing over 50% tissue pixels based on the masks were retained for further analysis. The tiles were stored as JPEG (quality 80).

## 2.4. Quality control

To ensure data quality, blurred tiles were excluded from analysis. To this end, Laplacian filtering was applied to the tiles to enhance edges and local variation, which are considered indicative of an in-focus image[4]. Tiles exhibiting variance lower than 500, calculated over all pixels after the filtering, were considered blurry. We used the implementation in the 'OpenCV' package with kernel size 1. At this stage, tiles containing large areas of adipose tissue were also removed as they exhibit low variance.

## 2.5. Colour normalisation

In order to reduce the colour variations due to differences in the histological sample preparation and scanning between different laboratories and batches of samples, we adopted the normalisation approach proposed by Macenko et al.[5] The same colour transformations were applied to all tiles from a WSI, that is, normalisation was performed on the slide-level. To achieve more stable stain vector estimates, we first corrected the luminosity of each WSI. We randomly sampled 100 tiles from each WSI and concatenated them to form a single image, followed by transformation from RGB to LAB colour space. To obtain a reference value, we computed the 95th percentile of the luminosity values of the image ($l_{ref95}$). We then processed all tiles in the WSI according to the following steps: (1) RGB to LAB transformation; (2) Luminosity values of pixels exceeding $l_{ref95}$ were set to 255 (white); (3) Luminosity values of all other pixels were linearly rescaled to a range from 0 to 255; (4) LAB to RGB transformation. The process was repeated for all WSIs.

When performing the actual colour normalisation, we applied stain deconvolution in optical density(OD) space to estimate the stain matrix, which represents the absorbance of the H and E stains in the RGB colour channels, and the pixel-wise saturation coefficients for each stain[5,6]. The saturation values represent the amount of H or E stain at each pixel, but are also affected by systematic differences in sample preparation and imaging. To estimate a reference, we randomly sampled 3000 luminosity-corrected tiles across all datasets and again concatenated them into a single image. From this image, that represents the average appearance of all the data, we then computed the HE stain matrix, and the 99th percentile of the saturation coefficients to obtain pseudo-maximum values for each stain ($S_{ref99}$).

After estimating the global reference values, we processed all the images by randomly sampling 100 luminosity-corrected tiles from each WSI, followed by computing the slide-specific stain matrices based on these tiles. Finally, to apply the normalisation procedure to all tiles from a WSI, we first extracted the pixel-wise saturation coefficients for a tile, based on the slide level stain matrix and the tile's pixel values in OD space. For improved robustness, we then excluded the top 1st percentile of saturation values for each stain and linearly rescaled the remaining values as described by Macenko et al.[5] such that the maximum value matches $S_{ref99}$. We then applied the global reference stain matrix on the corrected saturation values to obtain normalised pixel values in OD space. The values were then mapped back to RGB colour space for final use. The process was repeated for all tiles in each WSI of the datasets.

## 2.6. RNA-seq data preparation

We collected transcriptome-wide RNA-seq data representing mRNA expression for a total of 20,477 genes in the reference genome. RNA sequencing, pre-processing and normalisation was performed as described previously[7,8]. In total, 19,112 genes had non-zero gene expression variance. In addition, we hypothesised that genes with close to zero variance are less informative for expression-morphology analysis and the potential for extracting relevant morphological features gradually decreases with diminishing gene expression variance. The benefit of getting meaningful results is further limited after considering the computational cost of training each model. Hence, we chose to only include genes with a variance larger than 0.01 in further analysis, which resulted in 17, 695 genes as the final training targets.

As Clinseq-BC and TCGA-BC were merged together, to reduce batch effects associated with the RNA sequencing, the RNA-seq data from Clinseq-BC were normalised to have median value equal to TCGA-BC. In brief, we first calculate the median expression level of each gene for both TCGA-BC and Clinseq-BC data sources. Only data from the training and validation sets were included in this step. Next, the differences between median values of these two data sources were calculated. Finally, the Clinseq-BC expression values were normalised by gene-wise addition of the offsets computed in the previous step. TCGA-BC data remained unchanged.

During the testing phase, RNA-seq data in the test sets from Clinseq-BC, TCGA-BC and ABiM cohorts were all median normalised using the same procedure with TCGA-BC (training and validation data) as a reference.

# 3. Detection of invasive tumour regions

## 3.1. Data

Only invasive tumour regions were used for training and testing the expression-morphology models (see Section 4). On the Clinseq-BC dataset, the regions were manually annotated by pathologists, as described in Section 2.2. We used the annotations to train a cancer detection model, which we applied on the other datasets to obtain masks indicative of predicted tumour regions also for the non-annotated slides. The WSIs were tiled as described in Section 2.3. and the pathologists' annotations were used to infer tile-level labels of malignant and benign tissue for model training and evaluation.

For the purpose of developing the cancer detection model, we randomly sampled 20% (n=49) of the Clinseq-BC slides for testing and divided the remaining WSIs into training (n=154) and validation (n=39) sets. The validation set was used to select the tile size parameter, and the testing set was used for a final evaluation of cancer detection performance. Out of the approximately 2.3 million tiles in the training data, we used 80% for model training and 20% for evaluation to allow parameter tuning and model development.

## 3.2. Training

We applied transfer learning to speed up model convergence by training an Inception V3[9] deep convolutional neural network (CNN) initialised with weights pre-trained on ImageNet[10]. We used the ADAM[11] optimizer with the binary cross entropy loss function and the following parameters: learning rate = $1 \times 10^{-4}$, $\beta1 = 0.9$, $\beta2 = 0.999$, $\varepsilon$ = None and decay = 0. Random 90° degree rotations and flips of the tiles were applied as data augmentation. We used a minibatch of 48 tiles, sampling an equal number of malignant and benign tiles into each batch. The training was run for 50 steps per epoch. We used early stopping with a minimum change in loss of 0.003 and a patience of 10 epochs and continued training until the early stopping criteria were met or a maximum number of 2,500 steps had been taken.

### 3.3. Prediction

The trained cancer detection CNN was applied to predict the probability of cancer for each tile in the validation set and model performance was evaluated using Receiver Operating Characteristics (ROC) analysis. The CNN achieved an Area Under the Curve (AUC) value of 0.973. The performance of the final model was evaluated on the held-out test set, resulting in an AUC of 0.956.

Based on the validation set ROC curve, a suitable cut-off threshold was selected in order to obtain binary classification results for the tiles. The trained CNN was then used to produce predicted probabilities of cancer for all tissue tiles from the non-annotated datasets. The output probabilities were thresholded using the cut-off value to obtain a binary mask representing predicted tumour regions. The mask was post-processed by filling holes and removing objects consisting of less than 405 pixels, representing debris and noise. As a result, tumour label masks similar to those available for the manually annotated Clinseq-BC dataset were obtained for the TCGA-BC and ABiM datasets.

## 4. Expression-morphology analysis

### 4.1. Model optimisation

For each gene, we optimised one CNN model with image tiles as predictors and the sample-level gene expression level obtained from RNA-seq as a response variable. We used Inception V3[9] architecture, modified by switching the last layer to one neuron with a linear activation to build a regression model. Tiles from the training and tuning set were used to optimise the model. We employed the Adam[11] optimizer with the mean squared error loss function and default parameters as follows: learning rate = $1 \times 10^{-6}$, $\beta 1 = 0.9$, $\beta 2 = 0.999$, $\varepsilon$ = None and decay = 0. Random 90° degree rotations and flips of the tiles were applied as data augmentation.

We used a minibatch of 32 image tiles per step and ran the optimisation for 150 steps per epoch. We sampled 313 mini-batches from the tuning set to assess the validation loss on each epoch, and used early stopping with a minimum change in loss of 0.003 and a patience of 80 epochs, continuing optimisation until the early stopping criterion was met or a maximum number of 500 epochs (75 000 steps) had been completed. From each optimisation run, we stored the models from the 10 epochs resulting in the best performance on the tuning tiles and 10 models from randomly selected epochs. Depending on when the early stopping criterion was met, the optimisation runs took approximately 12-70 hours on a single GPU.

### 4.2. Model validation

For each gene, the model with the lowest loss recorded on the tuning set was applied on the combined Clinseq-BC and TCGA-BC validation set. We took the mean of all tile-level predictions across one slide to obtain a patient-level prediction. To evaluate model performance, we calculated the Spearman correlation between the predicted patient-level gene expression values and those measured using RNA-seq across the validation set. The associated p-values were adjusted for multiple testing using the Benjamini-Hochberg approach[12]. In an attempt to measure the proportion of variance that could be predicted using the CNN models, $R^2$ score was also calculated.

## 4.3. Model selection for testing

To further validate the generalisability of the CNNs, we selected a subset of promising genes (i.e. models) with predicted $R^2_{pred}$ higher than 0.2 and adjusted p-value lower than 0.001 according to the performance on the validation set. In total, 1,011 genes were brought forward for model testing on the internal test set. Out of these, 995 genes could be matched in the ABiM study, and were included in the evaluation on the external test data. As the variance is dataset dependent and the values of RNA-seq data can vary due to differences in protocols and profiling platforms, the $R^2_{pred}$ score was not calculated in the external ABiM test set.

The tile-level predictions were post-processed as described in Section 4.2. except that we applied Bonferroni correction for multiple testing, which is more conservative compared to the Benjamini-Hochberg method applied for the initial discovery analysis.

## 4.4. Gene set enrichment analysis

With the aim of understanding if genes associated with particular molecular mechanisms were enriched among the genes that were predicted well, we conducted a pathway analysis to identify enriched pathways based on the results on the validation set. To do this, instead of arbitrarily selecting a cutoff threshold for assigning significance among all available genes, and performing analysis based on the subset of genes, we considered a rank based algorithm[13] to avoid potential bias in such selection.

In brief, the 17,695 genes ranked by adjusted p-value from the Spearman analysis were used as input. We followed the procedures described in[13] and conducted the analysis with the 'SetRank' R package and Reactome[14] was selected as the database for pathway annotations.

## 5. Spatial transcriptomic analysis

From an additional independent collection of 168 tumours with both FFPE blocks and WSIs available, 24 tumours were selected for spatial transcriptomic (ST) profiling using the oncology and immune oriented gene panel for the GeoMx DSP platform (GeoMx Immune Pathways Panel, NanoString Technologies, Seattle, WA). The 24 slides were selected to have predicted (by the CNN models) spatially varying expression levels, assessed by visual inspection, across a number of genes (BCL2, CD4, GZMB, HIF1A, HLA-DQA1, ITGB2 and VEGFA). These genes represent a diverse set from the panel in that they belong to different pathways and were also among the best performing genes ($R^2$ >0.15, adjusted p-value <0.0001) in the validation set (EMO-average). Regions of interest (ROI) (600μm × 600μm) were manually selected from the HE stained WSIs based on EMO-spatial predictions, to cover a range of predicted gene expression values across a variety of genes. For each tumour, two consecutive sections were produced. The first section was stained with HE and used to generate a routine WSI (used for prediction of expression level in the EMO-spatial workflow). The second section was used for ST profiling in a standard workflow for the GeoMX DSP platform. This slide was stained with four fluorescent stains targeting PanCK, SMA, CD45 and DNA to outline general morphological structures (HE stains was not an option on the platform. Manual registration of the selected ROIs the first section, which was used to produce the HE-stained slide and associated EMO-spatial predictions, and the corresponding locations on the second consecutive section, used for ST profiling. Two slides were damaged during fluorescent staining and discarded, resulting in 22 slides remaining for ST analysis. Finally, gene expression values within each ROI were quantified by the GeoMx DSP platform by counting the unique indexing oligos assigned to each target with the NanoString nCounter® instrument. Gene expression values were normalised by dividing each value with the average expression levels across 6 negative controls, to account for any non-specific binding, and subsequently $\log_2$-transformed before further analysis.

To estimate the EMO-spatial predictions, we calculated the mean of tile-level predictions within each ROI per gene. The gene panel consists of 84 RNA probes (See Supplementary Table 4 for full list of genes), out of which 6 served as negative controls, and 2 (CCL5 and PECAM1) had a variance of gene expression lower than 0.001 and were therefore excluded prior to any further analyses. Furthermore, the probe named 'multi-krt' include probes against a group of genes: KRT18, KRT6B, KRT6C, KRT6A, KRT19, KRT17, KRT7, KRT10 and KRT14; and the probe named 'pan-Melanocyte' contains probes against SOX10, PMEL and S100B. The predicted gene expression values for these two targets were calculated by summing the predictions for the respective sets of genes.

We then measured the performance of the CNN models by first comparing the predictions with ST-measured gene expression using a linear mixed effect model (LME) with results displayed with a bar plot and a line plot (Figure 3. b-c). The model was fitted (maximum likelihood) with the log transformed ST estimated expression as response, the EMO-spatial prediction as a fixed effect and slide id as a random effect (accounting for variability between slides). A likelihood ratio test was applied to test the significance of the fixed effect parameter. P-values were adjusted for multiple testing using the Benjamini-Hochberg method, and FDR < 0.05 was considered significant. Furthermore, for each slide and gene, Spearman's $\rho$ was also calculated, between EMO-spatial predictions and ST expression estimates, across 12 ROIs per tumour. The empirical distribution of Spearman's $\rho$ estimates across the 22 individuals, and for each gene, was summarised as boxplots (Supplementary Figure 1d).

## 6. Software and hardware

All image preprocessing steps were conducted with Python (v. 3.6) packages, including scikit-image (v. 1.14.2), OpenCV (v. 3.4.1), OpenSlide (v.3.4.1 and API v. 1.1.1). Training of CNN models was carried out using Keras (v. 2.2.4) with Tensorflow[15] backend (v. 1.12). Colour normalisation was performed using Python, adapted from StainTools ( https://github.com/Peter554/StainTools) and 'Staining Unmixing and Normalisation in Python' (https://github.com/schaugf/HEnorm_python). Spearman correlation was calculated using the SciPy package (v. 1.2.0) in Python, $R^2$ was calculated with Python package scikit-learn (v. 0.20.2). Statistical testing, multiple testing correction fitting were performed with the statsmodels Python package (v. 0.9.0). LME model was fitted using R (3.6.3) with R package 'lme4' and 'lmerTest'. Model training and predictions were run on the GPU partition of the Puhti compute cluster (CSC IT Center for Science, Kajaani, Finland), consisting of 80 compute nodes. Each node is equipped with two 20-core Xeon Gold 6230 CPUs (Intel, Santa Clara, CA, USA), 384 GB of memory, four Tesla V100 32 GB GPU accelerators (Nvidia, Santa Clara, CA, USA) and 3.6 TB of local NVME storage. The GPUs were running Nvidia driver version 440.33.01.

Since models for different genes are fully independent of each other, transcriptome-wide training and prediction represent 'an embarrassingly parallel problem'. We therefore ran each model as a separate computation job on a single V100 GPU, and automated job submission through the SLURM scheduler system, resulting in 50-300 models being trained in parallel at any given time over a period of several months. At the beginning of each computation run, the input image tiles were copied from the central file system to the local NVME disk to avoid I/O bottlenecks due to the large number of parallel runs relying on the same data. In parallel with the GPU computation, minibatches were prepared using multi-threading on 4 CPU cores to maintain an in-memory data buffer equal in size to two minibatches.

# 7. Supplementary Results

## 7.1. Gene set enrichment analysis

To determine if genes involved in particular molecular mechanisms or processes were enriched in the set of transcripts that could be predicted from histopathology images, we conducted a gene set enrichment analysis (GSEA[16]) across all 17,695 genes[13]. 16 pathways (Table S2) were significantly enriched in the GSEA[16] analysis. 'Sema4D induced cell migration and growth-cone collapse' had the strongest association. Sema4D has previously been reported to be overexpressed in breast cancer[17]. Sema4D, together with the small GTPase Rho gene family (i.e. RhoA, RhoB, RhoC), which are encoded by genes that belong to the same pathway and were well-predicted by the CNN model, are associated with tumour angiogenesis[18]. Ranking second was 'Signaling by Retinoic Acid'. Retinoic acid (RA) has been reported as being associated with down-regulating genes that relate to breast cancer cell proliferation and up-regulating pro-apoptotic genes which induces cell death[19]. These events are also associated with morphological changes. In addition, four pathways relating to cell cycle were also identified, including, 'Cyclin A/B1 associated events during G2/M transition'. Genes well predicted and belonging to this pathway (CCNA2, CCNB1) encode Cyclin A2 and B1 respectively. These proteins have been reported to be associated with breast cancer histological grade[20] and prognosis[21,22]. Another well predicted gene, CDK1, is a member of the enriched pathway 'G2/M DNA replication checkpoint', and the protein it encodes is associated with prognosis in breast cancer patients[23]. Moreover, in the pathway relating to 'Loss of Nlp from mitotic centrosomes', Nlp (ninein-like protein) has been recently recognised as an oncogenic protein, whose centrosomal localisation and stability could be disturbed in case of BRCA1 mutations, and eventually lead to abnormal mitotic progression as well as tumorigenesis[24,25]. The remaining significant pathways are involved in biological processes such as 'signal transduction', 'metabolism of proteins', 'cell cycle' and 'immune system'.

# Supplementary Tables

**Table S1. Clinicopathological characteristics of patients per cohort.**

|  | Clinseq-BC | TCGA-BC | ABiM |
|---|---|---|---|
| N | 270 | 721 | 350 |
| Age | 59.54 ± 11.42 | 57.57 ± 12.95 | 62.63 ± 13.96 |
| Histological grade | | | |
|     NHG 1 (%) | 38 (14.07) | 67 (9.29) | 42 (12.00) |
|     NHG 2 (%) | 108 (40.00) | 264 (36.62) | 151 (43.14) |
|     NHG 3 (%) | 110 (40.74) | 228 (31.62) | 157 (44.86) |
|     NA | 14 (5.19) | 162 (22.47) | 0 |
| Tumour size | | | |
|     ≥20 mm (%) | 127 (47.04) | 495 (68.65) | 160 (45.71) |
|     <20 mm (%) | 105 (38.89) | 206 (28.57) | 177 (50.57) |
|     NA | 38 (14.07) | 20 (2.77) | 13 (3.71) |
| Lymph node | | | |
|     Positive (%) | 33 (12.22) | 355 (49.24) | 124 (35.43) |
|     Negative (%) | 199 (73.70) | 346 (47.99) | 213 (60.86) |
|     NA | 38 (14.07) | 20 (2.77) | 13 (3.71) |
| Distant metastasis | | | |
|     Positive (%) | 1 (0.37) | 9 (1.25) | 1 (0.29) |
|     Negative (%) | 253 (93.70) | 605 (83.91) | 300 (85.71) |
|     NA | 16 (5.93) | 107 (14.84) | 49 (14.00) |
| ER status[a] | | | |
|     Positive (%) | 228 (84.44) | 507 (70.32) | 293 (83.71) |
|     Negative (%) | 41 (15.19) | 161 (22.33) | 57 (16.29) |
|     NA | 1 (0.37) | 53 (7.35) | 0 |
| PR status[a] | | | |
|     Positive (%) | 176 (65.19) | 440 (61.03) | 265 (75.71) |
|     Negative (%) | 92 (34.07) | 225 (31.21) | 85 (24.29) |
|     NA | 2 (0.74) | 56 (7.77) | 0 |
| HER2 status | | | |
|     Positive (%) | 41 (15.19) | 80 (11.10) | 69 (19.71) |
|     Negative (%) | 221 (81.85) | 461 (63.94) | 281 (80.29) |
|     NA | 8 (2.96) | 180 (24.97) | 0 |
| KI67 status[b] | | | |
|     High (%) | 132 (48.89) | - | 207 (59.14) |
|     Low (%) | 98 (36.30) | - | 143 (40.86) |
|     NA | 40 (14.81) | - | 0 |

NHG Nottingham histological grade, NA not applicable, ER estrogen receptor, PR progesterone receptor,
a Positive if >= 10% positive stained cells. b High if >= 20% positive stained cells within hotspot regions.

**Table S2. Summarization of enriched pathways.**

| Name | Description | Main | Size | Adjusted P-value |
|---|---|---|---|---|
| R-HSA-416572 | Sema4D induced cell migration and growth-cone collapse | developmental biology | 24 | 0.0012 |
| R-HSA-5362517 | Signaling by Retinoic Acid | Signal transduction | 42 | 0.0025 |
| R-HSA-373080 | Class B/2 (Secretin family receptors) | Signal transduction | 84 | 0.0205 |
| R-HSA-389960 | Formation of tubulin folding intermediates by CCT/TriC | metabolism of proteins | 25 | 0.0213 |
| R-HSA-389957 | Prefoldin mediated transfer of substrate to CCT/TriC | metabolism of proteins | 28 | 0.0213 |
| R-HSA-380259 | Loss of Nlp from mitotic centrosomes | cell cycle | 69 | 0.0213 |
| R-HSA-2168880 | Scavenging of heme from plasma | vesicle-mediated transport | 13 | 0.0235 |
| R-HSA-69273 | Cyclin A/B1 associated events during G2/M transition | cell cycle | 22 | 0.0368 |
| R-HSA-156711 | Polo-like kinase mediated events | cell cycle | 16 | 0.0368 |
| R-HSA-157881 | Cyclin B2 mediated events | cell cycle | 5 | 0.0368 |
| R-HSA-69478 | G2/M DNA replication checkpoint | cell cycle | 5 | 0.0368 |
| R-HSA-202427 | Phosphorylation of CD3 and TCR zeta chains | Signal transduction | 19 | 0.0376 |
| R-HSA-202430 | Translocation of ZAP-70 to Immunological synapse | immune system | 16 | 0.0376 |
| R-HSA-202433 | Generation of second messenger molecules | immune system | 30 | 0.0376 |
| R-HSA-389948 | PD-1 signaling | immune system | 20 | 0.0376 |
| R-HSA-75896 | Plasmalogen biosynthesis | metabolism | 2 | 0.0421 |

**Table S3. Prediction performance for the 1011 genes on validation and test sets.**

| Ensembl_gene_id | Hgnc symbol | Coef (valid) | Rsq (valid) | P-val (valid) | adj.pval (valid) | Coef (internal) | adj.pval (internal) | Coef (external) | adj.pval (external) |
|---|---|---|---|---|---|---|---|---|---|
| ENSG00000091831 | ESR1 | 4.62E-01 | 3.10E-01 | 8.70E-08 | 1.25E-06 | 4.78E-01 | 3.23E-08 | 3.93E-01 | 2.06E-11 |
| ENSG00000175356 | SCUBE2 | 5.47E-01 | 3.08E-01 | 6.82E-11 | 3.20E-09 | 4.17E-01 | 1.31E-05 | 3.92E-01 | 2.64E-11 |
| ENSG00000082175 | PGR | 5.73E-01 | 3.26E-01 | 5.43E-12 | 4.05E-10 | 4.92E-01 | 7.33E-09 | 3.90E-01 | 3.95E-11 |
| ENSG00000148513 | ANKRD30A | 5.74E-01 | 3.41E-01 | 4.87E-12 | 3.68E-10 | 3.90E-01 | 1.28E-04 | 3.31E-01 | 2.27E-07 |
| ENSG00000173467 | AGR3 | 5.86E-01 | 4.00E-01 | 1.35E-12 | 1.28E-10 | 5.04E-01 | 1.92E-09 | 3.81E-01 | 1.47E-10 |
| ENSG00000171428 | NAT1 | 5.54E-01 | 2.64E-01 | 3.60E-11 | 1.87E-09 | 4.75E-01 | 4.74E-08 | 3.64E-01 | 1.94E-09 |
| ENSG00000158258 | CLSTN2 | 4.93E-01 | 2.54E-01 | 8.14E-09 | 1.74E-07 | 4.58E-01 | 2.76E-07 | 4.13E-01 | 8.13E-13 |
| ENSG00000074410 | CA12 | 4.43E-01 | 2.99E-01 | 3.17E-07 | 3.77E-06 | 5.31E-01 | 7.01E-11 | 3.07E-01 | 4.62E-06 |
| ENSG00000129514 | FOXA1 | 4.95E-01 | 3.29E-01 | 6.92E-09 | 1.52E-07 | 4.85E-01 | 1.64E-08 | 3.37E-01 | 9.58E-08 |
| ENSG00000144218 | AFF3 | 4.74E-01 | 2.90E-01 | 3.38E-08 | 5.69E-07 | 3.85E-01 | 1.80E-04 | 2.64E-01 | 5.58E-04 |
| ENSG00000186868 | MAPT | 4.63E-01 | 2.43E-01 | 8.00E-08 | 1.17E-06 | 4.38E-01 | 1.90E-06 | 3.93E-01 | 2.18E-11 |
| ENSG00000138755 | CXCL9 | 5.82E-01 | 2.80E-01 | 2.10E-12 | 1.85E-10 | 5.13E-01 | 6.04E-10 | 4.65E-01 | 3.52E-17 |
| ENSG00000204385 | SLC44A4 | 4.61E-01 | 3.12E-01 | 8.97E-08 | 1.28E-06 | 4.41E-01 | 1.43E-06 | 2.53E-01 | 1.65E-03 |
| ENSG00000115648 | MLPH | 5.23E-01 | 3.43E-01 | 6.47E-10 | 2.11E-08 | 4.77E-01 | 3.83E-08 | 3.19E-01 | 1.05E-06 |
| ENSG00000178568 | ERBB4 | 4.30E-01 | 2.24E-01 | 7.86E-07 | 8.06E-06 | 3.27E-01 | 1.17E-02 | 3.86E-01 | 7.43E-11 |
| ENSG00000169083 | AR | 4.53E-01 | 2.55E-01 | 1.56E-07 | 2.05E-06 | 2.62E-01 | 5.32E-01 | 3.07E-01 | 4.52E-06 |
| ENSG00000107485 | GATA3 | 4.83E-01 | 3.39E-01 | 1.73E-08 | 3.28E-07 | 5.61E-01 | 1.18E-12 | 3.85E-01 | 8.08E-11 |
| ENSG00000138615 | CILP | 4.73E-01 | 2.23E-01 | 3.88E-08 | 6.39E-07 | 4.08E-01 | 2.79E-05 | 4.20E-01 | 2.09E-13 |
| ENSG00000112936 | C7 | 4.94E-01 | 2.16E-01 | 7.51E-09 | 1.63E-07 | 5.54E-01 | 3.07E-12 | 3.61E-01 | 3.06E-09 |
| ENSG00000152049 | KCNE4 | 5.20E-01 | 2.18E-01 | 8.06E-10 | 2.56E-08 | 5.20E-01 | 2.79E-10 | 3.78E-01 | 2.65E-10 |
| ENSG00000109436 | TBC1D9 | 5.95E-01 | 4.27E-01 | 5.06E-13 | 5.70E-11 | 4.97E-01 | 4.28E-09 | 4.43E-01 | 2.99E-15 |
| ENSG00000113296 | THBS4 | 5.68E-01 | 3.11E-01 | 8.80E-12 | 6.08E-10 | 4.73E-01 | 6.00E-08 | 4.35E-01 | 1.28E-14 |
| ENSG00000106819 | ASPN | 5.54E-01 | 2.80E-01 | 3.52E-11 | 1.84E-09 | 6.06E-01 | 1.35E-15 | 5.12E-01 | 9.70E-22 |
| ENSG00000049540 | ELN | 5.68E-01 | 2.95E-01 | 8.82E-12 | 6.08E-10 | 5.09E-01 | 9.63E-10 | 4.32E-01 | 2.62E-14 |
| ENSG00000074527 | NTN4 | 5.55E-01 | 2.38E-01 | 3.35E-11 | 1.78E-09 | 5.21E-01 | 2.32E-10 | 3.32E-01 | 1.97E-07 |
| ENSG00000157303 | SUSD3 | 4.91E-01 | 2.44E-01 | 9.43E-09 | 1.97E-07 | 5.05E-01 | 1.72E-09 | 4.01E-01 | 5.44E-12 |
| ENSG00000108176 | DNAJC12 | 5.83E-01 | 2.42E-01 | 1.89E-12 | 1.70E-10 | 4.08E-01 | 2.71E-05 | 5.12E-01 | 9.63E-22 |
| ENSG00000123500 | COL10A1 | 5.24E-01 | 2.41E-01 | 5.70E-10 | 1.89E-08 | 5.93E-01 | 1.11E-14 | 5.19E-01 | 1.62E-22 |
| ENSG00000187720 | THSD4 | 6.17E-01 | 3.70E-01 | 3.73E-14 | 6.88E-12 | 4.72E-01 | 6.23E-08 | 3.85E-01 | 8.17E-11 |
| ENSG00000166535 | A2ML1 | 4.60E-01 | 2.48E-01 | 1.00E-07 | 1.40E-06 | 4.39E-01 | 1.65E-06 | 3.27E-01 | 3.76E-07 |
| ENSG00000187955 | COL14A1 | 6.01E-01 | 3.23E-01 | 2.49E-13 | 3.27E-11 | 6.06E-01 | 1.23E-15 | 4.83E-01 | 7.89E-19 |
| ENSG00000169245 | CXCL10 | 6.48E-01 | 3.19E-01 | 7.11E-16 | 2.86E-13 | 4.06E-01 | 3.24E-05 | 5.67E-01 | 3.75E-28 |
| ENSG00000090382 | LYZ | 6.23E-01 | 2.59E-01 | 1.77E-14 | 3.87E-12 | 3.49E-01 | 2.75E-03 | 2.54E-01 | 1.45E-03 |
| ENSG00000168350 | DEGS2 | 5.69E-01 | 3.19E-01 | 7.96E-12 | 5.59E-10 | 5.05E-01 | 1.60E-09 | 4.79E-01 | 1.86E-18 |
| ENSG00000167244 | IGF2 | 6.37E-01 | 3.27E-01 | 3.10E-15 | 1.06E-12 | 5.59E-01 | 1.71E-12 | 5.85E-01 | 1.78E-30 |
| ENSG00000163491 | NEK10 | 5.89E-01 | 2.48E-01 | 9.26E-13 | 9.15E-11 | 4.53E-01 | 4.63E-07 | 3.99E-01 | 8.78E-12 |
| ENSG00000106809 | OGN | 6.08E-01 | 3.17E-01 | 1.10E-13 | 1.76E-11 | 6.03E-01 | 2.16E-15 | 4.93E-01 | 7.36E-20 |
| ENSG00000176887 | SOX11 | 3.55E-01 | 2.21E-01 | 5.93E-05 | 3.11E-04 | 3.06E-01 | 4.64E-02 | 3.77E-01 | 2.73E-10 |
| ENSG00000170500 | LONRF2 | 5.56E-01 | 3.16E-01 | 2.87E-11 | 1.57E-09 | 4.37E-01 | 2.09E-06 | 4.02E-01 | 4.99E-12 |
| ENSG00000115255 | REEP6 | 4.96E-01 | 2.23E-01 | 6.24E-09 | 1.40E-07 | 4.09E-01 | 2.62E-05 | 2.88E-01 | 4.02E-05 |
| ENSG00000150667 | FSIP1 | 4.59E-01 | 2.02E-01 | 1.04E-07 | 1.45E-06 | 2.57E-01 | 6.61E-01 | 2.91E-01 | 3.08E-05 |
| ENSG00000145423 | SFRP2 | 5.01E-01 | 2.09E-01 | 4.11E-09 | 9.93E-08 | 6.29E-01 | 2.37E-17 | 3.72E-01 | 6.44E-10 |
| ENSG00000103257 | SLC7A5 | 6.28E-01 | 3.60E-01 | 9.95E-15 | 2.48E-12 | 4.07E-01 | 3.03E-05 | 4.32E-01 | 2.43E-14 |
| ENSG00000039068 | CDH1 | 4.55E-01 | 2.27E-01 | 1.44E-07 | 1.91E-06 | 4.39E-01 | 1.74E-06 | 3.25E-01 | 4.48E-07 |
| ENSG00000101057 | MYBL2 | 6.28E-01 | 3.39E-01 | 9.82E-15 | 2.48E-12 | 6.13E-01 | 3.83E-16 | 5.50E-01 | 4.39E-26 |
| ENSG00000183044 | ABAT | 5.44E-01 | 3.08E-01 | 9.05E-11 | 4.04E-09 | 5.11E-01 | 8.05E-10 | 2.19E-01 | 3.60E-02 |
| ENSG00000183779 | ZNF703 | 3.46E-01 | 2.14E-01 | 9.29E-05 | 4.56E-04 | 3.37E-01 | 6.21E-03 | 3.06E-01 | 5.00E-06 |
| ENSG00000186910 | SERPINA11 | 5.95E-01 | 2.48E-01 | 4.78E-13 | 5.45E-11 | 4.29E-01 | 4.47E-06 | 3.13E-01 | 2.16E-06 |
| ENSG00000050628 | PTGER3 | 5.15E-01 | 2.30E-01 | 1.26E-09 | 3.69E-08 | 4.76E-01 | 3.99E-08 | 4.26E-01 | 6.74E-14 |
| ENSG00000168542 | COL3A1 | 5.18E-01 | 2.16E-01 | 1.02E-09 | 3.08E-08 | 5.60E-01 | 1.32E-12 | 3.38E-01 | 7.89E-08 |
| ENSG00000133110 | POSTN | 4.90E-01 | 2.25E-01 | 1.01E-08 | 2.08E-07 | 5.73E-01 | 2.11E-13 | 3.39E-01 | 7.44E-08 |
| ENSG00000135069 | PSAT1 | 5.16E-01 | 3.63E-01 | 1.23E-09 | 3.62E-08 | 5.64E-01 | 7.92E-13 | 3.92E-01 | 2.48E-11 |
| ENSG00000164692 | COL1A2 | 5.15E-01 | 2.21E-01 | 1.25E-09 | 3.67E-08 | 5.79E-01 | 8.71E-14 | 3.98E-01 | 1.00E-11 |
| ENSG00000100219 | XBP1 | 3.87E-01 | 2.37E-01 | 1.04E-05 | 6.99E-05 | 5.33E-01 | 5.21E-11 | 2.52E-01 | 1.82E-03 |
| ENSG00000154451 | GBP5 | 6.53E-01 | 3.27E-01 | 3.39E-16 | 1.76E-13 | 4.68E-01 | 9.70E-08 | 5.32E-01 | 5.11E-24 |
| ENSG00000131203 | IDO1 | 6.28E-01 | 3.55E-01 | 1.02E-14 | 2.50E-12 | 4.56E-01 | 3.27E-07 | 5.41E-01 | 5.08E-25 |
| ENSG00000134533 | RERG | 4.69E-01 | 2.24E-01 | 4.95E-08 | 7.84E-07 | 3.73E-01 | 4.78E-04 | 2.93E-01 | 2.45E-05 |
| ENSG00000166482 | MFAP4 | 6.02E-01 | 3.03E-01 | 2.17E-13 | 3.00E-11 | 5.93E-01 | 1.08E-14 | 5.09E-01 | 1.77E-21 |
| ENSG00000011465 | DCN | 5.04E-01 | 2.24E-01 | 3.24E-09 | 8.02E-08 | 5.94E-01 | 9.36E-15 | 3.98E-01 | 9.52E-12 |
| ENSG00000071967 | CYBRD1 | 6.52E-01 | 3.50E-01 | 4.22E-16 | 2.02E-13 | 5.40E-01 | 2.14E-11 | 5.77E-01 | 1.67E-29 |
| ENSG00000163879 | DNALI1 | 4.49E-01 | 2.31E-01 | 2.21E-07 | 2.77E-06 | 3.97E-01 | 7.10E-05 | 3.99E-01 | 8.63E-12 |
| ENSG00000092068 | SLC7A8 | 5.08E-01 | 3.53E-01 | 2.39E-09 | 6.24E-08 | 4.91E-01 | 8.03E-09 | 3.70E-01 | 8.00E-10 |
| ENSG00000154864 | PIEZO2 | 5.27E-01 | 2.35E-01 | 4.34E-10 | 1.53E-08 | 3.28E-01 | 1.12E-02 | 2.88E-01 | 4.31E-05 |
| ENSG00000054598 | FOXC1 | 3.70E-01 | 2.62E-01 | 2.71E-05 | 1.58E-04 | 4.75E-01 | 4.79E-08 | 3.66E-01 | 1.49E-09 |
| ENSG00000164125 | FAM198B | 5.55E-01 | 2.64E-01 | 3.26E-11 | 1.74E-09 | 4.91E-01 | 7.93E-09 | 3.49E-01 | 1.94E-08 |
| ENSG00000170743 | SYT9 | 5.34E-01 | 2.01E-01 | 2.42E-10 | 9.32E-09 | 3.23E-01 | 1.57E-02 | 3.47E-01 | 2.34E-08 |
| ENSG00000169248 | CXCL11 | 5.61E-01 | 2.59E-01 | 1.74E-11 | 1.04E-09 | 4.11E-01 | 2.10E-05 | 5.09E-01 | 1.99E-21 |
| ENSG00000198944 | SOWAHA | 5.11E-01 | 2.17E-01 | 1.77E-09 | 4.85E-08 | 3.33E-01 | 7.90E-03 | 3.90E-01 | 3.93E-11 |
| ENSG00000134352 | IL6ST | 6.42E-01 | 4.10E-01 | 1.54E-15 | 5.58E-13 | 5.31E-01 | 7.18E-11 | 3.88E-01 | 4.70E-11 |

**Table S3. Prediction performance for the 1011 genes on validation and test sets.**

| Ensembl_gene_id | Hgnc symbol | Coef (valid) | Rsq (valid) | P-val (valid) | adj.pval (valid) | Coef (internal) | adj.pval (internal) | Coef (external) | adj.pval (external) |
|---|---|---|---|---|---|---|---|---|---|
| ENSG00000111716 | LDHB | 4.73E-01 | 2.62E-01 | 3.69E-08 | 6.12E-07 | 3.81E-01 | 2.47E-04 | 2.32E-01 | 1.12E-02 |
| ENSG00000173894 | CBX2 | 6.36E-01 | 3.73E-01 | 3.56E-15 | 1.13E-12 | 5.72E-01 | 2.64E-13 | 5.17E-01 | 2.84E-22 |
| ENSG00000196932 | TMEM26 | 5.08E-01 | 2.43E-01 | 2.28E-09 | 6.01E-08 | 4.03E-01 | 4.30E-05 | 3.96E-01 | 1.31E-11 |
| ENSG00000154027 | AK5 | 4.78E-01 | 2.08E-01 | 2.56E-08 | 4.56E-07 | 3.33E-01 | 8.18E-03 | 3.28E-01 | 3.03E-07 |
| ENSG00000186340 | THBS2 | 5.20E-01 | 2.12E-01 | 8.53E-10 | 2.67E-08 | 5.83E-01 | 5.24E-14 | 3.37E-01 | 9.60E-08 |
| ENSG00000196177 | ACADSB | 4.63E-01 | 2.30E-01 | 8.17E-08 | 1.19E-06 | 4.03E-01 | 4.26E-05 | 3.61E-01 | 3.42E-09 |
| ENSG00000171791 | BCL2 | 6.03E-01 | 4.14E-01 | 1.93E-13 | 2.75E-11 | 4.55E-01 | 3.55E-07 | 4.19E-01 | 2.44E-13 |
| ENSG00000164220 | F2RL2 | 5.64E-01 | 2.78E-01 | 1.30E-11 | 8.28E-10 | 5.17E-01 | 3.96E-10 | 4.87E-01 | 2.92E-19 |
| ENSG00000114805 | PLCH1 | 5.42E-01 | 2.53E-01 | 1.19E-10 | 5.10E-09 | 3.79E-01 | 2.91E-04 | 4.24E-01 | 1.05E-13 |
| ENSG00000107159 | CA9 | 5.26E-01 | 3.18E-01 | 4.78E-10 | 1.65E-08 | 2.51E-01 | 8.95E-01 | 2.90E-01 | 3.36E-05 |
| ENSG00000139865 | TTC6 | 5.15E-01 | 3.06E-01 | 1.26E-09 | 3.68E-08 | 4.45E-01 | 1.01E-06 | | |
| ENSG00000127084 | FGD3 | 5.30E-01 | 2.69E-01 | 3.44E-10 | 1.25E-08 | 4.52E-01 | 5.00E-07 | 3.20E-01 | 8.93E-07 |
| ENSG00000187244 | BCAM | 5.42E-01 | 2.95E-01 | 1.09E-10 | 4.73E-09 | 4.88E-01 | 1.13E-08 | 4.36E-01 | 1.13E-14 |
| ENSG00000117228 | GBP1 | 6.48E-01 | 3.28E-01 | 7.01E-16 | 2.86E-13 | 4.07E-01 | 3.11E-05 | 5.09E-01 | 1.85E-21 |
| ENSG00000091986 | CCDC80 | 5.03E-01 | 2.13E-01 | 3.51E-09 | 8.63E-08 | 5.33E-01 | 5.48E-11 | 4.48E-01 | 1.17E-15 |
| ENSG00000017427 | IGF1 | 4.66E-01 | 2.08E-01 | 6.45E-08 | 9.75E-07 | 5.17E-01 | 3.76E-10 | 3.95E-01 | 1.66E-11 |
| ENSG00000159388 | BTG2 | 5.95E-01 | 3.61E-01 | 4.71E-13 | 5.42E-11 | 5.00E-01 | 2.78E-09 | 4.68E-01 | 2.04E-17 |
| ENSG00000175287 | PHYHD1 | 5.16E-01 | 2.61E-01 | 1.16E-09 | 3.44E-08 | 4.11E-01 | 2.22E-05 | 3.92E-01 | 2.64E-11 |
| ENSG00000112562 | SMOC2 | 4.38E-01 | 2.10E-01 | 4.41E-07 | 4.94E-06 | 4.58E-01 | 2.81E-07 | 4.86E-01 | 3.59E-19 |
| ENSG00000090539 | CHRD | 5.54E-01 | 2.45E-01 | 3.53E-11 | 1.84E-09 | 4.87E-01 | 1.26E-08 | 3.90E-01 | 3.58E-11 |
| ENSG00000163362 | C1orf106 | 5.53E-01 | 2.65E-01 | 4.14E-11 | 2.12E-09 | 4.79E-01 | 3.00E-08 | 4.89E-01 | 1.92E-19 |
| ENSG00000120262 | CCDC170 | 5.09E-01 | 2.42E-01 | 2.14E-09 | 5.67E-08 | 4.83E-01 | 2.02E-08 | 4.20E-01 | 2.36E-13 |
| ENSG00000165816 | VWA2 | 5.75E-01 | 2.98E-01 | 4.27E-12 | 3.31E-10 | 3.66E-01 | 7.92E-04 | 3.87E-01 | 6.27E-11 |
| ENSG00000109452 | INPP4B | 5.26E-01 | 2.84E-01 | 4.86E-10 | 1.67E-08 | 3.80E-01 | 2.75E-04 | 3.34E-01 | 1.40E-07 |
| ENSG00000188153 | COL4A5 | 5.52E-01 | 2.36E-01 | 4.31E-11 | 2.19E-09 | 2.97E-01 | 7.82E-02 | 2.34E-01 | 9.83E-03 |
| ENSG00000148488 | ST8SIA6 | 5.40E-01 | 2.78E-01 | 1.40E-10 | 5.85E-09 | 4.28E-01 | 4.71E-06 | 2.13E-01 | 5.75E-02 |
| ENSG00000087245 | MMP2 | 5.15E-01 | 2.31E-01 | 1.30E-09 | 3.78E-08 | 6.10E-01 | 6.50E-16 | 3.96E-01 | 1.43E-11 |
| ENSG00000075275 | CELSR1 | 5.31E-01 | 3.13E-01 | 3.09E-10 | 1.15E-08 | 4.45E-01 | 9.54E-07 | 2.80E-01 | 1.04E-04 |
| ENSG00000254726 | MEX3A | 3.76E-01 | 2.03E-01 | 1.95E-05 | 1.20E-04 | 3.99E-01 | 5.89E-05 | 3.33E-01 | 1.56E-07 |
| ENSG00000213886 | UBD | 4.87E-01 | 2.85E-01 | 1.26E-08 | 2.50E-07 | 4.36E-01 | 2.30E-06 | 3.46E-01 | 2.79E-08 |
| ENSG00000163359 | COL6A3 | 5.37E-01 | 2.28E-01 | 1.73E-10 | 7.04E-09 | 5.80E-01 | 7.80E-14 | 3.58E-01 | 5.19E-09 |
| ENSG00000134028 | ADAMDEC1 | 6.00E-01 | 3.35E-01 | 2.67E-13 | 3.45E-11 | 4.13E-01 | 1.88E-05 | 4.41E-01 | 4.54E-15 |
| ENSG00000132031 | MATN3 | 5.60E-01 | 2.85E-01 | 1.98E-11 | 1.16E-09 | 4.62E-01 | 1.77E-07 | 4.55E-01 | 2.79E-16 |
| ENSG00000196405 | EVL | 4.92E-01 | 2.24E-01 | 8.60E-09 | 1.81E-07 | 4.49E-01 | 6.53E-07 | 2.96E-01 | 1.63E-05 |
| ENSG00000143412 | ANXA9 | 4.35E-01 | 2.27E-01 | 5.37E-07 | 5.80E-06 | 4.58E-01 | 2.66E-07 | 2.58E-01 | 1.03E-03 |
| ENSG00000196569 | LAMA2 | 6.05E-01 | 3.17E-01 | 1.49E-13 | 2.26E-11 | 5.35E-01 | 4.18E-11 | 4.70E-01 | 1.32E-17 |
| ENSG00000148773 | MKI67 | 5.95E-01 | 2.97E-01 | 4.91E-13 | 5.57E-11 | 5.64E-01 | 7.94E-13 | 4.31E-01 | 2.89E-14 |
| ENSG00000039139 | DNAH5 | 4.37E-01 | 2.12E-01 | 4.80E-07 | 5.30E-06 | 4.30E-01 | 3.93E-06 | 2.12E-01 | 6.38E-02 |
| ENSG00000121753 | BAI2 | 4.11E-01 | 2.20E-01 | 2.51E-06 | 2.16E-05 | 3.85E-01 | 1.90E-04 | 3.09E-01 | 3.61E-06 |
| ENSG00000141338 | ABCA8 | 4.78E-01 | 2.18E-01 | 2.65E-08 | 4.68E-07 | 4.79E-01 | 3.07E-08 | 4.60E-01 | 9.07E-17 |
| ENSG00000067057 | PFKP | 5.39E-01 | 2.44E-01 | 1.53E-10 | 6.32E-09 | 4.52E-01 | 4.91E-07 | 3.40E-01 | 6.77E-08 |
| ENSG00000175063 | UBE2C | 6.04E-01 | 3.35E-01 | 1.84E-13 | 2.68E-11 | 5.93E-01 | 1.02E-14 | 4.52E-01 | 5.46E-16 |
| ENSG00000089685 | BIRC5 | 6.58E-01 | 3.32E-01 | 1.87E-16 | 1.12E-13 | 6.53E-01 | 2.71E-19 | 4.64E-01 | 4.62E-17 |
| ENSG00000162654 | GBP4 | 5.88E-01 | 2.31E-01 | 1.09E-12 | 1.06E-10 | 3.64E-01 | 9.20E-04 | 4.68E-01 | 2.06E-17 |
| ENSG00000100234 | TIMP3 | 4.80E-01 | 2.14E-01 | 2.25E-08 | 4.15E-07 | 3.78E-01 | 3.19E-04 | 3.12E-01 | 2.55E-06 |
| ENSG00000178935 | ZNF552 | 4.12E-01 | 2.66E-01 | 2.44E-06 | 2.11E-05 | 4.33E-01 | 3.07E-06 | 3.43E-01 | 4.24E-08 |
| ENSG00000152953 | STK32B | 5.81E-01 | 2.49E-01 | 2.21E-12 | 1.93E-10 | 4.56E-01 | 3.43E-07 | 4.55E-01 | 2.87E-16 |
| ENSG00000174348 | PODN | 6.66E-01 | 3.70E-01 | 5.71E-17 | 4.59E-14 | 6.25E-01 | 5.13E-17 | 4.95E-01 | 5.14E-20 |
| ENSG00000173210 | ABLIM3 | 4.49E-01 | 2.16E-01 | 2.10E-07 | 2.65E-06 | 3.51E-01 | 2.34E-03 | 3.95E-01 | 1.49E-11 |
| ENSG00000120332 | TNN | 5.33E-01 | 2.45E-01 | 2.51E-10 | 9.60E-09 | 3.90E-01 | 1.23E-04 | 4.72E-01 | 8.57E-18 |
| ENSG00000185133 | INPP5J | 4.28E-01 | 2.15E-01 | 8.75E-07 | 8.78E-06 | 4.26E-01 | 5.68E-06 | 2.78E-01 | 1.26E-04 |
| ENSG00000107742 | SPOCK2 | 5.41E-01 | 2.12E-01 | 1.19E-10 | 5.10E-09 | 4.71E-01 | 6.93E-08 | 4.10E-01 | 1.26E-12 |
| ENSG00000164626 | KCNK5 | 4.73E-01 | 2.05E-01 | 3.73E-08 | 6.16E-07 | 1.91E-01 | 1.00E+00 | 2.54E-01 | 1.51E-03 |
| ENSG00000078081 | LAMP3 | 6.10E-01 | 3.02E-01 | 9.11E-14 | 1.48E-11 | 5.06E-01 | 1.37E-09 | 4.44E-01 | 2.24E-15 |
| ENSG00000088325 | TPX2 | 6.05E-01 | 3.25E-01 | 1.61E-13 | 2.39E-11 | 6.09E-01 | 8.19E-16 | 5.72E-01 | 9.08E-29 |
| ENSG00000152377 | SPOCK1 | 5.10E-01 | 2.38E-01 | 2.05E-09 | 5.45E-08 | 4.97E-01 | 4.05E-09 | 4.97E-01 | 2.77E-20 |
| ENSG00000131781 | FMO5 | 4.91E-01 | 2.12E-01 | 9.53E-09 | 1.98E-07 | 3.43E-01 | 4.22E-03 | 3.38E-01 | 8.14E-08 |
| ENSG00000143341 | HMCN1 | 5.00E-01 | 2.38E-01 | 4.41E-09 | 1.04E-07 | 5.65E-01 | 6.86E-13 | 4.36E-01 | 1.10E-14 |
| ENSG00000112414 | GPR126 | 5.24E-01 | 2.36E-01 | 5.81E-10 | 1.92E-08 | 4.51E-01 | 5.65E-07 | 5.10E-01 | 1.60E-21 |
| ENSG00000171848 | RRM2 | 5.93E-01 | 2.97E-01 | 6.30E-13 | 6.88E-11 | 5.70E-01 | 3.31E-13 | 5.94E-01 | 1.05E-31 |
| ENSG00000081237 | PTPRC | 5.22E-01 | 2.05E-01 | 7.27E-10 | 2.34E-08 | 3.92E-01 | 1.10E-04 | 3.55E-01 | 7.69E-09 |
| ENSG00000162543 | UBXN10 | 4.90E-01 | 2.02E-01 | 1.00E-08 | 2.06E-07 | 4.95E-01 | 5.06E-09 | 4.13E-01 | 6.96E-13 |
| ENSG00000019169 | MARCO | 5.34E-01 | 3.48E-01 | 2.33E-10 | 9.08E-09 | 4.13E-01 | 1.84E-05 | 2.52E-01 | 1.79E-03 |
| ENSG00000119866 | BCL11A | 4.02E-01 | 2.41E-01 | 4.49E-06 | 3.48E-05 | 5.16E-01 | 4.53E-10 | 2.98E-01 | 1.31E-05 |
| ENSG00000122861 | PLAU | 5.81E-01 | 2.01E-01 | 2.35E-12 | 2.02E-10 | 4.12E-01 | 2.03E-05 | 4.02E-01 | 5.35E-12 |
| ENSG00000169047 | IRS1 | 5.01E-01 | 2.40E-01 | 4.18E-09 | 1.00E-07 | 2.96E-01 | 7.96E-02 | 2.63E-01 | 6.19E-04 |
| ENSG00000157388 | CACNA1D | 5.62E-01 | 2.96E-01 | 1.57E-11 | 9.51E-10 | 5.35E-01 | 4.16E-11 | 4.12E-01 | 8.35E-13 |
| ENSG00000011426 | ANLN | 5.58E-01 | 2.82E-01 | 2.50E-11 | 1.40E-09 | 4.60E-01 | 2.15E-07 | 5.32E-01 | 6.40E-24 |
| ENSG00000111206 | FOXM1 | 6.51E-01 | 4.32E-01 | 4.55E-16 | 2.06E-13 | 6.10E-01 | 6.32E-16 | 5.79E-01 | 9.64E-30 |

**Table S3. Prediction performance for the 1011 genes on validation and test sets.**

| Ensembl_gene_id | Hgnc symbol | Coef (valid) | Rsq (valid) | P-val (valid) | adj.pval (valid) | Coef (internal) | adj.pval (internal) | Coef (external) | adj.pval (external) |
|---|---|---|---|---|---|---|---|---|---|
| ENSG00000117399 | CDC20 | 6.36E-01 | 3.96E-01 | 3.68E-15 | 1.14E-12 | 6.38E-01 | 5.04E-18 | 4.61E-01 | 8.08E-17 |
| ENSG00000163072 | NOSTRIN | 5.95E-01 | 2.77E-01 | 5.15E-13 | 5.73E-11 | 5.17E-01 | 3.73E-10 | 4.46E-01 | 1.62E-15 |
| ENSG00000131831 | RAI2 | 5.67E-01 | 3.30E-01 | 1.00E-11 | 6.70E-10 | 4.24E-01 | 6.71E-06 | 5.37E-01 | 1.54E-24 |
| ENSG00000007402 | CACNA2D2 | 4.91E-01 | 2.73E-01 | 9.12E-09 | 1.91E-07 | 5.25E-01 | 1.52E-10 | 4.51E-01 | 6.30E-16 |
| ENSG00000148154 | UGCG | 5.74E-01 | 2.79E-01 | 4.76E-12 | 3.61E-10 | 4.61E-01 | 1.92E-07 | 3.69E-01 | 9.69E-10 |
| ENSG00000026751 | SLAMF7 | 5.52E-01 | 2.43E-01 | 4.48E-11 | 2.25E-09 | 4.69E-01 | 9.19E-08 | 4.18E-01 | 2.83E-13 |
| ENSG00000165124 | SVEP1 | 5.50E-01 | 2.62E-01 | 5.55E-11 | 2.68E-09 | 4.77E-01 | 3.58E-08 | 3.43E-01 | 4.01E-08 |
| ENSG00000166033 | HTRA1 | 5.18E-01 | 2.82E-01 | 1.03E-09 | 3.10E-08 | 5.99E-01 | 3.81E-15 | 4.90E-01 | 1.72E-19 |
| ENSG00000147168 | IL2RG | 5.99E-01 | 2.55E-01 | 3.28E-13 | 4.06E-11 | 4.56E-01 | 3.28E-07 | 3.68E-01 | 1.10E-09 |
| ENSG00000168394 | TAP1 | 6.21E-01 | 3.45E-01 | 2.20E-14 | 4.42E-12 | 3.31E-01 | 9.18E-03 | 4.33E-01 | 1.91E-14 |
| ENSG00000100167 | SEPT3 | 4.73E-01 | 2.57E-01 | 3.66E-08 | 6.10E-07 | 4.77E-01 | 3.61E-08 | 3.11E-01 | 2.75E-06 |
| ENSG00000127083 | OMD | 5.00E-01 | 2.41E-01 | 4.38E-09 | 1.04E-07 | 5.77E-01 | 1.19E-13 | 4.43E-01 | 2.98E-15 |
| ENSG00000066279 | ASPM | 5.15E-01 | 2.76E-01 | 1.30E-09 | 3.78E-08 | 5.40E-01 | 2.09E-11 | 5.16E-01 | 3.56E-22 |
| ENSG00000029993 | HMGB3 | 5.45E-01 | 2.61E-01 | 8.53E-11 | 3.86E-09 | 3.85E-01 | 1.82E-04 | 2.85E-01 | 6.03E-05 |
| ENSG00000094804 | CDC6 | 5.99E-01 | 2.34E-01 | 2.99E-13 | 3.79E-11 | 5.37E-01 | 2.97E-11 | 4.74E-01 | 4.93E-18 |
| ENSG00000140511 | HAPLN3 | 5.21E-01 | 2.20E-01 | 7.58E-10 | 2.43E-08 | 5.65E-01 | 7.38E-13 | 3.74E-01 | 4.79E-10 |
| ENSG00000144354 | CDCA7 | 6.46E-01 | 3.56E-01 | 9.62E-16 | 3.62E-13 | 5.04E-01 | 1.94E-09 | 3.87E-01 | 5.60E-11 |
| ENSG00000152583 | SPARCL1 | 6.18E-01 | 2.72E-01 | 3.57E-14 | 6.64E-12 | 5.77E-01 | 1.25E-13 | 5.54E-01 | 1.68E-26 |
| ENSG00000197892 | KIF13B | 6.04E-01 | 2.79E-01 | 1.71E-13 | 2.51E-11 | 4.75E-01 | 4.63E-08 | 4.15E-01 | 5.56E-13 |
| ENSG00000169604 | ANTXR1 | 5.07E-01 | 2.13E-01 | 2.56E-09 | 6.61E-08 | 5.16E-01 | 4.40E-10 | 3.78E-01 | 2.46E-10 |
| ENSG00000140937 | CDH11 | 4.66E-01 | 2.03E-01 | 6.51E-08 | 9.82E-07 | 5.75E-01 | 1.72E-13 | 3.86E-01 | 6.83E-11 |
| ENSG00000116824 | CD2 | 5.65E-01 | 2.55E-01 | 1.27E-11 | 8.14E-10 | 4.71E-01 | 7.27E-08 | 3.86E-01 | 6.80E-11 |
| ENSG00000102316 | MAGED2 | 4.77E-01 | 2.12E-01 | 2.77E-08 | 4.81E-07 | 4.01E-01 | 4.97E-05 | 2.59E-01 | 8.57E-04 |
| ENSG00000182175 | RGMA | 3.56E-01 | 2.06E-01 | 5.65E-05 | 2.98E-04 | 4.57E-01 | 3.09E-07 | 3.53E-01 | 1.05E-08 |
| ENSG00000128606 | LRRC17 | 6.53E-01 | 3.78E-01 | 3.58E-16 | 1.76E-13 | 6.04E-01 | 1.70E-15 | 5.26E-01 | 2.61E-23 |
| ENSG00000168329 | CX3CR1 | 6.90E-01 | 3.61E-01 | 1.41E-18 | 3.12E-15 | 4.64E-01 | 1.40E-07 | 4.75E-01 | 4.39E-18 |
| ENSG00000113140 | SPARC | 5.04E-01 | 2.21E-01 | 3.26E-09 | 8.06E-08 | 5.64E-01 | 7.98E-13 | 4.08E-01 | 1.65E-12 |
| ENSG00000106537 | TSPAN13 | 4.73E-01 | 2.17E-01 | 3.87E-08 | 6.39E-07 | 3.86E-01 | 1.68E-04 | 4.14E-01 | 6.34E-13 |
| ENSG00000162576 | MXRA8 | 4.67E-01 | 2.28E-01 | 5.87E-08 | 9.02E-07 | 4.34E-01 | 2.88E-06 | 4.39E-01 | 6.27E-15 |
| ENSG00000198851 | CD3E | 5.28E-01 | 2.19E-01 | 4.05E-10 | 1.44E-08 | 4.56E-01 | 3.21E-07 | 3.84E-01 | 9.78E-11 |
| ENSG00000177575 | CD163 | 4.94E-01 | 2.18E-01 | 7.28E-09 | 1.59E-07 | 1.54E-01 | 1.00E+00 | 2.12E-01 | 6.27E-02 |
| ENSG00000204642 | HLA-F | 4.99E-01 | 2.60E-01 | 4.87E-09 | 1.13E-07 | 3.03E-01 | 5.36E-02 | 3.17E-01 | 1.29E-06 |
| ENSG00000143452 | HORMAD1 | 3.41E-01 | 2.57E-01 | 1.23E-04 | 5.77E-04 | 3.13E-01 | 2.98E-02 | 1.92E-01 | 3.06E-01 |
| ENSG00000115415 | STAT1 | 5.84E-01 | 3.02E-01 | 1.74E-12 | 1.59E-10 | 2.87E-01 | 1.38E-01 | 4.30E-01 | 3.18E-14 |
| ENSG00000141753 | IGFBP4 | 4.67E-01 | 2.35E-01 | 5.81E-08 | 8.97E-07 | 4.33E-01 | 3.00E-06 | 4.17E-01 | 3.91E-13 |
| ENSG00000172986 | GXYLT2 | 5.76E-01 | 2.69E-01 | 3.93E-12 | 3.09E-10 | 5.72E-01 | 2.62E-13 | 4.01E-01 | 5.50E-12 |
| ENSG00000091490 | SEL1L3 | 5.03E-01 | 2.16E-01 | 3.53E-09 | 8.65E-08 | 4.89E-01 | 9.62E-09 | 3.62E-01 | 2.79E-09 |
| ENSG00000148468 | FAM171A1 | 3.40E-01 | 2.18E-01 | 1.29E-04 | 6.04E-04 | 4.38E-01 | 1.84E-06 | 3.34E-01 | 1.47E-07 |
| ENSG00000101447 | FAM83D | 6.09E-01 | 3.49E-01 | 9.90E-14 | 1.59E-11 | 5.66E-01 | 6.34E-13 | 4.69E-01 | 1.47E-17 |
| ENSG00000076382 | SPAG5 | 5.05E-01 | 2.26E-01 | 3.12E-09 | 7.81E-08 | 5.35E-01 | 4.31E-11 | 4.79E-01 | 1.85E-18 |
| ENSG00000234745 | HLA-B | 5.42E-01 | 2.39E-01 | 1.12E-10 | 4.85E-09 | 2.10E-01 | 1.00E+00 | 3.39E-01 | 7.13E-08 |
| ENSG00000145555 | MYO10 | 4.62E-01 | 2.07E-01 | 8.29E-08 | 1.20E-06 | 3.21E-01 | 1.77E-02 | 2.43E-01 | 4.26E-03 |
| ENSG00000131471 | AOC3 | 5.18E-01 | 2.19E-01 | 1.02E-09 | 3.09E-08 | 4.75E-01 | 4.63E-08 | 3.66E-01 | 1.44E-09 |
| ENSG00000183160 | TMEM119 | 4.53E-01 | 2.03E-01 | 1.62E-07 | 2.12E-06 | 4.88E-01 | 1.16E-08 | 4.09E-01 | 1.47E-12 |
| ENSG00000026559 | KCNG1 | 4.19E-01 | 2.37E-01 | 1.58E-06 | 1.46E-05 | 3.37E-01 | 6.19E-03 | 2.92E-01 | 2.51E-05 |
| ENSG00000131016 | AKAP12 | 5.08E-01 | 2.39E-01 | 2.35E-09 | 6.14E-08 | 5.02E-01 | 2.31E-09 | 5.89E-01 | 4.56E-31 |
| ENSG00000160161 | CILP2 | 4.76E-01 | 2.13E-01 | 3.09E-08 | 5.27E-07 | 5.08E-01 | 1.16E-09 | 3.82E-01 | 1.29E-10 |
| ENSG00000134830 | C5AR2 | 4.35E-01 | 2.13E-01 | 5.48E-07 | 5.91E-06 | 4.46E-01 | 8.76E-07 | 3.56E-01 | 6.69E-09 |
| ENSG00000156103 | MMP16 | 4.99E-01 | 2.15E-01 | 4.94E-09 | 1.15E-07 | 3.66E-01 | 8.28E-04 | 3.57E-01 | 5.56E-09 |
| ENSG00000142945 | KIF2C | 6.60E-01 | 4.24E-01 | 1.36E-16 | 9.40E-14 | 6.42E-01 | 2.24E-18 | 5.11E-01 | 1.04E-21 |
| ENSG00000130005 | GAMT | 4.64E-01 | 2.16E-01 | 7.41E-08 | 1.10E-06 | 4.06E-01 | 3.36E-05 | 2.75E-01 | 1.73E-04 |
| ENSG00000119681 | LTBP2 | 5.10E-01 | 2.45E-01 | 1.97E-09 | 5.28E-08 | 4.77E-01 | 3.85E-08 | 4.70E-01 | 1.30E-17 |
| ENSG00000165304 | MELK | 6.20E-01 | 3.60E-01 | 2.63E-14 | 5.12E-12 | 5.82E-01 | 5.81E-14 | 4.31E-01 | 2.98E-14 |
| ENSG00000177409 | SAMD9L | 4.86E-01 | 2.08E-01 | 1.35E-08 | 2.64E-07 | 1.51E-01 | 1.00E+00 | 2.05E-01 | 1.15E-01 |
| ENSG00000068489 | PRR11 | 6.36E-01 | 3.06E-01 | 3.43E-15 | 1.12E-12 | 6.28E-01 | 2.92E-17 | 5.49E-01 | 6.83E-26 |
| ENSG00000125869 | LAMP5 | 6.00E-01 | 3.18E-01 | 2.70E-13 | 3.45E-11 | 4.89E-01 | 1.02E-08 | 5.58E-01 | 4.31E-27 |
| ENSG00000136546 | SCN7A | 6.75E-01 | 4.09E-01 | 1.42E-17 | 1.71E-14 | 4.68E-01 | 9.24E-08 | 4.82E-01 | 8.70E-19 |
| ENSG00000182481 | KPNA2 | 5.45E-01 | 2.37E-01 | 8.37E-11 | 3.82E-09 | 6.19E-01 | 1.53E-16 | 5.63E-01 | 1.25E-27 |
| ENSG00000117215 | PLA2G2D | 5.39E-01 | 2.52E-01 | 1.55E-10 | 6.38E-09 | 4.24E-01 | 7.17E-06 | 3.25E-01 | 4.59E-07 |
| ENSG00000187474 | FPR3 | 6.02E-01 | 2.66E-01 | 2.28E-13 | 3.10E-11 | 3.68E-01 | 7.14E-04 | 4.65E-01 | 3.43E-17 |
| ENSG00000081985 | IL12RB2 | 6.26E-01 | 3.12E-01 | 1.21E-14 | 2.81E-12 | 4.96E-01 | 4.78E-09 | 3.45E-01 | 3.18E-08 |
| ENSG00000128510 | CPA4 | 4.57E-01 | 2.40E-01 | 1.23E-07 | 1.68E-06 | 3.51E-01 | 2.33E-03 | 3.12E-01 | 2.30E-06 |
| ENSG00000170312 | CDK1 | 6.26E-01 | 3.45E-01 | 1.19E-14 | 2.81E-12 | 5.64E-01 | 8.49E-13 | 5.18E-01 | 1.84E-22 |
| ENSG00000119699 | TGFB3 | 5.96E-01 | 3.38E-01 | 4.32E-13 | 5.03E-11 | 4.95E-01 | 5.27E-09 | 3.98E-01 | 9.71E-12 |
| ENSG00000109861 | CTSC | 6.50E-01 | 3.21E-01 | 5.59E-16 | 2.47E-13 | 3.91E-01 | 1.11E-04 | 3.69E-01 | 1.04E-09 |
| ENSG00000146072 | TNFRSF21 | 4.97E-01 | 2.65E-01 | 5.59E-09 | 1.28E-07 | 2.50E-01 | 9.58E-01 | 3.42E-01 | 4.92E-08 |
| ENSG00000166851 | PLK1 | 6.34E-01 | 3.64E-01 | 4.42E-15 | 1.30E-12 | 6.15E-01 | 3.09E-16 | 5.32E-01 | 6.33E-24 |
| ENSG00000065328 | MCM10 | 7.07E-01 | 4.49E-01 | 8.27E-20 | 2.93E-16 | 6.27E-01 | 3.29E-17 | 6.13E-01 | 1.70E-34 |
| ENSG00000116678 | LEPR | 4.76E-01 | 2.03E-01 | 3.07E-08 | 5.26E-07 | 3.34E-01 | 7.86E-03 | 4.30E-01 | 3.78E-14 |

**Table S3. Prediction performance for the 1011 genes on validation and test sets.**

| Ensembl_gene_id | Hgnc symbol | Coef (valid) | Rsq (valid) | P-val (valid) | adj.pval (valid) | Coef (internal) | adj.pval (internal) | Coef (external) | adj.pval (external) |
|---|---|---|---|---|---|---|---|---|---|
| ENSG00000169851 | PCDH7 | 5.05E-01 | 2.50E-01 | 3.01E-09 | 7.61E-08 | 5.26E-01 | 1.31E-10 | 3.42E-01 | 4.72E-08 |
| ENSG00000153563 | CD8A | 5.18E-01 | 2.19E-01 | 1.00E-09 | 3.03E-08 | 4.91E-01 | 8.19E-09 | 3.65E-01 | 1.69E-09 |
| ENSG00000240065 | PSMB9 | 5.60E-01 | 2.63E-01 | 2.09E-11 | 1.20E-09 | 2.69E-01 | 3.63E-01 | 3.18E-01 | 1.22E-06 |
| ENSG00000138180 | CEP55 | 6.72E-01 | 4.40E-01 | 2.23E-17 | 2.32E-14 | 5.89E-01 | 1.90E-14 | 5.70E-01 | 1.47E-28 |
| ENSG00000163683 | SMIM14 | 5.26E-01 | 3.19E-01 | 4.88E-10 | 1.67E-08 | 3.51E-01 | 2.45E-03 | 1.88E-01 | 4.16E-01 |
| ENSG00000168497 | SDPR | 5.24E-01 | 2.59E-01 | 5.72E-10 | 1.89E-08 | 5.32E-01 | 5.86E-11 | 3.97E-01 | 1.24E-11 |
| ENSG00000163131 | CTSS | 5.49E-01 | 2.46E-01 | 6.04E-11 | 2.88E-09 | 3.15E-01 | 2.59E-02 | 3.10E-01 | 2.94E-06 |
| ENSG00000127663 | KDM4B | 4.86E-01 | 2.24E-01 | 1.39E-08 | 2.72E-07 | 4.09E-01 | 2.51E-05 | 3.90E-01 | 3.82E-11 |
| ENSG00000160255 | ITGB2 | 4.79E-01 | 2.06E-01 | 2.38E-08 | 4.30E-07 | 2.46E-01 | 1.00E+00 | 2.54E-01 | 1.49E-03 |
| ENSG00000004468 | CD38 | 6.04E-01 | 3.17E-01 | 1.86E-13 | 2.68E-11 | 5.29E-01 | 9.20E-11 | 5.32E-01 | 5.99E-24 |
| ENSG00000095637 | SORBS1 | 5.92E-01 | 2.57E-01 | 6.68E-13 | 7.21E-11 | 4.14E-01 | 1.69E-05 | 4.00E-01 | 6.69E-12 |
| ENSG00000181847 | TIGIT | 5.76E-01 | 2.47E-01 | 3.83E-12 | 3.06E-10 | 4.99E-01 | 3.13E-09 | 3.99E-01 | 7.83E-12 |
| ENSG00000123384 | LRP1 | 5.11E-01 | 2.40E-01 | 1.80E-09 | 4.90E-08 | 4.83E-01 | 1.87E-08 | 4.48E-01 | 1.11E-15 |
| ENSG00000090889 | KIF4A | 6.64E-01 | 3.46E-01 | 7.80E-17 | 5.75E-14 | 6.01E-01 | 3.03E-15 | 5.12E-01 | 9.16E-22 |
| ENSG00000179954 | SSC5D | 4.75E-01 | 2.30E-01 | 3.34E-08 | 5.63E-07 | 4.25E-01 | 6.25E-06 | 3.98E-01 | 9.90E-12 |
| ENSG00000198901 | PRC1 | 5.96E-01 | 3.00E-01 | 4.47E-13 | 5.17E-11 | 5.11E-01 | 7.72E-10 | 4.59E-01 | 1.34E-16 |
| ENSG00000189057 | FAM111B | 5.03E-01 | 2.19E-01 | 3.66E-09 | 8.93E-08 | 4.34E-01 | 2.74E-06 | 3.47E-01 | 2.46E-08 |
| ENSG00000187486 | KCNJ11 | 4.34E-01 | 2.05E-01 | 5.74E-07 | 6.13E-06 | 4.65E-01 | 1.35E-07 | 3.02E-01 | 8.49E-06 |
| ENSG00000024526 | DEPDC1 | 6.12E-01 | 3.72E-01 | 6.97E-14 | 1.14E-11 | 5.01E-01 | 2.52E-09 | 5.24E-01 | 3.95E-23 |
| ENSG00000050165 | DKK3 | 4.72E-01 | 2.06E-01 | 4.03E-08 | 6.60E-07 | 4.63E-01 | 1.61E-07 | 4.23E-01 | 1.33E-13 |
| ENSG00000105173 | CCNE1 | 6.50E-01 | 3.03E-01 | 5.83E-16 | 2.52E-13 | 5.69E-01 | 3.77E-13 | 4.95E-01 | 4.54E-20 |
| ENSG00000237649 | KIFC1 | 5.31E-01 | 2.80E-01 | 3.24E-10 | 1.19E-08 | 5.66E-01 | 6.14E-13 | 4.36E-01 | 1.09E-14 |
| ENSG00000143228 | NUF2 | 5.69E-01 | 3.15E-01 | 8.08E-12 | 5.63E-10 | 5.34E-01 | 4.87E-11 | 5.31E-01 | 7.40E-24 |
| ENSG00000143297 | FCRL5 | 5.23E-01 | 2.24E-01 | 6.58E-10 | 2.14E-08 | 4.76E-01 | 4.01E-08 | 4.00E-01 | 7.43E-12 |
| ENSG00000169679 | BUB1 | 6.15E-01 | 3.92E-01 | 4.57E-14 | 8.26E-12 | 5.89E-01 | 1.93E-14 | 5.83E-01 | 2.64E-30 |
| ENSG00000029153 | ARNTL2 | 4.22E-01 | 2.30E-01 | 1.32E-06 | 1.25E-05 | 3.70E-01 | 5.78E-04 | 3.09E-01 | 3.67E-06 |
| ENSG00000112984 | KIF20A | 5.63E-01 | 2.67E-01 | 1.48E-11 | 9.07E-10 | 5.40E-01 | 2.18E-11 | 4.60E-01 | 1.01E-16 |
| ENSG00000170624 | SGCD | 6.40E-01 | 3.25E-01 | 2.07E-15 | 7.17E-13 | 5.39E-01 | 2.32E-11 | 4.32E-01 | 2.19E-14 |
| ENSG00000118193 | KIF14 | 5.67E-01 | 3.28E-01 | 1.00E-11 | 6.70E-10 | 4.66E-01 | 1.16E-07 | 5.40E-01 | 7.07E-25 |
| ENSG00000131386 | GALNT15 | 5.25E-01 | 2.56E-01 | 5.37E-10 | 1.81E-08 | 5.20E-01 | 2.76E-10 | | |
| ENSG00000087586 | AURKA | 6.76E-01 | 4.09E-01 | 1.29E-17 | 1.71E-14 | 6.60E-01 | 6.50E-20 | 5.68E-01 | 3.14E-28 |
| ENSG00000183943 | PRKX | 5.45E-01 | 3.02E-01 | 8.36E-11 | 3.82E-09 | 4.95E-01 | 4.88E-09 | 3.41E-01 | 5.31E-08 |
| ENSG00000164687 | FABP5 | 3.29E-01 | 2.21E-01 | 2.14E-04 | 9.20E-04 | 3.55E-01 | 1.83E-03 | 1.82E-01 | 6.26E-01 |
| ENSG00000117091 | CD48 | 5.40E-01 | 2.07E-01 | 1.37E-10 | 5.80E-09 | 4.13E-01 | 1.88E-05 | 2.38E-01 | 6.71E-03 |
| ENSG00000236699 | ARHGEF38 | 4.79E-01 | 2.47E-01 | 2.30E-08 | 4.21E-07 | 4.65E-01 | 1.33E-07 | 2.22E-01 | 2.71E-02 |
| ENSG00000103569 | AQP9 | 6.00E-01 | 2.78E-01 | 2.67E-13 | 3.45E-11 | 3.38E-01 | 5.71E-03 | 4.81E-01 | 1.24E-18 |
| ENSG00000117650 | NEK2 | 4.77E-01 | 2.01E-01 | 2.72E-08 | 4.77E-07 | 3.82E-01 | 2.30E-04 | 4.19E-01 | 2.51E-13 |
| ENSG00000184985 | SORCS2 | 6.02E-01 | 2.81E-01 | 2.36E-13 | 3.18E-11 | 5.31E-01 | 6.49E-11 | 5.74E-01 | 5.27E-29 |
| ENSG00000102362 | SYTL4 | 4.50E-01 | 2.03E-01 | 1.94E-07 | 2.48E-06 | 4.02E-01 | 4.57E-05 | 2.52E-01 | 1.84E-03 |
| ENSG00000155886 | SLC24A2 | 6.75E-01 | 3.65E-01 | 1.45E-17 | 1.71E-14 | 5.80E-01 | 7.26E-14 | 5.40E-01 | 6.65E-25 |
| ENSG00000182866 | LCK | 5.58E-01 | 2.46E-01 | 2.51E-11 | 1.40E-09 | 5.46E-01 | 9.63E-12 | 3.84E-01 | 9.48E-11 |
| ENSG00000170959 | DCDC1 | 5.76E-01 | 2.07E-01 | 3.84E-12 | 3.06E-10 | 4.53E-01 | 4.37E-07 | 3.72E-01 | 6.60E-10 |
| ENSG00000143401 | ANP32E | 4.33E-01 | 2.20E-01 | 6.37E-07 | 6.71E-06 | 4.23E-01 | 7.21E-06 | 2.86E-01 | 5.01E-05 |
| ENSG00000157456 | CCNB2 | 6.53E-01 | 4.04E-01 | 3.47E-16 | 1.76E-13 | 6.22E-01 | 9.13E-17 | 5.19E-01 | 1.50E-22 |
| ENSG00000159674 | SPON2 | 5.09E-01 | 2.06E-01 | 2.23E-09 | 5.90E-08 | 3.66E-01 | 7.73E-04 | 3.27E-01 | 3.80E-07 |
| ENSG00000164574 | GALNT10 | 5.53E-01 | 2.95E-01 | 4.17E-11 | 2.13E-09 | 4.66E-01 | 1.22E-07 | 2.99E-01 | 1.20E-05 |
| ENSG00000146670 | CDCA5 | 6.46E-01 | 3.40E-01 | 9.12E-16 | 3.59E-13 | 6.18E-01 | 1.66E-16 | 4.20E-01 | 2.10E-13 |
| ENSG00000013297 | CLDN11 | 5.60E-01 | 2.50E-01 | 2.01E-11 | 1.17E-09 | 4.16E-01 | 1.35E-05 | 5.03E-01 | 7.50E-21 |
| ENSG00000101439 | CST3 | 4.27E-01 | 2.23E-01 | 9.41E-07 | 9.37E-06 | 4.49E-01 | 6.44E-07 | 3.55E-01 | 7.59E-09 |
| ENSG00000137804 | NUSAP1 | 5.76E-01 | 3.03E-01 | 3.86E-12 | 3.07E-10 | 5.66E-01 | 6.10E-13 | 4.91E-01 | 1.13E-19 |
| ENSG00000115956 | PLEK | 5.52E-01 | 2.31E-01 | 4.49E-11 | 2.25E-09 | 3.08E-01 | 3.88E-02 | 3.28E-01 | 3.30E-07 |
| ENSG00000007237 | GAS7 | 5.18E-01 | 2.42E-01 | 9.78E-10 | 2.97E-08 | 5.46E-01 | 1.00E-11 | 4.16E-01 | 4.29E-13 |
| ENSG00000111344 | RASAL1 | 3.95E-01 | 2.11E-01 | 6.65E-06 | 4.82E-05 | 3.14E-01 | 2.78E-02 | 3.39E-01 | 7.16E-08 |
| ENSG00000071539 | TRIP13 | 7.03E-01 | 4.07E-01 | 1.70E-19 | 5.03E-16 | 5.83E-01 | 5.09E-14 | 5.09E-01 | 1.99E-21 |
| ENSG00000174371 | EXO1 | 5.50E-01 | 2.68E-01 | 5.23E-11 | 2.56E-09 | 5.03E-01 | 1.95E-09 | 5.28E-01 | 1.65E-23 |
| ENSG00000010319 | SEMA3G | 5.04E-01 | 2.26E-01 | 3.14E-09 | 7.85E-08 | 5.46E-01 | 9.58E-12 | 4.99E-01 | 1.83E-20 |
| ENSG00000129946 | SHC2 | 5.14E-01 | 2.20E-01 | 1.41E-09 | 4.03E-08 | 4.14E-01 | 1.60E-05 | 2.36E-01 | 8.43E-03 |
| ENSG00000168675 | LDLRAD4 | 5.13E-01 | 2.30E-01 | 1.56E-09 | 4.38E-08 | 3.56E-01 | 1.61E-03 | 3.00E-01 | 9.85E-06 |
| ENSG00000140853 | NLRC5 | 6.01E-01 | 2.84E-01 | 2.49E-13 | 3.27E-11 | 2.86E-01 | 1.45E-01 | 4.20E-01 | 2.25E-13 |
| ENSG00000106780 | MEGF9 | 4.56E-01 | 2.26E-01 | 1.34E-07 | 1.80E-06 | 4.04E-01 | 4.07E-05 | 2.60E-01 | 8.52E-04 |
| ENSG00000005469 | CROT | 4.81E-01 | 2.01E-01 | 2.07E-08 | 3.87E-07 | 4.36E-01 | 2.18E-06 | 3.43E-01 | 4.09E-08 |
| ENSG00000099864 | PALM | 5.68E-01 | 2.84E-01 | 9.13E-12 | 6.24E-10 | 4.14E-01 | 1.69E-05 | 3.74E-01 | 4.39E-10 |
| ENSG00000112742 | TTK | 6.23E-01 | 3.69E-01 | 1.89E-14 | 3.93E-12 | 5.74E-01 | 1.83E-13 | 5.69E-01 | 2.00E-28 |
| ENSG00000126787 | DLGAP5 | 6.23E-01 | 3.99E-01 | 1.87E-14 | 3.93E-12 | 6.07E-01 | 1.09E-15 | 5.42E-01 | 3.67E-25 |
| ENSG00000138778 | CENPE | 6.44E-01 | 3.66E-01 | 1.17E-15 | 4.32E-13 | 5.79E-01 | 9.27E-14 | 4.22E-01 | 1.60E-13 |
| ENSG00000146070 | PLA2G7 | 7.13E-01 | 3.61E-01 | 3.28E-20 | 1.46E-16 | 4.38E-01 | 1.96E-06 | 3.87E-01 | 6.05E-11 |
| ENSG00000163431 | LMOD1 | 5.25E-01 | 2.18E-01 | 5.42E-10 | 1.82E-08 | 4.73E-01 | 5.68E-08 | 4.22E-01 | 1.61E-13 |
| ENSG00000153283 | CD96 | 5.35E-01 | 2.36E-01 | 2.20E-10 | 8.68E-09 | 5.19E-01 | 2.94E-10 | 4.29E-01 | 4.37E-14 |
| ENSG00000158714 | SLAMF8 | 6.24E-01 | 3.18E-01 | 1.69E-14 | 3.74E-12 | 4.52E-01 | 4.88E-07 | 4.29E-01 | 3.89E-14 |

**Table S3. Prediction performance for the 1011 genes on validation and test sets.**

| Ensembl_gene_id | Hgnc symbol | Coef (valid) | Rsq (valid) | P-val (valid) | adj.pval (valid) | Coef (internal) | adj.pval (internal) | Coef (external) | adj.pval (external) |
|---|---|---|---|---|---|---|---|---|---|
| ENSG00000020633 | RUNX3 | 5.23E-01 | 2.48E-01 | 6.16E-10 | 2.02E-08 | 4.75E-01 | 4.54E-08 | 3.51E-01 | 1.45E-08 |
| ENSG00000144749 | LRIG1 | 4.42E-01 | 2.88E-01 | 3.35E-07 | 3.96E-06 | 5.21E-01 | 2.33E-10 | 3.97E-01 | 1.09E-11 |
| ENSG00000174099 | MSRB3 | 5.31E-01 | 2.56E-01 | 3.23E-10 | 1.19E-08 | 5.94E-01 | 8.41E-15 | 4.81E-01 | 1.14E-18 |
| ENSG00000101463 | SYNDIG1 | 5.33E-01 | 2.16E-01 | 2.58E-10 | 9.79E-09 | 5.31E-01 | 6.44E-11 | 5.15E-01 | 4.19E-22 |
| ENSG00000104951 | IL4I1 | 5.78E-01 | 3.24E-01 | 3.06E-12 | 2.58E-10 | 3.80E-01 | 2.83E-04 | 4.17E-01 | 3.63E-13 |
| ENSG00000100453 | GZMB | 6.22E-01 | 3.83E-01 | 1.95E-14 | 3.99E-12 | 5.09E-01 | 1.08E-09 | 4.49E-01 | 9.92E-16 |
| ENSG00000143119 | CD53 | 5.37E-01 | 2.26E-01 | 1.82E-10 | 7.32E-09 | 3.23E-01 | 1.52E-02 | 3.28E-01 | 3.39E-07 |
| ENSG00000101224 | CDC25B | 6.06E-01 | 3.16E-01 | 1.37E-13 | 2.11E-11 | 5.33E-01 | 5.44E-11 | 3.78E-01 | 2.50E-10 |
| ENSG00000171451 | DSEL | 4.89E-01 | 2.02E-01 | 1.13E-08 | 2.28E-07 | 3.81E-01 | 2.48E-04 | 3.64E-01 | 2.00E-09 |
| ENSG00000145386 | CCNA2 | 5.73E-01 | 3.44E-01 | 5.19E-12 | 3.89E-10 | 5.80E-01 | 8.14E-14 | 5.32E-01 | 5.45E-24 |
| ENSG00000164611 | PTTG1 | 5.98E-01 | 3.46E-01 | 3.36E-13 | 4.10E-11 | 5.30E-01 | 7.74E-11 | 3.85E-01 | 7.71E-11 |
| ENSG00000137309 | HMGA1 | 5.30E-01 | 2.84E-01 | 3.33E-10 | 1.22E-08 | 4.99E-01 | 3.19E-09 | 3.42E-01 | 4.73E-08 |
| ENSG00000073910 | FRY | 5.10E-01 | 2.62E-01 | 1.96E-09 | 5.24E-08 | 5.29E-01 | 8.73E-11 | 3.98E-01 | 9.83E-12 |
| ENSG00000113721 | PDGFRB | 4.80E-01 | 2.40E-01 | 2.27E-08 | 4.16E-07 | 5.18E-01 | 3.53E-10 | 4.21E-01 | 1.70E-13 |
| ENSG00000134057 | CCNB1 | 5.29E-01 | 2.67E-01 | 3.83E-10 | 1.37E-08 | 5.08E-01 | 1.17E-09 | 4.09E-01 | 1.57E-12 |
| ENSG00000129038 | LOXL1 | 5.10E-01 | 2.35E-01 | 1.98E-09 | 5.28E-08 | 5.18E-01 | 3.52E-10 | 4.66E-01 | 3.08E-17 |
| ENSG00000164056 | SPRY1 | 4.77E-01 | 2.26E-01 | 2.86E-08 | 4.95E-07 | 5.20E-01 | 2.61E-10 | 3.34E-01 | 1.36E-07 |
| ENSG00000149582 | TMEM25 | 4.29E-01 | 2.07E-01 | 8.02E-07 | 8.18E-06 | 5.17E-01 | 4.02E-10 | 3.58E-01 | 4.98E-09 |
| ENSG00000178999 | AURKB | 6.52E-01 | 4.57E-01 | 4.36E-16 | 2.03E-13 | 6.12E-01 | 5.05E-16 | 4.72E-01 | 8.06E-18 |
| ENSG00000169515 | CCDC8 | 6.12E-01 | 2.67E-01 | 6.88E-14 | 1.14E-11 | 4.27E-01 | 5.26E-06 | 4.17E-01 | 3.41E-13 |
| ENSG00000080986 | NDC80 | 6.05E-01 | 3.32E-01 | 1.56E-13 | 2.33E-11 | 6.42E-01 | 2.39E-18 | 4.86E-01 | 3.67E-19 |
| ENSG00000184661 | CDCA2 | 5.78E-01 | 3.18E-01 | 3.15E-12 | 2.63E-10 | 5.21E-01 | 2.28E-10 | 3.46E-01 | 2.79E-08 |
| ENSG00000170962 | PDGFD | 4.96E-01 | 2.08E-01 | 6.07E-09 | 1.37E-07 | 4.30E-01 | 3.99E-06 | 4.11E-01 | 9.86E-13 |
| ENSG00000156970 | BUB1B | 6.33E-01 | 3.52E-01 | 5.06E-15 | 1.47E-12 | 5.59E-01 | 1.58E-12 | 5.03E-01 | 7.59E-21 |
| ENSG00000124942 | AHNAK | 5.60E-01 | 3.51E-01 | 2.04E-11 | 1.18E-09 | 3.23E-01 | 1.60E-02 | 3.72E-01 | 6.15E-10 |
| ENSG00000134690 | CDCA8 | 6.35E-01 | 3.55E-01 | 4.18E-15 | 1.28E-12 | 6.32E-01 | 1.59E-17 | 5.31E-01 | 7.07E-24 |
| ENSG00000108700 | CCL8 | 5.20E-01 | 2.16E-01 | 8.60E-10 | 2.68E-08 | 2.30E-01 | 1.00E+00 | 3.86E-01 | 6.62E-11 |
| ENSG00000105374 | NKG7 | 5.70E-01 | 2.33E-01 | 7.21E-12 | 5.12E-10 | 4.81E-01 | 2.45E-08 | 3.94E-01 | 1.77E-11 |
| ENSG00000138160 | KIF11 | 5.55E-01 | 3.22E-01 | 3.22E-11 | 1.72E-09 | 5.41E-01 | 1.78E-11 | 4.69E-01 | 1.41E-17 |
| ENSG00000109805 | NCAPG | 6.31E-01 | 3.53E-01 | 6.79E-15 | 1.79E-12 | 6.32E-01 | 1.46E-17 | 5.81E-01 | 4.91E-30 |
| ENSG00000101412 | E2F1 | 5.92E-01 | 3.01E-01 | 6.85E-13 | 7.33E-11 | 5.06E-01 | 1.37E-09 | 3.59E-01 | 4.53E-09 |
| ENSG00000091656 | ZFHX4 | 5.35E-01 | 2.95E-01 | 2.28E-10 | 8.90E-09 | 4.35E-01 | 2.60E-06 | 4.54E-01 | 3.61E-16 |
| ENSG00000140379 | BCL2A1 | 5.87E-01 | 2.75E-01 | 1.13E-12 | 1.09E-10 | 3.67E-01 | 7.28E-04 | 4.74E-01 | 5.20E-18 |
| ENSG00000157193 | LRP8 | 5.91E-01 | 2.88E-01 | 7.86E-13 | 8.14E-11 | 5.99E-01 | 4.19E-15 | 5.60E-01 | 2.47E-27 |
| ENSG00000160791 | CCR5 | 5.75E-01 | 2.36E-01 | 4.24E-12 | 3.30E-10 | 3.29E-01 | 1.08E-02 | 4.33E-01 | 2.09E-14 |
| ENSG00000123485 | HJURP | 5.82E-01 | 3.64E-01 | 1.97E-12 | 1.75E-10 | 5.97E-01 | 5.52E-15 | 4.59E-01 | 1.34E-16 |
| ENSG00000107551 | RASSF4 | 5.78E-01 | 2.74E-01 | 3.15E-12 | 2.63E-10 | 3.37E-01 | 6.04E-03 | 4.43E-01 | 2.92E-15 |
| ENSG00000172216 | CEBPB | 4.84E-01 | 2.15E-01 | 1.66E-08 | 3.16E-07 | 3.69E-01 | 6.28E-04 | 2.40E-01 | 5.60E-03 |
| ENSG00000162739 | SLAMF6 | 5.47E-01 | 2.28E-01 | 7.21E-11 | 3.36E-09 | 4.77E-01 | 3.93E-08 | 3.70E-01 | 8.71E-10 |
| ENSG00000137807 | KIF23 | 6.31E-01 | 3.31E-01 | 6.65E-15 | 1.78E-12 | 5.00E-01 | 2.83E-09 | 5.03E-01 | 6.94E-21 |
| ENSG00000242574 | HLA-DMB | 5.28E-01 | 2.02E-01 | 4.17E-10 | 1.47E-08 | 2.61E-01 | 5.43E-01 | 2.21E-01 | 3.12E-02 |
| ENSG00000204267 | TAP2 | 6.42E-01 | 3.86E-01 | 1.64E-15 | 5.81E-13 | 3.86E-01 | 1.75E-04 | 5.00E-01 | 1.52E-20 |
| ENSG00000069482 | GAL | 3.34E-01 | 2.76E-01 | 1.68E-04 | 7.52E-04 | 4.08E-01 | 2.77E-05 | 3.39E-01 | 6.95E-08 |
| ENSG00000128591 | FLNC | 5.66E-01 | 2.22E-01 | 1.05E-11 | 6.95E-10 | 5.21E-01 | 2.25E-10 | 4.52E-01 | 4.66E-16 |
| ENSG00000113368 | LMNB1 | 4.94E-01 | 2.72E-01 | 7.62E-09 | 1.65E-07 | 6.01E-01 | 3.12E-15 | 3.48E-01 | 2.21E-08 |
| ENSG00000166710 | B2M | 5.15E-01 | 2.24E-01 | 1.29E-09 | 3.77E-08 | 6.70E-02 | 1.00E+00 | 3.39E-01 | 7.56E-08 |
| ENSG00000102096 | PIM2 | 5.35E-01 | 2.14E-01 | 2.17E-10 | 8.58E-09 | 4.23E-01 | 7.28E-06 | 3.73E-01 | 4.96E-10 |
| ENSG00000167286 | CD3D | 5.16E-01 | 2.14E-01 | 1.15E-09 | 3.43E-08 | 4.90E-01 | 8.77E-09 | 3.55E-01 | 7.29E-09 |
| ENSG00000104738 | MCM4 | 4.72E-01 | 2.38E-01 | 4.08E-08 | 6.65E-07 | 4.86E-01 | 1.34E-08 | 2.62E-01 | 6.47E-04 |
| ENSG00000186818 | LILRB4 | 4.87E-01 | 2.03E-01 | 1.25E-08 | 2.49E-07 | 2.87E-01 | 1.39E-01 | 4.09E-01 | 1.61E-12 |
| ENSG00000011422 | PLAUR | 4.41E-01 | 2.08E-01 | 3.66E-07 | 4.23E-06 | 3.46E-01 | 3.44E-03 | 4.15E-01 | 5.11E-13 |
| ENSG00000140105 | WARS | 6.46E-01 | 3.52E-01 | 9.40E-16 | 3.62E-13 | 4.54E-01 | 4.15E-07 | 4.41E-01 | 4.65E-15 |
| ENSG00000160213 | CSTB | 5.38E-01 | 2.82E-01 | 1.67E-10 | 6.81E-09 | 2.97E-01 | 7.61E-02 | 1.96E-01 | 2.31E-01 |
| ENSG00000196159 | FAT4 | 4.78E-01 | 2.17E-01 | 2.57E-08 | 4.57E-07 | 3.82E-01 | 2.36E-04 | 4.67E-01 | 2.37E-17 |
| ENSG00000165424 | ZCCHC24 | 6.79E-01 | 3.61E-01 | 7.70E-18 | 1.36E-14 | 6.05E-01 | 1.57E-15 | 6.36E-01 | 4.48E-38 |
| ENSG00000077152 | UBE2T | 5.01E-01 | 2.60E-01 | 4.31E-09 | 1.03E-07 | 5.07E-01 | 1.24E-09 | 5.01E-01 | 1.11E-20 |
| ENSG00000135476 | ESPL1 | 5.01E-01 | 2.29E-01 | 4.16E-09 | 1.00E-07 | 5.51E-01 | 4.61E-12 | 4.63E-01 | 5.63E-17 |
| ENSG00000173068 | BNC2 | 5.86E-01 | 2.92E-01 | 1.26E-12 | 1.21E-10 | 5.82E-01 | 5.65E-14 | 4.37E-01 | 9.07E-15 |
| ENSG00000120820 | GLT8D2 | 4.90E-01 | 2.26E-01 | 1.03E-08 | 2.10E-07 | 5.67E-01 | 5.32E-13 | 3.63E-01 | 2.57E-09 |
| ENSG00000105639 | JAK3 | 5.29E-01 | 2.09E-01 | 3.66E-10 | 1.32E-08 | 4.72E-01 | 6.64E-08 | 3.60E-01 | 3.62E-09 |
| ENSG00000198826 | ARHGAP11A | 6.06E-01 | 3.41E-01 | 1.33E-13 | 2.07E-11 | 5.63E-01 | 9.84E-13 | 4.77E-01 | 2.88E-18 |
| ENSG00000172243 | CLEC7A | 4.79E-01 | 2.23E-01 | 2.39E-08 | 4.33E-07 | 3.50E-01 | 2.63E-03 | 2.99E-01 | 1.21E-05 |
| ENSG00000198759 | EGFL6 | 5.71E-01 | 2.32E-01 | 6.31E-12 | 4.60E-10 | 3.09E-01 | 3.66E-02 | 3.55E-01 | 7.68E-09 |
| ENSG00000111640 | GAPDH | 4.96E-01 | 2.58E-01 | 6.31E-09 | 1.41E-07 | 2.33E-01 | 1.00E+00 | 3.96E-01 | 1.34E-11 |
| ENSG00000167771 | RCOR2 | 4.60E-01 | 2.35E-01 | 9.87E-08 | 1.39E-06 | 3.25E-01 | 1.35E-02 | 2.74E-01 | 1.96E-04 |
| ENSG00000187240 | DYNC2H1 | 5.15E-01 | 2.41E-01 | 1.32E-09 | 3.83E-08 | 4.37E-01 | 2.08E-06 | 3.71E-01 | 7.08E-10 |
| ENSG00000137507 | LRRC32 | 5.90E-01 | 3.15E-01 | 8.15E-13 | 8.34E-11 | 6.02E-01 | 2.33E-15 | 4.37E-01 | 9.05E-15 |
| ENSG00000203760 | CENPW | 5.28E-01 | 2.23E-01 | 4.25E-10 | 1.50E-08 | 5.16E-01 | 4.25E-10 | 3.61E-01 | 3.31E-09 |
| ENSG00000151790 | TDO2 | 5.98E-01 | 2.96E-01 | 3.45E-13 | 4.13E-11 | 3.43E-01 | 4.17E-03 | 4.93E-01 | 7.14E-20 |

**Table S3. Prediction performance for the 1011 genes on validation and test sets.**

| Ensembl_gene_id | Hgnc symbol | Coef (valid) | Rsq (valid) | P-val (valid) | adj.pval (valid) | Coef (internal) | adj.pval (internal) | Coef (external) | adj.pval (external) |
|---|---|---|---|---|---|---|---|---|---|
| ENSG00000163637 | PRICKLE2 | 4.64E-01 | 2.58E-01 | 7.53E-08 | 1.11E-06 | 5.60E-01 | 1.43E-12 | 4.11E-01 | 1.01E-12 |
| ENSG00000135451 | TROAP | 5.54E-01 | 2.90E-01 | 3.75E-11 | 1.94E-09 | 6.12E-01 | 4.82E-16 | 4.89E-01 | 2.07E-19 |
| ENSG00000241644 | INMT | 4.80E-01 | 2.15E-01 | 2.17E-08 | 4.02E-07 | 5.14E-01 | 5.32E-10 | 3.85E-01 | 8.71E-11 |
| ENSG00000174804 | FZD4 | 4.86E-01 | 2.03E-01 | 1.41E-08 | 2.75E-07 | 3.14E-01 | 2.75E-02 | 3.51E-01 | 1.45E-08 |
| ENSG00000111665 | CDCA3 | 6.08E-01 | 4.15E-01 | 1.12E-13 | 1.78E-11 | 5.70E-01 | 3.18E-13 | 4.79E-01 | 1.96E-18 |
| ENSG00000163697 | APBB2 | 5.57E-01 | 2.97E-01 | 2.69E-11 | 1.48E-09 | 4.16E-01 | 1.42E-05 | 3.30E-01 | 2.58E-07 |
| ENSG00000154330 | PGM5 | 6.28E-01 | 3.17E-01 | 9.95E-15 | 2.48E-12 | 5.21E-01 | 2.49E-10 | 4.86E-01 | 3.82E-19 |
| ENSG00000122952 | ZWINT | 4.74E-01 | 2.10E-01 | 3.59E-08 | 6.01E-07 | 5.22E-01 | 2.16E-10 | 3.58E-01 | 4.80E-09 |
| ENSG00000172037 | LAMB2 | 5.05E-01 | 3.16E-01 | 3.01E-09 | 7.61E-08 | 4.95E-01 | 4.94E-09 | 4.49E-01 | 9.79E-16 |
| ENSG00000121152 | NCAPH | 6.59E-01 | 4.04E-01 | 1.49E-16 | 9.74E-14 | 6.18E-01 | 1.69E-16 | 6.10E-01 | 4.44E-34 |
| ENSG00000166278 | C2 | 5.68E-01 | 2.17E-01 | 9.33E-12 | 6.33E-10 | 1.88E-01 | 1.00E+00 | 3.44E-01 | 3.68E-08 |
| ENSG00000178467 | P4HTM | 5.15E-01 | 2.04E-01 | 1.31E-09 | 3.81E-08 | 4.74E-01 | 5.33E-08 | 3.95E-01 | 1.68E-11 |
| ENSG00000164109 | MAD2L1 | 6.17E-01 | 3.43E-01 | 3.82E-14 | 6.96E-12 | 5.36E-01 | 3.37E-11 | 4.89E-01 | 1.73E-19 |
| ENSG00000104213 | PDGFRL | 5.40E-01 | 2.61E-01 | 1.39E-10 | 5.85E-09 | 5.41E-01 | 1.86E-11 | 4.49E-01 | 9.23E-16 |
| ENSG00000184113 | CLDN5 | 5.54E-01 | 2.62E-01 | 3.46E-11 | 1.82E-09 | 4.78E-01 | 3.42E-08 | 4.64E-01 | 4.08E-17 |
| ENSG00000175305 | CCNE2 | 4.97E-01 | 2.53E-01 | 5.76E-09 | 1.31E-07 | 4.49E-01 | 6.40E-07 | 4.64E-01 | 4.45E-17 |
| ENSG00000125347 | IRF1 | 5.57E-01 | 2.38E-01 | 2.59E-11 | 1.44E-09 | 3.19E-01 | 2.07E-02 | 2.77E-01 | 1.34E-04 |
| ENSG00000115523 | GNLY | 6.26E-01 | 2.87E-01 | 1.28E-14 | 2.95E-12 | 3.90E-01 | 1.29E-04 | 3.49E-01 | 1.76E-08 |
| ENSG00000024422 | EHD2 | 3.97E-01 | 2.25E-01 | 6.14E-06 | 4.49E-05 | 5.16E-01 | 4.34E-10 | 4.37E-01 | 9.60E-15 |
| ENSG00000168268 | NT5DC2 | 4.16E-01 | 2.16E-01 | 1.85E-06 | 1.66E-05 | 3.37E-01 | 6.20E-03 | 1.98E-01 | 1.86E-01 |
| ENSG00000072571 | HMMR | 5.42E-01 | 2.65E-01 | 1.10E-10 | 4.76E-09 | 4.94E-01 | 6.01E-09 | 4.69E-01 | 1.51E-17 |
| ENSG00000073111 | MCM2 | 5.15E-01 | 2.15E-01 | 1.33E-09 | 3.85E-08 | 5.23E-01 | 1.91E-10 | 2.39E-01 | 6.35E-03 |
| ENSG00000167513 | CDT1 | 6.00E-01 | 2.83E-01 | 2.71E-13 | 3.45E-11 | 5.63E-01 | 8.58E-13 | 3.11E-01 | 2.79E-06 |
| ENSG00000180644 | PRF1 | 6.20E-01 | 2.89E-01 | 2.55E-14 | 5.02E-12 | 5.06E-01 | 1.52E-09 | 4.39E-01 | 6.29E-15 |
| ENSG00000120254 | MTHFD1L | 5.01E-01 | 3.10E-01 | 4.13E-09 | 9.97E-08 | 4.54E-01 | 4.05E-07 | 3.40E-01 | 6.51E-08 |
| ENSG00000074800 | ENO1 | 5.67E-01 | 3.05E-01 | 9.56E-12 | 6.46E-10 | 4.50E-01 | 5.93E-07 | 4.07E-01 | 2.01E-12 |
| ENSG00000093009 | CDC45 | 6.76E-01 | 4.18E-01 | 1.25E-17 | 1.71E-14 | 6.31E-01 | 1.77E-17 | 5.64E-01 | 9.90E-28 |
| ENSG00000171241 | SHCBP1 | 7.13E-01 | 4.40E-01 | 3.30E-20 | 1.46E-16 | 5.10E-01 | 8.96E-10 | 4.40E-01 | 5.00E-15 |
| ENSG00000144642 | RBMS3 | 6.24E-01 | 3.26E-01 | 1.60E-14 | 3.58E-12 | 5.40E-01 | 2.08E-11 | 4.57E-01 | 1.65E-16 |
| ENSG00000139734 | DIAPH3 | 5.59E-01 | 2.51E-01 | 2.17E-11 | 1.23E-09 | 4.36E-01 | 2.22E-06 | 4.57E-01 | 1.72E-16 |
| ENSG00000112981 | NME5 | 4.85E-01 | 2.05E-01 | 1.52E-08 | 2.94E-07 | 3.62E-01 | 1.08E-03 | 3.49E-01 | 1.75E-08 |
| ENSG00000161800 | RACGAP1 | 5.18E-01 | 2.42E-01 | 9.79E-10 | 2.97E-08 | 5.11E-01 | 8.05E-10 | 5.17E-01 | 2.48E-22 |
| ENSG00000123975 | CKS2 | 4.42E-01 | 2.13E-01 | 3.49E-07 | 4.08E-06 | 3.52E-01 | 2.25E-03 | 2.89E-01 | 3.48E-05 |
| ENSG00000143409 | FAM63A | 4.79E-01 | 2.03E-01 | 2.31E-08 | 4.21E-07 | 4.59E-01 | 2.35E-07 | 2.67E-01 | 3.82E-04 |
| ENSG00000072818 | ACAP1 | 5.01E-01 | 2.12E-01 | 4.28E-09 | 1.02E-07 | 4.18E-01 | 1.18E-05 | 2.80E-01 | 1.03E-04 |
| ENSG00000169607 | CKAP2L | 5.77E-01 | 3.11E-01 | 3.62E-12 | 2.94E-10 | 5.59E-01 | 1.70E-12 | 5.77E-01 | 1.72E-29 |
| ENSG00000137812 | CASC5 | 5.43E-01 | 3.01E-01 | 1.04E-10 | 4.51E-09 | 4.71E-01 | 7.29E-08 | 4.84E-01 | 5.38E-19 |
| ENSG00000034053 | APBA2 | 5.45E-01 | 3.01E-01 | 8.62E-11 | 3.88E-09 | 4.01E-01 | 4.98E-05 | 3.29E-01 | 2.88E-07 |
| ENSG00000176890 | TYMS | 5.41E-01 | 2.26E-01 | 1.25E-10 | 5.32E-09 | 4.46E-01 | 8.55E-07 | 2.58E-01 | 9.85E-04 |
| ENSG00000115085 | ZAP70 | 5.11E-01 | 2.02E-01 | 1.79E-09 | 4.88E-08 | 4.44E-01 | 1.12E-06 | 3.43E-01 | 4.38E-08 |
| ENSG00000157827 | FMNL2 | 5.41E-01 | 2.54E-01 | 1.20E-10 | 5.11E-09 | 4.65E-01 | 1.39E-07 | 3.49E-01 | 1.73E-08 |
| ENSG00000186185 | KIF18B | 5.54E-01 | 2.41E-01 | 3.72E-11 | 1.93E-09 | 5.59E-01 | 1.64E-12 | 4.48E-01 | 1.11E-15 |
| ENSG00000090006 | LTBP4 | 5.06E-01 | 2.12E-01 | 2.71E-09 | 6.95E-08 | 4.83E-01 | 1.89E-08 | 4.98E-01 | 2.43E-20 |
| ENSG00000108106 | UBE2S | 5.59E-01 | 2.34E-01 | 2.23E-11 | 1.26E-09 | 5.49E-01 | 6.09E-12 | 3.60E-01 | 3.79E-09 |
| ENSG00000183722 | LHFP | 5.16E-01 | 2.46E-01 | 1.22E-09 | 3.59E-08 | 5.26E-01 | 1.24E-10 | 4.61E-01 | 8.02E-17 |
| ENSG00000143851 | PTPN7 | 5.55E-01 | 2.30E-01 | 3.28E-11 | 1.75E-09 | 3.71E-01 | 5.43E-04 | 3.77E-01 | 2.84E-10 |
| ENSG00000112874 | NUDT12 | 5.63E-01 | 2.57E-01 | 1.44E-11 | 8.90E-10 | 4.54E-01 | 4.06E-07 | 4.43E-01 | 3.16E-15 |
| ENSG00000117480 | FAAH | 4.85E-01 | 2.04E-01 | 1.54E-08 | 2.97E-07 | 3.47E-01 | 3.13E-03 | 2.11E-01 | 6.91E-02 |
| ENSG00000115163 | CENPA | 6.54E-01 | 4.10E-01 | 2.95E-16 | 1.63E-13 | 6.16E-01 | 2.39E-16 | 5.06E-01 | 3.32E-21 |
| ENSG00000154262 | ABCA6 | 4.78E-01 | 2.19E-01 | 2.59E-08 | 4.60E-07 | 3.87E-01 | 1.55E-04 | 4.15E-01 | 5.69E-13 |
| ENSG00000160654 | CD3G | 4.96E-01 | 2.05E-01 | 6.37E-09 | 1.42E-07 | 4.72E-01 | 6.55E-08 | 4.02E-01 | 5.30E-12 |
| ENSG00000138764 | CCNG2 | 4.60E-01 | 2.01E-01 | 9.48E-08 | 1.35E-06 | 3.20E-01 | 1.85E-02 | 1.70E-01 | 1.00E+00 |
| ENSG00000092853 | CLSPN | 5.44E-01 | 2.38E-01 | 9.68E-11 | 4.25E-09 | 4.95E-01 | 5.45E-09 | 4.82E-01 | 9.29E-19 |
| ENSG00000254087 | LYN | 4.81E-01 | 2.14E-01 | 2.05E-08 | 3.83E-07 | 3.99E-01 | 6.10E-05 | 3.55E-01 | 7.36E-09 |
| ENSG00000132819 | RBM38 | 5.52E-01 | 2.78E-01 | 4.58E-11 | 2.28E-09 | 5.51E-01 | 4.84E-12 | 3.89E-01 | 4.41E-11 |
| ENSG00000101003 | GINS1 | 5.60E-01 | 3.11E-01 | 1.94E-11 | 1.14E-09 | 4.85E-01 | 1.57E-08 | 4.87E-01 | 3.34E-19 |
| ENSG00000075218 | GTSE1 | 6.13E-01 | 3.37E-01 | 6.19E-14 | 1.07E-11 | 5.98E-01 | 4.81E-15 | 4.32E-01 | 2.27E-14 |
| ENSG00000165617 | DACT1 | 5.30E-01 | 2.11E-01 | 3.39E-10 | 1.24E-08 | 5.68E-01 | 4.24E-13 | 4.29E-01 | 3.96E-14 |
| ENSG00000133466 | C1QTNF6 | 5.39E-01 | 2.57E-01 | 1.48E-10 | 6.15E-09 | 4.89E-01 | 9.88E-09 | 3.81E-01 | 1.46E-10 |
| ENSG00000166428 | PLD4 | 5.66E-01 | 2.24E-01 | 1.06E-11 | 6.95E-10 | 3.95E-01 | 8.42E-05 | 4.09E-01 | 1.53E-12 |
| ENSG00000163808 | KIF15 | 5.62E-01 | 3.08E-01 | 1.62E-11 | 9.74E-10 | 5.47E-01 | 8.24E-12 | 4.25E-01 | 8.47E-14 |
| ENSG00000163823 | CCR1 | 5.38E-01 | 2.14E-01 | 1.59E-10 | 6.52E-09 | 2.71E-01 | 3.32E-01 | 4.24E-01 | 1.01E-13 |
| ENSG00000109501 | WFS1 | 3.80E-01 | 2.00E-01 | 1.56E-05 | 1.00E-04 | 4.65E-01 | 1.34E-07 | 3.34E-01 | 1.42E-07 |
| ENSG00000103522 | IL21R | 4.87E-01 | 2.11E-01 | 1.32E-08 | 2.60E-07 | 3.97E-01 | 6.91E-05 | 3.02E-01 | 7.89E-06 |
| ENSG00000073464 | CLCN4 | 4.23E-01 | 2.38E-01 | 1.18E-06 | 1.14E-05 | 4.64E-01 | 1.54E-07 | 3.47E-01 | 2.43E-08 |
| ENSG00000148516 | ZEB1 | 5.71E-01 | 2.47E-01 | 6.50E-12 | 4.69E-10 | 5.51E-01 | 4.88E-12 | 4.78E-01 | 2.43E-18 |
| ENSG00000139626 | ITGB7 | 4.84E-01 | 2.45E-01 | 1.62E-08 | 3.09E-07 | 3.15E-01 | 2.62E-02 | 3.79E-01 | 2.18E-10 |
| ENSG00000106462 | EZH2 | 5.50E-01 | 2.96E-01 | 5.52E-11 | 2.67E-09 | 5.28E-01 | 1.01E-10 | 4.85E-01 | 4.86E-19 |
| ENSG00000123473 | STIL | 6.27E-01 | 3.65E-01 | 1.08E-14 | 2.61E-12 | 5.78E-01 | 9.84E-14 | 6.06E-01 | 2.01E-33 |

**Table S3. Prediction performance for the 1011 genes on validation and test sets.**

| Ensembl_gene_id | Hgnc symbol | Coef (valid) | Rsq (valid) | P-val (valid) | adj.pval (valid) | Coef (internal) | adj.pval (internal) | Coef (external) | adj.pval (external) |
|---|---|---|---|---|---|---|---|---|---|
| ENSG00000135912 | TTLL4 | 5.22E-01 | 3.03E-01 | 7.22E-10 | 2.33E-08 | 4.96E-01 | 4.64E-09 | 3.80E-01 | 1.78E-10 |
| ENSG00000127329 | PTPRB | 4.52E-01 | 2.02E-01 | 1.73E-07 | 2.26E-06 | 4.56E-01 | 3.28E-07 | 3.99E-01 | 8.29E-12 |
| ENSG00000013810 | TACC3 | 4.91E-01 | 2.02E-01 | 9.47E-09 | 1.98E-07 | 5.46E-01 | 9.08E-12 | 2.68E-01 | 3.54E-04 |
| ENSG00000154839 | SKA1 | 6.70E-01 | 4.61E-01 | 3.09E-17 | 2.86E-14 | 6.46E-01 | 1.18E-18 | 4.93E-01 | 8.42E-20 |
| ENSG00000091651 | ORC6 | 7.23E-01 | 4.91E-01 | 5.52E-21 | 4.88E-17 | 5.97E-01 | 5.71E-15 | 5.72E-01 | 9.63E-29 |
| ENSG00000165633 | VSTM4 | 5.74E-01 | 2.49E-01 | 4.55E-12 | 3.47E-10 | 4.57E-01 | 3.04E-07 | 4.82E-01 | 9.58E-19 |
| ENSG00000186517 | ARHGAP30 | 5.19E-01 | 2.15E-01 | 8.73E-10 | 2.70E-08 | 2.73E-01 | 2.89E-01 | 2.87E-01 | 4.57E-05 |
| ENSG00000124256 | ZBP1 | 5.67E-01 | 2.46E-01 | 9.76E-12 | 6.56E-10 | 2.64E-01 | 4.72E-01 | 3.08E-01 | 3.93E-06 |
| ENSG00000166508 | MCM7 | 5.56E-01 | 2.57E-01 | 3.04E-11 | 1.64E-09 | 5.00E-01 | 2.96E-09 | 2.88E-01 | 4.01E-05 |
| ENSG00000105011 | ASF1B | 5.19E-01 | 2.43E-01 | 8.95E-10 | 2.75E-08 | 5.12E-01 | 7.54E-10 | 2.82E-01 | 7.96E-05 |
| ENSG00000170011 | MYRIP | 4.95E-01 | 2.09E-01 | 6.98E-09 | 1.53E-07 | 2.96E-01 | 7.93E-02 | 3.68E-01 | 1.11E-09 |
| ENSG00000130045 | NXNL2 | 5.39E-01 | 2.43E-01 | 1.48E-10 | 6.15E-09 | 5.02E-01 | 2.22E-09 | 4.21E-01 | 1.83E-13 |
| ENSG00000154127 | UBASH3B | 5.11E-01 | 2.17E-01 | 1.82E-09 | 4.95E-08 | 4.01E-01 | 4.91E-05 | 2.82E-01 | 8.42E-05 |
| ENSG00000167208 | SNX20 | 5.88E-01 | 2.02E-01 | 1.10E-12 | 1.07E-10 | 4.27E-01 | 5.11E-06 | 3.32E-01 | 1.96E-07 |
| ENSG00000143815 | LBR | 5.05E-01 | 2.67E-01 | 3.06E-09 | 7.72E-08 | 5.43E-01 | 1.40E-11 | 3.01E-01 | 8.84E-06 |
| ENSG00000149554 | CHEK1 | 6.37E-01 | 3.80E-01 | 3.18E-15 | 1.06E-12 | 5.32E-01 | 5.82E-11 | 2.97E-01 | 1.52E-05 |
| ENSG00000123505 | AMD1 | 5.61E-01 | 2.48E-01 | 1.77E-11 | 1.05E-09 | 4.69E-01 | 9.18E-08 | 4.30E-01 | 3.53E-14 |
| ENSG00000173269 | MMRN2 | 5.11E-01 | 2.19E-01 | 1.87E-09 | 5.05E-08 | 5.06E-01 | 1.53E-09 | 4.33E-01 | 2.08E-14 |
| ENSG00000007968 | E2F2 | 6.28E-01 | 3.61E-01 | 9.94E-15 | 2.48E-12 | 6.16E-01 | 2.29E-16 | 4.44E-01 | 2.65E-15 |
| ENSG00000008277 | ADAM22 | 5.66E-01 | 2.14E-01 | 1.12E-11 | 7.26E-10 | 2.42E-01 | 1.00E+00 | 2.82E-01 | 8.42E-05 |
| ENSG00000269113 | TRABD2B | 5.33E-01 | 2.36E-01 | 2.57E-10 | 9.79E-09 | 4.73E-01 | 6.01E-08 | | |
| ENSG00000101347 | SAMHD1 | 5.10E-01 | 2.41E-01 | 1.95E-09 | 5.24E-08 | 1.42E-01 | 1.00E+00 | 3.77E-01 | 3.07E-10 |
| ENSG00000051341 | POLQ | 6.75E-01 | 3.99E-01 | 1.42E-17 | 1.71E-14 | 5.51E-01 | 4.59E-12 | 5.17E-01 | 2.30E-22 |
| ENSG00000172543 | CTSW | 5.64E-01 | 2.33E-01 | 1.27E-11 | 8.14E-10 | 4.03E-01 | 4.18E-05 | 3.35E-01 | 1.23E-07 |
| ENSG00000171793 | CTPS1 | 6.22E-01 | 3.14E-01 | 1.96E-14 | 3.99E-12 | 4.41E-01 | 1.42E-06 | 4.78E-01 | 2.29E-18 |
| ENSG00000188820 | FAM26F | 5.64E-01 | 2.25E-01 | 1.40E-11 | 8.73E-10 | 3.88E-01 | 1.47E-04 | 3.31E-01 | 2.06E-07 |
| ENSG00000065978 | YBX1 | 6.12E-01 | 3.73E-01 | 6.92E-14 | 1.14E-11 | 5.19E-01 | 2.89E-10 | 2.46E-01 | 3.12E-03 |
| ENSG00000140534 | TICRR | 6.59E-01 | 3.92E-01 | 1.62E-16 | 1.02E-13 | 4.89E-01 | 1.03E-08 | 4.70E-01 | 1.12E-17 |
| ENSG00000046889 | PREX2 | 5.66E-01 | 2.95E-01 | 1.05E-11 | 6.95E-10 | 4.20E-01 | 9.84E-06 | 3.69E-01 | 1.00E-09 |
| ENSG00000099622 | CIRBP | 5.70E-01 | 3.16E-01 | 7.55E-12 | 5.34E-10 | 4.71E-01 | 7.23E-08 | 5.23E-01 | 5.25E-23 |
| ENSG00000043462 | LCP2 | 5.60E-01 | 2.09E-01 | 2.09E-11 | 1.20E-09 | 2.89E-01 | 1.23E-01 | 3.84E-01 | 1.02E-10 |
| ENSG00000187535 | IFT140 | 4.68E-01 | 2.16E-01 | 5.55E-08 | 8.63E-07 | 3.64E-01 | 9.38E-04 | 2.72E-01 | 2.28E-04 |
| ENSG00000116194 | ANGPTL1 | 4.83E-01 | 2.04E-01 | 1.76E-08 | 3.34E-07 | 4.87E-01 | 1.27E-08 | 4.51E-01 | 6.45E-16 |
| ENSG00000077063 | CTTNBP2 | 5.20E-01 | 2.06E-01 | 8.47E-10 | 2.66E-08 | 2.98E-01 | 7.46E-02 | 3.45E-01 | 3.16E-08 |
| ENSG00000152256 | PDK1 | 5.16E-01 | 2.56E-01 | 1.22E-09 | 3.59E-08 | 4.62E-01 | 1.75E-07 | 4.11E-01 | 1.15E-12 |
| ENSG00000145604 | SKP2 | 5.68E-01 | 3.29E-01 | 9.17E-12 | 6.24E-10 | 5.11E-01 | 8.43E-10 | 4.45E-01 | 2.19E-15 |
| ENSG00000010030 | ETV7 | 5.90E-01 | 2.73E-01 | 8.97E-13 | 8.92E-11 | 3.12E-01 | 3.08E-02 | 3.61E-01 | 3.13E-09 |
| ENSG00000130653 | PNPLA7 | 5.38E-01 | 2.23E-01 | 1.64E-10 | 6.69E-09 | 3.41E-01 | 4.70E-03 | 3.40E-01 | 6.61E-08 |
| ENSG00000089692 | LAG3 | 5.91E-01 | 2.78E-01 | 7.46E-13 | 7.76E-11 | 4.57E-01 | 2.95E-07 | 4.40E-01 | 5.05E-15 |
| ENSG00000163145 | C1QTNF7 | 5.98E-01 | 3.55E-01 | 3.33E-13 | 4.10E-11 | 5.67E-01 | 4.97E-13 | 5.86E-01 | 1.04E-30 |
| ENSG00000149451 | ADAM33 | 6.60E-01 | 3.47E-01 | 1.38E-16 | 9.40E-14 | 6.23E-01 | 7.42E-17 | 5.20E-01 | 1.17E-22 |
| ENSG00000137310 | TCF19 | 4.90E-01 | 2.68E-01 | 9.78E-09 | 2.02E-07 | 5.56E-01 | 2.59E-12 | 3.72E-01 | 5.79E-10 |
| ENSG00000158869 | FCER1G | 5.24E-01 | 2.05E-01 | 5.77E-10 | 1.91E-08 | 2.64E-01 | 4.76E-01 | 3.23E-01 | 5.71E-07 |
| ENSG00000177098 | SCN4B | 5.29E-01 | 2.44E-01 | 3.64E-10 | 1.32E-08 | 5.10E-01 | 9.50E-10 | 5.93E-01 | 1.19E-31 |
| ENSG00000196872 | KIAA1211L | 4.79E-01 | 2.10E-01 | 2.41E-08 | 4.35E-07 | 4.61E-01 | 1.97E-07 | 3.77E-01 | 2.89E-10 |
| ENSG00000186810 | CXCR3 | 4.92E-01 | 2.14E-01 | 8.37E-09 | 1.78E-07 | 4.81E-01 | 2.33E-08 | 3.69E-01 | 9.97E-10 |
| ENSG00000164045 | CDC25A | 6.13E-01 | 4.00E-01 | 6.50E-14 | 1.11E-11 | 6.09E-01 | 7.70E-16 | 5.42E-01 | 4.09E-25 |
| ENSG00000111247 | RAD51AP1 | 5.92E-01 | 3.74E-01 | 6.54E-13 | 7.10E-11 | 5.27E-01 | 1.12E-10 | 4.96E-01 | 4.24E-20 |
| ENSG00000172215 | CXCR6 | 5.50E-01 | 2.39E-01 | 5.22E-11 | 2.56E-09 | 4.55E-01 | 3.82E-07 | 3.76E-01 | 3.63E-10 |
| ENSG00000165480 | SKA3 | 5.98E-01 | 3.45E-01 | 3.40E-13 | 4.12E-11 | 5.41E-01 | 1.90E-11 | 5.24E-01 | 4.16E-23 |
| ENSG00000135047 | CTSL | 5.42E-01 | 2.24E-01 | 1.15E-10 | 4.96E-09 | 2.83E-01 | 1.73E-01 | 4.80E-01 | 1.35E-18 |
| ENSG00000198805 | PNP | 5.07E-01 | 2.29E-01 | 2.63E-09 | 6.75E-08 | 4.72E-01 | 6.06E-08 | 4.18E-01 | 3.38E-13 |
| ENSG00000129173 | E2F8 | 5.61E-01 | 2.94E-01 | 1.76E-11 | 1.04E-09 | 5.27E-01 | 1.06E-10 | 5.57E-01 | 6.60E-27 |
| ENSG00000100307 | CBX7 | 5.20E-01 | 2.45E-01 | 8.26E-10 | 2.62E-08 | 3.98E-01 | 6.51E-05 | 5.53E-01 | 1.77E-26 |
| ENSG00000102935 | ZNF423 | 5.95E-01 | 2.33E-01 | 5.14E-13 | 5.73E-11 | 5.01E-01 | 2.53E-09 | 5.65E-01 | 6.10E-28 |
| ENSG00000006652 | IFRD1 | 4.31E-01 | 2.50E-01 | 7.44E-07 | 7.68E-06 | 2.95E-01 | 8.64E-02 | 3.27E-01 | 3.73E-07 |
| ENSG00000113810 | SMC4 | 4.92E-01 | 2.21E-01 | 8.51E-09 | 1.80E-07 | 3.69E-01 | 6.59E-04 | 2.76E-01 | 1.51E-04 |
| ENSG00000111077 | TENC1 | 6.20E-01 | 3.86E-01 | 2.50E-14 | 4.97E-12 | 5.84E-01 | 4.21E-14 | 5.12E-01 | 7.80E-22 |
| ENSG00000116701 | NCF2 | 5.47E-01 | 2.35E-01 | 7.01E-11 | 3.28E-09 | 2.84E-01 | 1.64E-01 | 3.24E-01 | 5.41E-07 |
| ENSG00000185532 | PRKG1 | 3.91E-01 | 2.01E-01 | 8.59E-06 | 5.97E-05 | 4.01E-01 | 5.21E-05 | 2.17E-01 | 4.14E-02 |
| ENSG00000134460 | IL2RA | 6.23E-01 | 2.39E-01 | 1.84E-14 | 3.92E-12 | 5.23E-01 | 1.83E-10 | 3.89E-01 | 4.62E-11 |
| ENSG00000167779 | IGFBP6 | 4.37E-01 | 2.12E-01 | 4.86E-07 | 5.35E-06 | 4.49E-01 | 6.82E-07 | 4.09E-01 | 1.54E-12 |
| ENSG00000079337 | RAPGEF3 | 4.67E-01 | 2.40E-01 | 6.04E-08 | 9.24E-07 | 4.93E-01 | 6.75E-09 | 4.20E-01 | 2.26E-13 |
| ENSG00000050405 | LIMA1 | 4.28E-01 | 2.27E-01 | 8.64E-07 | 8.69E-06 | 4.38E-01 | 1.91E-06 | 2.33E-01 | 1.02E-02 |
| ENSG00000164924 | YWHAZ | 4.73E-01 | 2.28E-01 | 3.68E-08 | 6.11E-07 | 5.01E-01 | 2.62E-09 | 4.04E-01 | 3.48E-12 |
| ENSG00000114698 | PLSCR4 | 5.27E-01 | 2.39E-01 | 4.67E-10 | 1.62E-08 | 4.89E-01 | 1.01E-08 | 5.12E-01 | 8.29E-22 |
| ENSG00000173762 | CD7 | 5.41E-01 | 2.13E-01 | 1.24E-10 | 5.28E-09 | 5.06E-01 | 1.51E-09 | 4.45E-01 | 2.16E-15 |
| ENSG00000154721 | JAM2 | 5.43E-01 | 2.22E-01 | 9.99E-11 | 4.36E-09 | 5.32E-01 | 6.31E-11 | 4.20E-01 | 2.06E-13 |
| ENSG00000054392 | HHAT | 4.39E-01 | 2.42E-01 | 4.30E-07 | 4.83E-06 | 3.80E-01 | 2.65E-04 | 3.16E-01 | 1.46E-06 |

**Table S3. Prediction performance for the 1011 genes on validation and test sets.**

| Ensembl_gene_id | Hgnc symbol | Coef (valid) | Rsq (valid) | P-val (valid) | adj.pval (valid) | Coef (internal) | adj.pval (internal) | Coef (external) | adj.pval (external) |
|---|---|---|---|---|---|---|---|---|---|
| ENSG00000197375 | SLC22A5 | 4.46E-01 | 2.07E-01 | 2.54E-07 | 3.10E-06 | 3.74E-01 | 4.23E-04 | 3.90E-01 | 3.68E-11 |
| ENSG00000158164 | TMSB15A | 3.38E-01 | 2.08E-01 | 1.42E-04 | 6.55E-04 | 3.63E-01 | 9.96E-04 | 1.38E-01 | 1.00E+00 |
| ENSG00000205268 | PDE7A | 5.34E-01 | 3.11E-01 | 2.44E-10 | 9.35E-09 | 4.73E-01 | 5.95E-08 | 3.28E-01 | 2.99E-07 |
| ENSG00000167613 | LAIR1 | 5.63E-01 | 2.17E-01 | 1.44E-11 | 8.90E-10 | 2.32E-01 | 1.00E+00 | 2.68E-01 | 3.40E-04 |
| ENSG00000173207 | CKS1B | 4.85E-01 | 2.58E-01 | 1.50E-08 | 2.89E-07 | 4.38E-01 | 1.89E-06 | 3.68E-01 | 1.20E-09 |
| ENSG00000105792 | CFAP69 | 5.99E-01 | 2.32E-01 | 3.25E-13 | 4.04E-11 | 4.64E-01 | 1.48E-07 | 3.21E-01 | 8.23E-07 |
| ENSG00000121621 | KIF18A | 5.75E-01 | 3.07E-01 | 4.33E-12 | 3.34E-10 | 5.02E-01 | 2.19E-09 | 6.24E-01 | 3.47E-36 |
| ENSG00000138613 | APH1B | 5.65E-01 | 2.58E-01 | 1.23E-11 | 7.93E-10 | 5.38E-01 | 2.70E-11 | 3.89E-01 | 4.11E-11 |
| ENSG00000187398 | LUZP2 | 5.44E-01 | 2.32E-01 | 9.61E-11 | 4.23E-09 | 2.90E-01 | 1.18E-01 | 3.62E-01 | 2.76E-09 |
| ENSG00000056558 | TRAF1 | 5.52E-01 | 2.19E-01 | 4.34E-11 | 2.20E-09 | 3.86E-01 | 1.73E-04 | 3.42E-01 | 5.17E-08 |
| ENSG00000100526 | CDKN3 | 5.13E-01 | 2.29E-01 | 1.51E-09 | 4.28E-08 | 4.55E-01 | 3.62E-07 | 3.57E-01 | 5.43E-09 |
| ENSG00000133121 | STARD13 | 5.71E-01 | 2.28E-01 | 6.55E-12 | 4.71E-10 | 4.29E-01 | 4.22E-06 | 4.16E-01 | 4.23E-13 |
| ENSG00000010292 | NCAPD2 | 5.31E-01 | 3.55E-01 | 3.21E-10 | 1.18E-08 | 5.29E-01 | 9.18E-11 | 3.65E-01 | 1.93E-09 |
| ENSG00000163507 | KIAA1524 | 5.90E-01 | 3.08E-01 | 8.31E-13 | 8.40E-11 | 5.23E-01 | 1.83E-10 | 4.53E-01 | 3.84E-16 |
| ENSG00000163600 | ICOS | 5.94E-01 | 2.68E-01 | 5.31E-13 | 5.87E-11 | 4.62E-01 | 1.76E-07 | 4.26E-01 | 6.96E-14 |
| ENSG00000241106 | HLA-DOB | 4.99E-01 | 2.01E-01 | 4.86E-09 | 1.13E-07 | 4.77E-01 | 3.68E-08 | 3.75E-01 | 3.66E-10 |
| ENSG00000160883 | HK3 | 6.31E-01 | 3.09E-01 | 6.54E-15 | 1.78E-12 | 3.02E-01 | 5.86E-02 | 4.42E-01 | 3.68E-15 |
| ENSG00000097046 | CDC7 | 4.55E-01 | 2.31E-01 | 1.36E-07 | 1.83E-06 | 4.48E-01 | 7.15E-07 | 3.84E-01 | 1.02E-10 |
| ENSG00000136982 | DSCC1 | 5.92E-01 | 3.92E-01 | 7.29E-13 | 7.68E-11 | 5.76E-01 | 1.43E-13 | 5.33E-01 | 4.74E-24 |
| ENSG00000143256 | PFDN2 | 4.45E-01 | 2.06E-01 | 2.83E-07 | 3.42E-06 | 1.85E-01 | 1.00E+00 | 1.61E-01 | 1.00E+00 |
| ENSG00000111328 | CDK2AP1 | 3.76E-01 | 2.63E-01 | 2.02E-05 | 1.23E-04 | 3.19E-01 | 2.01E-02 | 2.94E-01 | 2.15E-05 |
| ENSG00000066855 | MTFR1 | 4.68E-01 | 2.23E-01 | 5.27E-08 | 8.26E-07 | 4.78E-01 | 3.44E-08 | 4.01E-01 | 6.05E-12 |
| ENSG00000065911 | MTHFD2 | 4.85E-01 | 2.11E-01 | 1.48E-08 | 2.86E-07 | 4.05E-01 | 3.72E-05 | 4.07E-01 | 2.04E-12 |
| ENSG00000129195 | FAM64A | 6.18E-01 | 3.81E-01 | 3.53E-14 | 6.64E-12 | 5.40E-01 | 2.08E-11 | 4.51E-01 | 6.54E-16 |
| ENSG00000104814 | MAP4K1 | 5.26E-01 | 2.10E-01 | 4.95E-10 | 1.69E-08 | 4.04E-01 | 3.77E-05 | 3.06E-01 | 5.13E-06 |
| ENSG00000088305 | DNMT3B | 5.07E-01 | 2.58E-01 | 2.55E-09 | 6.57E-08 | 4.78E-01 | 3.29E-08 | 3.60E-01 | 3.59E-09 |
| ENSG00000100139 | MICALL1 | 4.52E-01 | 2.09E-01 | 1.73E-07 | 2.26E-06 | 3.28E-01 | 1.13E-02 | 3.54E-01 | 8.41E-09 |
| ENSG00000079263 | SP140 | 5.73E-01 | 2.21E-01 | 5.11E-12 | 3.85E-10 | 3.83E-01 | 2.21E-04 | 3.71E-01 | 7.18E-10 |
| ENSG00000020181 | GPR124 | 5.14E-01 | 2.50E-01 | 1.40E-09 | 4.01E-08 | 4.41E-01 | 1.46E-06 | 5.00E-01 | 1.46E-20 |
| ENSG00000110934 | BIN2 | 5.75E-01 | 2.13E-01 | 4.52E-12 | 3.47E-10 | 3.69E-01 | 6.32E-04 | 3.36E-01 | 1.05E-07 |
| ENSG00000073008 | PVR | 4.70E-01 | 2.36E-01 | 4.86E-08 | 7.73E-07 | 3.08E-01 | 3.91E-02 | 2.35E-01 | 9.06E-03 |
| ENSG00000167553 | TUBA1C | 4.93E-01 | 2.15E-01 | 7.86E-09 | 1.70E-07 | 5.14E-01 | 5.92E-10 | 4.18E-01 | 2.99E-13 |
| ENSG00000187741 | FANCA | 6.58E-01 | 4.37E-01 | 1.90E-16 | 1.12E-13 | 5.08E-01 | 1.18E-09 | 4.01E-01 | 5.71E-12 |
| ENSG00000164983 | TMEM65 | 4.78E-01 | 2.60E-01 | 2.54E-08 | 4.55E-07 | 3.50E-01 | 2.56E-03 | 3.43E-01 | 3.97E-08 |
| ENSG00000099139 | PCSK5 | 5.45E-01 | 2.58E-01 | 8.55E-11 | 3.86E-09 | 5.18E-01 | 3.33E-10 | 4.18E-01 | 3.06E-13 |
| ENSG00000085840 | ORC1 | 6.70E-01 | 3.95E-01 | 3.23E-17 | 2.86E-14 | 6.17E-01 | 2.06E-16 | 4.75E-01 | 4.18E-18 |
| ENSG00000123136 | DDX39A | 5.61E-01 | 2.54E-01 | 1.72E-11 | 1.03E-09 | 5.10E-01 | 9.57E-10 | 2.55E-01 | 1.27E-03 |
| ENSG00000124795 | DEK | 4.95E-01 | 2.29E-01 | 6.87E-09 | 1.51E-07 | 3.59E-01 | 1.34E-03 | 1.62E-01 | 1.00E+00 |
| ENSG00000129810 | SGOL1 | 6.13E-01 | 3.31E-01 | 5.99E-14 | 1.06E-11 | 5.95E-01 | 7.72E-15 | 5.49E-01 | 5.86E-26 |
| ENSG00000155465 | SLC7A7 | 5.35E-01 | 2.12E-01 | 2.26E-10 | 8.86E-09 | 2.87E-01 | 1.38E-01 | 3.53E-01 | 1.04E-08 |
| ENSG00000140525 | FANCI | 5.31E-01 | 2.13E-01 | 3.15E-10 | 1.17E-08 | 4.37E-01 | 1.98E-06 | 4.92E-01 | 1.06E-19 |
| ENSG00000154263 | ABCA10 | 5.37E-01 | 2.40E-01 | 1.80E-10 | 7.25E-09 | 3.80E-01 | 2.77E-04 | 3.33E-01 | 1.67E-07 |
| ENSG00000164934 | DCAF13 | 4.86E-01 | 2.33E-01 | 1.44E-08 | 2.79E-07 | 4.91E-01 | 7.86E-09 | 3.56E-01 | 6.69E-09 |
| ENSG00000123131 | PRDX4 | 5.45E-01 | 2.39E-01 | 8.27E-11 | 3.79E-09 | 3.82E-01 | 2.38E-04 | 3.25E-01 | 4.98E-07 |
| ENSG00000138670 | RASGEF1B | 5.00E-01 | 2.09E-01 | 4.63E-09 | 1.09E-07 | 3.68E-01 | 7.12E-04 | 3.86E-01 | 6.65E-11 |
| ENSG00000166086 | JAM3 | 5.44E-01 | 2.73E-01 | 9.09E-11 | 4.04E-09 | 5.38E-01 | 2.70E-11 | 4.35E-01 | 1.29E-14 |
| ENSG00000117090 | SLAMF1 | 5.63E-01 | 2.32E-01 | 1.54E-11 | 9.38E-10 | 4.60E-01 | 2.16E-07 | 3.54E-01 | 9.34E-09 |
| ENSG00000147443 | DOK2 | 5.71E-01 | 2.09E-01 | 6.35E-12 | 4.61E-10 | 3.65E-01 | 8.40E-04 | 2.42E-01 | 4.86E-03 |
| ENSG00000139330 | KERA | 6.94E-01 | 3.92E-01 | 8.45E-19 | 2.13E-15 | 5.26E-01 | 1.21E-10 | 4.67E-01 | 2.28E-17 |
| ENSG00000102144 | PGK1 | 5.65E-01 | 2.80E-01 | 1.17E-11 | 7.60E-10 | 4.72E-01 | 6.26E-08 | 4.14E-01 | 5.88E-13 |
| ENSG00000105220 | GPI | 4.83E-01 | 2.41E-01 | 1.77E-08 | 3.35E-07 | 3.42E-01 | 4.38E-03 | 2.83E-01 | 7.37E-05 |
| ENSG00000197299 | BLM | 6.71E-01 | 4.25E-01 | 2.83E-17 | 2.78E-14 | 5.72E-01 | 2.45E-13 | 5.32E-01 | 6.19E-24 |
| ENSG00000143882 | ATP6V1C2 | 4.09E-01 | 2.71E-01 | 2.95E-06 | 2.46E-05 | 4.64E-01 | 1.44E-07 | 3.55E-01 | 7.80E-09 |
| ENSG00000168496 | FEN1 | 5.64E-01 | 2.67E-01 | 1.37E-11 | 8.68E-10 | 5.29E-01 | 8.51E-11 | 2.93E-01 | 2.22E-05 |
| ENSG00000166002 | SMCO4 | 4.57E-01 | 2.46E-01 | 1.19E-07 | 1.64E-06 | 4.40E-01 | 1.57E-06 | 3.16E-01 | 1.40E-06 |
| ENSG00000163694 | RBM47 | 4.67E-01 | 2.39E-01 | 5.94E-08 | 9.11E-07 | 2.55E-01 | 7.39E-01 | 2.63E-01 | 5.69E-04 |
| ENSG00000145194 | ECE2 | 5.57E-01 | 3.14E-01 | 2.65E-11 | 1.46E-09 | 4.56E-01 | 3.15E-07 | 4.06E-01 | 2.58E-12 |
| ENSG00000165323 | FAT3 | 5.66E-01 | 2.68E-01 | 1.05E-11 | 6.95E-10 | 4.71E-01 | 7.26E-08 | 5.45E-01 | 1.98E-25 |
| ENSG00000117013 | KCNQ4 | 4.43E-01 | 2.54E-01 | 3.17E-07 | 3.77E-06 | 2.85E-01 | 1.55E-01 | 1.68E-01 | 1.00E+00 |
| ENSG00000164032 | H2AFZ | 4.50E-01 | 2.01E-01 | 1.94E-07 | 2.48E-06 | 4.83E-01 | 1.98E-08 | 3.46E-01 | 2.79E-08 |
| ENSG00000182636 | NDN | 4.87E-01 | 2.16E-01 | 1.25E-08 | 2.49E-07 | 5.43E-01 | 1.50E-11 | 3.15E-01 | 1.61E-06 |
| ENSG00000163599 | CTLA4 | 5.84E-01 | 2.69E-01 | 1.58E-12 | 1.45E-10 | 4.72E-01 | 6.16E-08 | 3.61E-01 | 3.16E-09 |
| ENSG00000165490 | DDIAS | 5.26E-01 | 2.81E-01 | 4.71E-10 | 1.63E-08 | 3.88E-01 | 1.48E-04 | 2.92E-01 | 2.71E-05 |
| ENSG00000114861 | FOXP1 | 5.30E-01 | 2.17E-01 | 3.52E-10 | 1.27E-08 | 3.03E-01 | 5.49E-02 | 3.38E-01 | 7.99E-08 |
| ENSG00000148357 | HMCN2 | 5.27E-01 | 2.34E-01 | 4.53E-10 | 1.58E-08 | 3.94E-01 | 8.78E-05 | | |
| ENSG00000150764 | DIXDC1 | 4.98E-01 | 2.32E-01 | 5.54E-09 | 1.27E-07 | 4.27E-01 | 5.29E-06 | 4.47E-01 | 1.37E-15 |
| ENSG00000167749 | KLK4 | 5.82E-01 | 2.63E-01 | 2.11E-12 | 1.85E-10 | 4.67E-01 | 1.05E-07 | 4.57E-01 | 1.71E-16 |
| ENSG00000166451 | CENPN | 7.37E-01 | 4.96E-01 | 3.78E-22 | 6.68E-18 | 5.59E-01 | 1.62E-12 | 4.81E-01 | 1.23E-18 |
| ENSG00000155957 | TMBIM4 | 5.10E-01 | 2.31E-01 | 1.98E-09 | 5.28E-08 | 3.84E-01 | 2.08E-04 | 3.47E-01 | 2.38E-08 |

**Table S3. Prediction performance for the 1011 genes on validation and test sets.**

| Ensembl_gene_id | Hgnc symbol | Coef (valid) | Rsq (valid) | P-val (valid) | adj.pval (valid) | Coef (internal) | adj.pval (internal) | Coef (external) | adj.pval (external) |
|---|---|---|---|---|---|---|---|---|---|
| ENSG00000176619 | LMNB2 | 5.49E-01 | 2.38E-01 | 5.85E-11 | 2.80E-09 | 4.18E-01 | 1.14E-05 | 2.16E-01 | 4.76E-02 |
| ENSG00000150753 | CCT5 | 5.80E-01 | 2.48E-01 | 2.66E-12 | 2.26E-10 | 4.56E-01 | 3.20E-09 | 4.57E-01 | 1.97E-16 |
| ENSG00000189129 | PLAC9 | 5.38E-01 | 2.36E-01 | 1.60E-10 | 6.57E-09 | 4.82E-01 | 2.09E-08 | 4.44E-01 | 2.26E-15 |
| ENSG00000079102 | RUNX1T1 | 5.72E-01 | 3.27E-01 | 5.61E-12 | 4.15E-10 | 5.61E-01 | 1.23E-12 | 4.43E-01 | 3.09E-15 |
| ENSG00000144554 | FANCD2 | 4.70E-01 | 2.07E-01 | 4.69E-08 | 7.49E-07 | 4.56E-01 | 3.13E-07 | 5.02E-01 | 9.97E-21 |
| ENSG00000156471 | PTDSS1 | 5.22E-01 | 2.41E-01 | 6.76E-10 | 2.19E-08 | 4.70E-01 | 7.67E-08 | 3.42E-01 | 4.72E-08 |
| ENSG00000035499 | DEPDC1B | 5.77E-01 | 3.10E-01 | 3.36E-12 | 2.75E-10 | 5.10E-01 | 9.17E-10 | 4.17E-01 | 3.54E-13 |
| ENSG00000181374 | CCL13 | 5.08E-01 | 2.16E-01 | 2.26E-09 | 5.95E-08 | 3.89E-01 | 1.34E-04 | 2.46E-01 | 3.06E-03 |
| ENSG00000129625 | REEP5 | 4.45E-01 | 2.33E-01 | 2.75E-07 | 3.34E-06 | 3.46E-01 | 3.26E-03 | 3.19E-01 | 9.49E-07 |
| ENSG00000123908 | AGO2 | 5.28E-01 | 2.77E-01 | 4.17E-10 | 1.47E-08 | 4.67E-01 | 1.07E-07 | 4.68E-01 | 1.94E-17 |
| ENSG00000137547 | MRPL15 | 4.90E-01 | 2.04E-01 | 9.99E-09 | 2.06E-07 | 3.52E-01 | 2.22E-03 | 2.00E-01 | 1.63E-01 |
| ENSG00000111669 | TPI1 | 5.90E-01 | 3.25E-01 | 8.72E-13 | 8.72E-11 | 3.62E-01 | 1.05E-03 | 3.54E-01 | 8.48E-09 |
| ENSG00000181192 | DHTKD1 | 4.39E-01 | 2.49E-01 | 4.28E-07 | 4.82E-06 | 4.70E-01 | 8.06E-08 | 3.05E-01 | 5.54E-06 |
| ENSG00000182836 | PLCXD3 | 5.36E-01 | 2.40E-01 | 1.94E-10 | 7.76E-09 | 5.40E-01 | 2.07E-11 | 4.28E-01 | 5.11E-14 |
| ENSG00000198863 | RUNDC1 | 4.62E-01 | 2.22E-01 | 8.36E-08 | 1.21E-06 | 3.94E-01 | 8.66E-05 | 4.57E-01 | 1.73E-16 |
| ENSG00000118997 | DNAH7 | 5.18E-01 | 2.69E-01 | 9.56E-10 | 2.92E-08 | 4.76E-01 | 4.34E-08 | 2.16E-01 | 4.68E-02 |
| ENSG00000136108 | CKAP2 | 4.64E-01 | 2.08E-01 | 7.54E-08 | 1.11E-06 | 3.54E-01 | 1.97E-03 | 4.32E-01 | 2.55E-14 |
| ENSG00000176435 | CLEC14A | 5.24E-01 | 2.36E-01 | 5.96E-10 | 1.97E-08 | 5.03E-01 | 2.10E-09 | 4.48E-01 | 1.20E-15 |
| ENSG00000076003 | MCM6 | 5.80E-01 | 3.21E-01 | 2.55E-12 | 2.19E-10 | 4.79E-01 | 3.19E-08 | 3.58E-01 | 5.35E-09 |
| ENSG00000119787 | ATL2 | 3.39E-01 | 2.11E-01 | 1.35E-04 | 6.24E-04 | 3.17E-01 | 2.23E-02 | 3.48E-01 | 1.98E-08 |
| ENSG00000135838 | NPL | 5.90E-01 | 2.22E-01 | 8.43E-13 | 8.48E-11 | 2.52E-01 | 8.64E-01 | 2.66E-01 | 4.26E-04 |
| ENSG00000126264 | HCST | 5.32E-01 | 2.15E-01 | 2.96E-10 | 1.10E-08 | 3.98E-01 | 6.49E-05 | 2.34E-01 | 1.01E-02 |
| ENSG00000019991 | HGF | 5.50E-01 | 2.59E-01 | 5.41E-11 | 2.63E-09 | 5.34E-01 | 4.76E-11 | 2.87E-01 | 4.47E-05 |
| ENSG00000132646 | PCNA | 4.76E-01 | 2.26E-01 | 3.07E-08 | 5.26E-07 | 4.62E-01 | 1.81E-07 | 3.94E-01 | 1.91E-11 |
| ENSG00000096063 | SRPK1 | 6.01E-01 | 3.56E-01 | 2.45E-13 | 3.27E-11 | 4.54E-01 | 3.90E-07 | 5.35E-01 | 2.55E-24 |
| ENSG00000051180 | RAD51 | 6.55E-01 | 3.75E-01 | 2.79E-16 | 1.59E-13 | 5.62E-01 | 1.10E-12 | 4.38E-01 | 8.35E-15 |
| ENSG00000167925 | GHDC | 4.75E-01 | 2.45E-01 | 3.26E-08 | 5.53E-07 | 3.30E-01 | 1.02E-02 | 3.66E-01 | 1.54E-09 |
| ENSG00000223547 | ZNF844 | 4.82E-01 | 2.34E-01 | 1.88E-08 | 3.54E-07 | 3.19E-01 | 2.04E-02 | 1.61E-01 | 1.00E+00 |
| ENSG00000186871 | ERCC6L | 6.31E-01 | 3.29E-01 | 6.64E-15 | 1.78E-12 | 5.76E-01 | 1.34E-13 | 4.37E-01 | 1.02E-14 |
| ENSG00000163568 | AIM2 | 6.19E-01 | 3.28E-01 | 2.82E-14 | 5.41E-12 | 4.69E-01 | 8.92E-08 | 4.43E-01 | 3.04E-15 |
| ENSG00000144152 | FBLN7 | 5.36E-01 | 2.45E-01 | 1.92E-10 | 7.72E-09 | 3.15E-01 | 2.63E-02 | 5.18E-01 | 2.02E-22 |
| ENSG00000109576 | AADAT | 4.00E-01 | 2.22E-01 | 5.08E-06 | 3.84E-05 | 3.25E-01 | 1.37E-02 | 2.90E-01 | 3.14E-05 |
| ENSG00000142731 | PLK4 | 5.68E-01 | 2.72E-01 | 8.95E-12 | 6.14E-10 | 4.74E-01 | 4.98E-08 | 4.83E-01 | 7.97E-19 |
| ENSG00000112242 | E2F3 | 4.80E-01 | 2.82E-01 | 2.22E-08 | 4.10E-07 | 4.65E-01 | 1.35E-07 | 4.35E-01 | 1.25E-14 |
| ENSG00000120217 | CD274 | 4.99E-01 | 2.11E-01 | 4.82E-09 | 1.13E-07 | 2.77E-01 | 2.32E-01 | 4.20E-01 | 2.14E-13 |
| ENSG00000132437 | DDC | 4.59E-01 | 2.14E-01 | 1.09E-07 | 1.51E-06 | 3.47E-01 | 3.23E-03 | 2.34E-01 | 9.36E-03 |
| ENSG00000121297 | TSHZ3 | 5.38E-01 | 2.30E-01 | 1.65E-10 | 6.75E-09 | 5.23E-01 | 1.79E-10 | 3.39E-01 | 7.41E-08 |
| ENSG00000101916 | TLR8 | 5.90E-01 | 2.30E-01 | 8.26E-13 | 8.40E-11 | 4.00E-01 | 5.41E-05 | 2.81E-01 | 9.00E-05 |
| ENSG00000143179 | UCK2 | 4.07E-01 | 2.11E-01 | 3.34E-06 | 2.73E-05 | 3.03E-01 | 5.36E-02 | 3.34E-01 | 1.45E-07 |
| ENSG00000161888 | SPC24 | 4.98E-01 | 2.19E-01 | 5.45E-09 | 1.25E-07 | 4.88E-01 | 1.07E-08 | 4.08E-01 | 1.67E-12 |
| ENSG00000102384 | CENPI | 6.49E-01 | 3.89E-01 | 6.18E-16 | 2.60E-13 | 5.22E-01 | 2.02E-10 | 5.54E-01 | 1.68E-26 |
| ENSG00000169946 | ZFPM2 | 5.27E-01 | 2.24E-01 | 4.46E-10 | 1.56E-08 | 4.99E-01 | 3.22E-09 | 4.02E-01 | 4.76E-12 |
| ENSG00000198932 | GPRASP1 | 5.29E-01 | 2.41E-01 | 3.93E-10 | 1.40E-08 | 3.90E-01 | 1.23E-04 | 4.92E-01 | 1.08E-19 |
| ENSG00000090861 | AARS | 5.25E-01 | 2.09E-01 | 5.41E-10 | 1.82E-08 | 4.52E-01 | 4.88E-07 | 2.85E-01 | 5.45E-05 |
| ENSG00000188803 | SHISA6 | 4.99E-01 | 2.18E-01 | 4.73E-09 | 1.11E-07 | 4.06E-01 | 3.25E-05 | 3.44E-01 | 3.92E-08 |
| ENSG00000085999 | RAD54L | 5.65E-01 | 3.35E-01 | 1.18E-11 | 7.63E-10 | 5.44E-01 | 1.24E-11 | 4.94E-01 | 5.99E-20 |
| ENSG00000170871 | KIAA0232 | 4.61E-01 | 2.67E-01 | 9.25E-08 | 1.32E-06 | 3.75E-01 | 4.13E-04 | 2.06E-01 | 1.05E-01 |
| ENSG00000196584 | XRCC2 | 5.16E-01 | 2.79E-01 | 1.18E-09 | 3.50E-08 | 4.85E-01 | 1.61E-08 | 4.09E-01 | 1.60E-12 |
| ENSG00000027869 | SH2D2A | 5.64E-01 | 2.60E-01 | 1.38E-11 | 8.68E-10 | 5.22E-01 | 2.22E-10 | 3.84E-01 | 9.11E-11 |
| ENSG00000111641 | NOP2 | 4.77E-01 | 3.53E-01 | 2.75E-08 | 4.79E-07 | 3.95E-01 | 8.51E-05 | 3.88E-01 | 5.21E-11 |
| ENSG00000120526 | NUDCD1 | 5.07E-01 | 2.39E-01 | 2.64E-09 | 6.78E-08 | 5.07E-01 | 1.36E-09 | 4.33E-01 | 2.12E-14 |
| ENSG00000222014 | RAB6C | 4.93E-01 | 2.17E-01 | 7.81E-09 | 1.69E-07 | 4.20E-01 | 9.44E-06 | 3.79E-01 | 2.18E-10 |
| ENSG00000082212 | ME2 | 5.47E-01 | 2.86E-01 | 7.04E-11 | 3.29E-09 | 4.62E-01 | 1.88E-07 | 4.86E-01 | 4.17E-19 |
| ENSG00000143870 | PDIA6 | 6.27E-01 | 4.00E-01 | 1.17E-14 | 2.79E-12 | 5.37E-01 | 3.14E-11 | 4.96E-01 | 3.62E-20 |
| ENSG00000197256 | KANK2 | 5.85E-01 | 3.14E-01 | 1.42E-12 | 1.33E-10 | 5.05E-01 | 1.55E-09 | 5.80E-01 | 7.51E-30 |
| ENSG00000180198 | RCC1 | 5.13E-01 | 2.38E-01 | 1.54E-09 | 4.35E-08 | 4.60E-01 | 2.14E-07 | 2.64E-01 | 5.35E-04 |
| ENSG00000183527 | PSMG1 | 6.04E-01 | 3.16E-01 | 1.75E-13 | 2.56E-11 | 5.01E-01 | 2.45E-09 | 3.41E-01 | 5.56E-08 |
| ENSG00000140368 | PSTPIP1 | 4.97E-01 | 2.03E-01 | 5.71E-09 | 1.30E-07 | 3.52E-01 | 2.15E-03 | 2.64E-01 | 5.22E-04 |
| ENSG00000197472 | ZNF695 | 4.61E-01 | 2.56E-01 | 9.28E-08 | 1.32E-06 | 3.83E-01 | 2.18E-04 | 4.32E-01 | 2.50E-14 |
| ENSG00000101082 | SLA2 | 5.54E-01 | 2.45E-01 | 3.46E-11 | 1.82E-09 | 4.94E-01 | 5.52E-09 | 3.01E-01 | 9.10E-06 |
| ENSG00000124207 | CSE1L | 4.81E-01 | 2.21E-01 | 2.09E-08 | 3.89E-07 | 4.99E-01 | 3.31E-09 | 4.47E-01 | 1.29E-15 |
| ENSG00000065675 | PRKCQ | 5.51E-01 | 2.11E-01 | 4.78E-11 | 2.37E-09 | 4.96E-01 | 4.84E-09 | 4.75E-01 | 3.84E-18 |
| ENSG00000071051 | NCK2 | 3.67E-01 | 2.10E-01 | 3.22E-05 | 1.84E-04 | 3.13E-01 | 2.97E-02 | 2.96E-01 | 1.61E-05 |
| ENSG00000157554 | ERG | 5.29E-01 | 2.36E-01 | 3.80E-10 | 1.36E-08 | 4.96E-01 | 4.43E-09 | 4.06E-01 | 2.72E-12 |
| ENSG00000120802 | TMPO | 4.70E-01 | 2.44E-01 | 4.72E-08 | 7.53E-07 | 5.23E-01 | 1.81E-10 | 3.97E-01 | 1.11E-11 |
| ENSG00000138346 | DNA2 | 4.60E-01 | 2.77E-01 | 1.01E-07 | 1.42E-06 | 5.18E-01 | 3.37E-10 | 4.49E-01 | 8.27E-16 |
| ENSG00000171631 | P2RY6 | 5.71E-01 | 2.76E-01 | 6.22E-12 | 4.55E-10 | 3.62E-01 | 1.08E-03 | 5.22E-01 | 8.28E-23 |
| ENSG00000101182 | PSMA7 | 4.98E-01 | 2.41E-01 | 5.31E-09 | 1.23E-07 | 4.94E-01 | 5.60E-09 | 2.97E-01 | 1.49E-05 |
| ENSG00000068724 | TTC7A | 4.89E-01 | 2.04E-01 | 1.10E-08 | 2.22E-07 | 3.75E-01 | 3.98E-04 | 3.33E-01 | 1.56E-07 |

**Table S3. Prediction performance for the 1011 genes on validation and test sets.**

| Ensembl_gene_id | Hgnc symbol | Coef (valid) | Rsq (valid) | P-val (valid) | adj.pval (valid) | Coef (internal) | adj.pval (internal) | Coef (external) | adj.pval (external) |
|---|---|---|---|---|---|---|---|---|---|
| ENSG00000179218 | CALR | 4.30E-01 | 2.58E-01 | 7.72E-07 | 7.93E-06 | 4.26E-01 | 5.77E-06 | 3.14E-01 | 1.84E-06 |
| ENSG00000163918 | RFC4 | 5.12E-01 | 3.29E-01 | 1.73E-09 | 4.76E-08 | 4.12E-01 | 2.02E-05 | 3.61E-01 | 3.21E-09 |
| ENSG00000132780 | NASP | 5.34E-01 | 3.10E-01 | 2.31E-10 | 9.02E-09 | 4.66E-01 | 1.25E-07 | 2.51E-01 | 1.94E-03 |
| ENSG00000089012 | SIRPG | 5.55E-01 | 2.35E-01 | 3.33E-11 | 1.77E-09 | 4.66E-01 | 1.19E-07 | 3.94E-01 | 1.83E-11 |
| ENSG00000167984 | NLRC3 | 4.67E-01 | 2.02E-01 | 5.98E-08 | 9.15E-07 | 3.49E-01 | 2.68E-03 | 3.23E-01 | 5.89E-07 |
| ENSG00000144229 | THSD7B | 5.49E-01 | 2.32E-01 | 6.04E-11 | 2.88E-09 | 4.94E-01 | 5.75E-09 | 5.43E-01 | 3.15E-25 |
| ENSG00000183696 | UPP1 | 4.20E-01 | 2.29E-01 | 1.51E-06 | 1.41E-05 | 3.60E-01 | 1.28E-03 | 3.11E-01 | 2.73E-06 |
| ENSG00000114670 | NEK11 | 4.35E-01 | 2.47E-01 | 5.56E-07 | 5.98E-06 | 2.85E-01 | 1.49E-01 | 4.29E-01 | 4.14E-14 |
| ENSG00000100162 | CENPM | 5.02E-01 | 2.20E-01 | 3.72E-09 | 9.07E-08 | 5.20E-01 | 2.72E-10 | 3.14E-01 | 1.78E-06 |
| ENSG00000138190 | EXOC6 | 5.35E-01 | 2.19E-01 | 2.09E-10 | 8.30E-09 | 4.05E-01 | 3.76E-05 | 2.40E-01 | 5.59E-03 |
| ENSG00000139971 | C14orf37 | 5.72E-01 | 2.61E-01 | 6.14E-12 | 4.51E-10 | 5.14E-01 | 5.78E-10 | 4.07E-01 | 2.02E-12 |
| ENSG00000143321 | HDGF | 4.22E-01 | 2.09E-01 | 1.28E-06 | 1.22E-05 | 2.52E-01 | 8.67E-01 | 2.13E-01 | 5.70E-02 |
| ENSG00000112118 | MCM3 | 4.34E-01 | 2.46E-01 | 6.07E-07 | 6.44E-06 | 4.29E-01 | 4.38E-06 | 3.40E-01 | 6.24E-08 |
| ENSG00000178409 | BEND3 | 4.50E-01 | 2.34E-01 | 2.03E-07 | 2.57E-06 | 3.52E-01 | 2.25E-03 | 3.37E-01 | 1.01E-07 |
| ENSG00000132122 | SPATA6 | 5.23E-01 | 2.15E-01 | 6.27E-10 | 2.06E-08 | 3.65E-01 | 8.71E-04 | 3.37E-01 | 9.39E-08 |
| ENSG00000169744 | LDB2 | 4.71E-01 | 2.12E-01 | 4.26E-08 | 6.90E-07 | 4.98E-01 | 3.80E-09 | 4.13E-01 | 7.39E-13 |
| ENSG00000078070 | MCCC1 | 4.90E-01 | 2.91E-01 | 1.03E-08 | 2.10E-07 | 4.54E-01 | 4.07E-07 | 2.01E-01 | 1.51E-01 |
| ENSG00000154582 | TCEB1 | 4.96E-01 | 2.02E-01 | 6.42E-09 | 1.43E-07 | 3.45E-01 | 3.67E-03 | 3.09E-01 | 3.71E-06 |
| ENSG00000164430 | MB21D1 | 4.68E-01 | 2.45E-01 | 5.36E-08 | 8.37E-07 | 3.18E-01 | 2.19E-02 | 3.43E-01 | 4.38E-08 |
| ENSG00000019144 | PHLDB1 | 4.96E-01 | 2.51E-01 | 6.40E-09 | 1.43E-07 | 4.78E-01 | 3.43E-08 | 5.08E-01 | 2.43E-21 |
| ENSG00000114554 | PLXNA1 | 5.17E-01 | 2.13E-01 | 1.12E-09 | 3.36E-08 | 2.13E-01 | 1.00E+00 | 1.12E-01 | 1.00E+00 |
| ENSG00000149050 | ZNF214 | 5.06E-01 | 2.35E-01 | 2.73E-09 | 6.96E-08 | 4.26E-01 | 5.78E-06 | 4.14E-01 | 6.27E-13 |
| ENSG00000174132 | FAM174A | 4.27E-01 | 2.10E-01 | 9.38E-07 | 9.34E-06 | 4.85E-01 | 1.57E-08 | 4.67E-01 | 2.42E-17 |
| ENSG00000128944 | KNSTRN | 6.02E-01 | 3.56E-01 | 2.14E-13 | 2.99E-11 | 5.70E-01 | 3.59E-13 | 4.08E-01 | 1.86E-12 |
| ENSG00000132622 | HSPA12B | 6.05E-01 | 3.03E-01 | 1.49E-13 | 2.26E-11 | 4.96E-01 | 4.64E-09 | 6.06E-01 | 1.67E-33 |
| ENSG00000131042 | LILRB2 | 5.27E-01 | 2.35E-01 | 4.38E-10 | 1.54E-08 | 2.85E-01 | 1.51E-01 | 4.19E-01 | 2.38E-13 |
| ENSG00000117625 | RCOR3 | 4.50E-01 | 2.09E-01 | 1.95E-07 | 2.48E-06 | 3.55E-01 | 1.84E-03 | 1.80E-01 | 6.91E-01 |
| ENSG00000197646 | PDCD1LG2 | 4.98E-01 | 2.24E-01 | 5.52E-09 | 1.26E-07 | 4.20E-01 | 9.61E-06 | 4.09E-01 | 1.61E-12 |
| ENSG00000163820 | FYCO1 | 4.50E-01 | 2.34E-01 | 1.99E-07 | 2.53E-06 | 3.93E-01 | 9.93E-05 | 1.60E-01 | 1.00E+00 |
| ENSG00000100154 | TTC28 | 5.59E-01 | 3.29E-01 | 2.11E-11 | 1.20E-09 | 4.49E-01 | 6.36E-07 | 4.63E-01 | 5.43E-17 |
| ENSG00000104356 | POP1 | 4.57E-01 | 2.61E-01 | 1.21E-07 | 1.66E-06 | 5.18E-01 | 3.35E-10 | 3.89E-01 | 4.25E-11 |
| ENSG00000107937 | GTPBP4 | 5.77E-01 | 3.64E-01 | 3.47E-12 | 2.83E-10 | 5.09E-01 | 1.08E-09 | 4.52E-01 | 5.34E-16 |
| ENSG00000102898 | NUTF2 | 4.14E-01 | 2.05E-01 | 2.10E-06 | 1.86E-05 | 3.66E-01 | 7.82E-04 | 3.18E-01 | 1.14E-06 |
| ENSG00000196782 | MAML3 | 4.24E-01 | 2.25E-01 | 1.12E-06 | 1.08E-05 | 2.95E-01 | 8.65E-02 | 3.13E-01 | 2.07E-06 |
| ENSG00000153898 | MCOLN2 | 4.55E-01 | 2.06E-01 | 1.40E-07 | 1.88E-06 | 3.26E-01 | 1.32E-02 | 4.36E-01 | 1.17E-14 |
| ENSG00000172167 | MTBP | 5.35E-01 | 2.96E-01 | 2.24E-10 | 8.80E-09 | 4.00E-01 | 5.66E-05 | 4.08E-01 | 1.86E-12 |
| ENSG00000143621 | ILF2 | 3.96E-01 | 2.10E-01 | 6.39E-06 | 4.66E-05 | 3.76E-01 | 3.69E-04 | 3.73E-01 | 5.51E-10 |
| ENSG00000112419 | PHACTR2 | 5.35E-01 | 2.07E-01 | 2.27E-10 | 8.90E-09 | 1.47E-01 | 1.00E+00 | 1.36E-01 | 1.00E+00 |
| ENSG00000149380 | P4HA3 | 4.88E-01 | 2.21E-01 | 1.22E-08 | 2.44E-07 | 5.24E-01 | 1.72E-10 | 4.71E-01 | 1.05E-17 |
| ENSG00000130204 | TOMM40 | 4.26E-01 | 2.04E-01 | 9.90E-07 | 9.78E-06 | 3.17E-01 | 2.32E-02 | 1.10E-01 | 1.00E+00 |
| ENSG00000138182 | KIF20B | 4.47E-01 | 2.19E-01 | 2.45E-07 | 3.00E-06 | 4.72E-01 | 6.47E-08 | 3.00E-01 | 1.04E-05 |
| ENSG00000172935 | MRGPRF | 5.24E-01 | 2.06E-01 | 5.68E-10 | 1.89E-08 | 5.64E-01 | 8.06E-13 | 3.24E-01 | 5.07E-07 |
| ENSG00000115596 | WNT6 | 3.89E-01 | 2.21E-01 | 9.34E-06 | 6.40E-05 | 4.19E-01 | 1.12E-05 | 2.82E-01 | 7.86E-05 |
| ENSG00000156469 | MTERF3 | 5.14E-01 | 2.24E-01 | 1.42E-09 | 4.05E-08 | 5.01E-01 | 2.68E-09 | 3.79E-01 | 1.99E-10 |
| ENSG00000182199 | SHMT2 | 5.04E-01 | 2.54E-01 | 3.22E-09 | 8.00E-08 | 4.60E-01 | 2.19E-07 | 3.27E-01 | 3.50E-07 |
| ENSG00000121671 | CRY2 | 5.78E-01 | 3.17E-01 | 3.27E-12 | 2.70E-10 | 5.46E-01 | 1.00E-11 | 5.61E-01 | 2.24E-27 |
| ENSG00000176083 | ZNF683 | 5.20E-01 | 2.06E-01 | 8.33E-10 | 2.63E-08 | 4.34E-01 | 2.79E-06 | 3.89E-01 | 4.10E-11 |
| ENSG00000149503 | INCENP | 5.86E-01 | 2.72E-01 | 1.32E-12 | 1.26E-10 | 4.47E-01 | 8.02E-07 | 3.00E-01 | 1.08E-05 |
| ENSG00000164941 | INTS8 | 4.91E-01 | 2.44E-01 | 9.55E-09 | 1.98E-07 | 5.33E-01 | 5.21E-11 | 4.01E-01 | 5.52E-12 |
| ENSG00000149231 | CCDC82 | 3.93E-01 | 2.33E-01 | 7.68E-06 | 5.45E-05 | 4.19E-01 | 1.08E-05 | 2.26E-01 | 2.03E-02 |
| ENSG00000144741 | SLC25A26 | 3.89E-01 | 2.05E-01 | 9.63E-06 | 6.57E-05 | 4.95E-01 | 5.29E-09 | 2.25E-01 | 2.21E-02 |
| ENSG00000124571 | XPO5 | 5.33E-01 | 2.91E-01 | 2.71E-10 | 1.02E-08 | 4.62E-01 | 1.89E-07 | 4.55E-01 | 2.48E-16 |
| ENSG00000189091 | SF3B3 | 4.39E-01 | 2.48E-01 | 4.24E-07 | 4.79E-06 | 5.34E-01 | 4.86E-11 | 2.94E-01 | 2.00E-05 |
| ENSG00000112029 | FBXO5 | 5.11E-01 | 2.45E-01 | 1.75E-09 | 4.79E-08 | 5.58E-01 | 1.83E-12 | 4.30E-01 | 3.27E-14 |
| ENSG00000165801 | ARHGEF40 | 5.68E-01 | 2.84E-01 | 8.83E-12 | 6.08E-10 | 4.45E-01 | 9.59E-07 | 5.46E-01 | 1.40E-25 |
| ENSG00000143924 | EML4 | 6.03E-01 | 3.34E-01 | 2.07E-13 | 2.91E-11 | 5.64E-01 | 7.86E-13 | 4.27E-01 | 6.05E-14 |
| ENSG00000181513 | ACBD4 | 5.24E-01 | 2.38E-01 | 5.72E-10 | 1.89E-08 | 3.95E-01 | 8.27E-05 | 3.88E-01 | 4.72E-11 |
| ENSG00000106105 | GARS | 5.57E-01 | 3.03E-01 | 2.66E-11 | 1.46E-09 | 4.20E-01 | 9.62E-06 | 3.07E-01 | 4.43E-06 |
| ENSG00000155561 | NUP205 | 4.36E-01 | 2.26E-01 | 5.10E-07 | 5.56E-06 | 4.80E-01 | 2.76E-08 | 4.05E-01 | 3.12E-12 |
| ENSG00000184575 | XPOT | 5.18E-01 | 2.06E-01 | 9.53E-10 | 2.92E-08 | 4.63E-01 | 1.66E-07 | 4.52E-01 | 5.52E-16 |
| ENSG00000169814 | BTD | 4.66E-01 | 2.17E-01 | 6.47E-08 | 9.76E-07 | 3.12E-01 | 3.06E-02 | 4.01E-01 | 6.00E-12 |
| ENSG00000174292 | TNK1 | 5.57E-01 | 2.16E-01 | 2.82E-11 | 1.55E-09 | 2.94E-01 | 8.97E-02 | 2.41E-01 | 4.98E-03 |
| ENSG00000148429 | USP6NL | 3.98E-01 | 2.64E-01 | 5.53E-06 | 4.13E-05 | 3.88E-01 | 1.45E-04 | 2.19E-01 | 3.45E-02 |
| ENSG00000103005 | USB1 | 5.20E-01 | 3.00E-01 | 8.36E-10 | 2.63E-08 | 4.15E-01 | 1.57E-05 | 3.49E-01 | 1.93E-08 |
| ENSG00000165533 | TTC8 | 4.84E-01 | 2.36E-01 | 1.63E-08 | 3.12E-07 | 5.07E-01 | 1.24E-09 | 2.52E-01 | 1.80E-03 |
| ENSG00000163872 | YEATS2 | 5.21E-01 | 2.84E-01 | 7.53E-10 | 2.42E-08 | 4.09E-01 | 2.62E-05 | 2.98E-01 | 1.26E-05 |
| ENSG00000120539 | MASTL | 5.80E-01 | 3.21E-01 | 2.61E-12 | 2.23E-10 | 3.74E-01 | 4.23E-04 | 5.28E-01 | 1.56E-23 |
| ENSG00000110880 | CORO1C | 5.44E-01 | 2.19E-01 | 9.15E-11 | 4.06E-09 | 4.16E-01 | 1.34E-05 | 3.38E-01 | 8.76E-08 |
| ENSG00000166523 | CLEC4E | 5.78E-01 | 3.08E-01 | 3.20E-12 | 2.65E-10 | 3.63E-01 | 9.71E-04 | 2.73E-01 | 2.21E-04 |

**Table S3. Prediction performance for the 1011 genes on validation and test sets.**

| Ensembl_gene_id | Hgnc symbol | Coef (valid) | Rsq (valid) | P-val (valid) | adj.pval (valid) | Coef (internal) | adj.pval (internal) | Coef (external) | adj.pval (external) |
|---|---|---|---|---|---|---|---|---|---|
| ENSG00000088543 | C3orf18 | 5.32E-01 | 2.68E-01 | 2.83E-10 | 1.06E-08 | 4.92E-01 | 7.08E-09 | 4.64E-01 | 3.95E-17 |
| ENSG00000152455 | SUV39H2 | 5.73E-01 | 3.72E-01 | 5.58E-12 | 4.15E-10 | 5.37E-01 | 3.33E-11 | 4.84E-01 | 6.23E-19 |
| ENSG00000102900 | NUP93 | 6.36E-01 | 4.18E-01 | 3.47E-15 | 1.12E-12 | 4.99E-01 | 3.38E-09 | 3.82E-01 | 1.23E-10 |
| ENSG00000113407 | TARS | 5.39E-01 | 2.45E-01 | 1.52E-10 | 6.27E-09 | 3.70E-01 | 6.07E-04 | 3.77E-01 | 2.78E-10 |
| ENSG00000168411 | RFWD3 | 5.52E-01 | 3.25E-01 | 4.59E-11 | 2.28E-09 | 5.33E-01 | 5.12E-11 | 4.46E-01 | 1.69E-15 |
| ENSG00000146410 | MTFR2 | 6.33E-01 | 3.99E-01 | 5.26E-15 | 1.50E-12 | 6.00E-01 | 3.57E-15 | 5.76E-01 | 2.88E-29 |
| ENSG00000217128 | FNIP1 | 4.61E-01 | 2.21E-01 | 9.17E-08 | 1.31E-06 | 2.67E-01 | 3.95E-01 | 3.04E-01 | 6.56E-06 |
| ENSG00000099901 | RANBP1 | 5.15E-01 | 2.06E-01 | 1.25E-09 | 3.67E-08 | 4.91E-01 | 7.87E-09 | 2.96E-01 | 1.68E-05 |
| ENSG00000120438 | TCP1 | 4.56E-01 | 2.34E-01 | 1.29E-07 | 1.75E-06 | 4.93E-01 | 6.55E-09 | 3.37E-01 | 1.02E-07 |
| ENSG00000157796 | WDR19 | 5.52E-01 | 2.93E-01 | 4.56E-11 | 2.28E-09 | 3.62E-01 | 1.10E-03 | 3.72E-01 | 5.99E-10 |
| ENSG00000220205 | VAMP2 | 4.97E-01 | 2.07E-01 | 5.75E-09 | 1.31E-07 | 4.07E-01 | 3.06E-05 | 3.94E-01 | 1.76E-11 |
| ENSG00000177494 | ZBED2 | 5.70E-01 | 2.55E-01 | 7.13E-12 | 5.09E-10 | 4.18E-01 | 1.20E-05 | 3.99E-01 | 8.05E-12 |
| ENSG00000128655 | PDE11A | 4.65E-01 | 2.01E-01 | 6.78E-08 | 1.01E-06 | 3.30E-01 | 9.99E-03 | 1.71E-01 | 1.00E+00 |
| ENSG00000155100 | OTUD6B | 4.63E-01 | 2.01E-01 | 7.80E-08 | 1.14E-06 | 2.67E-01 | 3.97E-01 | 2.99E-01 | 1.13E-05 |
| ENSG00000196503 | ARL9 | 4.66E-01 | 2.63E-01 | 6.38E-08 | 9.67E-07 | 3.79E-01 | 3.01E-04 | 2.05E-01 | 1.09E-01 |
| ENSG00000111203 | ITFG2 | 3.44E-01 | 3.31E-01 | 1.05E-04 | 5.06E-04 | 3.81E-01 | 2.48E-04 | 1.85E-02 | 1.00E+00 |
| ENSG00000137571 | SLCO5A1 | 5.13E-01 | 2.06E-01 | 1.48E-09 | 4.21E-08 | 4.93E-01 | 6.18E-09 | 4.02E-01 | 4.67E-12 |
| ENSG00000130826 | DKC1 | 5.58E-01 | 2.82E-01 | 2.50E-11 | 1.40E-09 | 5.66E-01 | 5.77E-13 | 4.55E-01 | 2.87E-16 |
| ENSG00000151465 | CDC123 | 5.46E-01 | 3.90E-01 | 8.15E-11 | 3.77E-09 | 4.47E-01 | 8.07E-07 | 3.16E-01 | 1.44E-06 |
| ENSG00000214160 | ALG3 | 5.92E-01 | 2.83E-01 | 7.22E-13 | 7.65E-11 | 4.10E-01 | 2.44E-05 | 2.78E-01 | 1.25E-04 |
| ENSG00000064102 | ASUN | 4.05E-01 | 2.74E-01 | 3.74E-06 | 3.00E-05 | 3.96E-01 | 7.81E-05 | 2.44E-01 | 3.69E-03 |
| ENSG00000156787 | TBC1D31 | 5.48E-01 | 2.87E-01 | 6.62E-11 | 3.12E-09 | 4.58E-01 | 2.80E-07 | 4.56E-01 | 2.37E-16 |
| ENSG00000109674 | NEIL3 | 5.12E-01 | 2.65E-01 | 1.67E-09 | 4.63E-08 | 5.51E-01 | 4.70E-12 | 4.49E-01 | 8.67E-16 |
| ENSG00000116885 | OSCP1 | 5.26E-01 | 2.62E-01 | 4.83E-10 | 1.66E-08 | 2.69E-01 | 3.55E-01 | 4.16E-01 | 4.10E-13 |
| ENSG00000096996 | IL12RB1 | 5.51E-01 | 2.41E-01 | 5.05E-11 | 2.49E-09 | 3.74E-01 | 4.54E-04 | 2.66E-01 | 4.61E-04 |
| ENSG00000204397 | CARD16 | 5.50E-01 | 2.20E-01 | 5.34E-11 | 2.60E-09 | 3.07E-01 | 4.22E-02 | 2.00E-01 | 1.65E-01 |
| ENSG00000185880 | TRIM69 | 5.45E-01 | 2.62E-01 | 8.25E-11 | 3.79E-09 | 2.26E-01 | 1.00E+00 | 2.64E-01 | 5.52E-04 |
| ENSG00000159110 | IFNAR2 | 4.90E-01 | 2.20E-01 | 1.03E-08 | 2.11E-07 | 4.36E-01 | 2.28E-06 | 3.84E-01 | 1.02E-10 |
| ENSG00000112079 | STK38 | 5.04E-01 | 2.19E-01 | 3.28E-09 | 8.09E-08 | 3.47E-01 | 3.04E-03 | 3.88E-01 | 5.22E-11 |
| ENSG00000108688 | CCL7 | 5.07E-01 | 2.04E-01 | 2.61E-09 | 6.72E-08 | 2.55E-01 | 7.55E-01 | 3.42E-01 | 4.65E-08 |
| ENSG00000076685 | NT5C2 | 4.32E-01 | 2.01E-01 | 6.56E-07 | 6.88E-06 | 3.32E-01 | 8.71E-03 | 4.40E-01 | 4.91E-15 |
| ENSG00000143977 | SNRPG | 4.52E-01 | 2.40E-01 | 1.74E-07 | 2.26E-06 | 3.11E-01 | 3.43E-02 | 2.59E-01 | 8.81E-04 |
| ENSG00000197380 | DACT3 | 5.88E-01 | 2.71E-01 | 1.07E-12 | 1.05E-10 | 6.13E-01 | 4.24E-16 | 4.34E-01 | 1.78E-14 |
| ENSG00000188389 | PDCD1 | 5.26E-01 | 2.71E-01 | 4.73E-10 | 1.64E-08 | 3.73E-01 | 4.66E-04 | 3.38E-01 | 8.51E-08 |
| ENSG00000139722 | VPS37B | 4.63E-01 | 2.05E-01 | 7.96E-08 | 1.16E-06 | 3.27E-01 | 1.20E-02 | 2.14E-01 | 5.58E-02 |
| ENSG00000165609 | NUDT5 | 4.48E-01 | 2.76E-01 | 2.25E-07 | 2.80E-06 | 3.81E-01 | 2.62E-04 | 3.05E-01 | 5.60E-06 |
| ENSG00000130309 | COLGALT1 | 4.47E-01 | 2.18E-01 | 2.43E-07 | 2.99E-06 | 3.57E-01 | 1.50E-03 | 3.12E-01 | 2.42E-06 |
| ENSG00000103035 | PSMD7 | 5.99E-01 | 2.87E-01 | 3.19E-13 | 4.01E-11 | 4.56E-01 | 3.13E-07 | 4.32E-01 | 2.37E-14 |
| ENSG00000121211 | MND1 | 5.98E-01 | 3.50E-01 | 3.45E-13 | 4.13E-11 | 5.06E-01 | 1.38E-09 | 3.54E-01 | 9.52E-09 |
| ENSG00000145743 | FBXL17 | 4.41E-01 | 2.15E-01 | 3.79E-07 | 4.36E-06 | 3.31E-01 | 9.04E-03 | 3.01E-01 | 9.32E-06 |
| ENSG00000168404 | MLKL | 5.01E-01 | 2.42E-01 | 4.19E-09 | 1.00E-07 | 3.08E-01 | 4.02E-02 | 3.74E-01 | 4.36E-10 |
| ENSG00000105227 | PRX | 4.61E-01 | 2.05E-01 | 8.97E-08 | 1.28E-06 | 3.31E-01 | 9.30E-03 | 5.14E-01 | 5.90E-22 |
| ENSG00000103121 | CMC2 | 6.53E-01 | 4.11E-01 | 3.56E-16 | 1.76E-13 | 5.16E-01 | 4.36E-10 | 3.82E-01 | 1.37E-10 |
| ENSG00000166275 | C10orf32 | 6.18E-01 | 2.08E-01 | 3.50E-14 | 6.64E-12 | 4.86E-01 | 1.39E-08 | 5.20E-01 | 1.36E-22 |
| ENSG00000197324 | LRP10 | 4.27E-01 | 2.29E-01 | 9.38E-07 | 9.34E-06 | 3.70E-01 | 5.95E-04 | 2.61E-01 | 7.04E-04 |
| ENSG00000109686 | SH3D19 | 5.08E-01 | 2.06E-01 | 2.43E-09 | 6.30E-08 | 2.30E-01 | 1.00E+00 | 3.99E-01 | 8.48E-12 |
| ENSG00000165629 | ATP5C1 | 5.25E-01 | 3.12E-01 | 5.33E-10 | 1.80E-08 | 3.80E-01 | 2.67E-04 | 4.29E-01 | 4.53E-14 |
| ENSG00000134539 | KLRD1 | 5.82E-01 | 2.03E-01 | 2.06E-12 | 1.83E-10 | 4.42E-01 | 1.34E-06 | 3.92E-01 | 2.58E-11 |
| ENSG00000049249 | TNFRSF9 | 5.58E-01 | 2.46E-01 | 2.36E-11 | 1.33E-09 | 4.85E-01 | 1.57E-08 | 3.96E-01 | 1.32E-11 |
| ENSG00000101084 | C20orf24 | 6.13E-01 | 2.62E-01 | 5.86E-14 | 1.05E-11 | 4.67E-01 | 1.04E-07 | 2.94E-01 | 2.15E-05 |
| ENSG00000119403 | PHF19 | 4.25E-01 | 2.56E-01 | 1.05E-06 | 1.03E-05 | 3.28E-01 | 1.16E-02 | 2.45E-01 | 3.61E-03 |
| ENSG00000264324 |  | 3.33E-01 | 2.16E-01 | 1.77E-04 | 7.82E-04 | 4.13E-01 | 1.83E-05 |  |  |
| ENSG00000140451 | PIF1 | 6.65E-01 | 3.80E-01 | 6.73E-17 | 5.18E-14 | 5.66E-01 | 6.05E-13 | 4.12E-01 | 8.76E-13 |
| ENSG00000119686 | FLVCR2 | 5.75E-01 | 2.57E-01 | 4.54E-12 | 3.47E-10 | 3.68E-01 | 6.71E-04 | 3.78E-01 | 2.38E-10 |
| ENSG00000118729 | CASQ2 | 5.89E-01 | 2.68E-01 | 1.01E-12 | 9.95E-11 | 5.09E-01 | 1.04E-09 | 5.16E-01 | 3.45E-22 |
| ENSG00000189157 | FAM47E | 5.83E-01 | 3.09E-01 | 1.80E-12 | 1.63E-10 | 4.83E-01 | 1.88E-08 | 2.94E-01 | 2.04E-05 |
| ENSG00000006634 | DBF4 | 5.48E-01 | 2.88E-01 | 6.60E-11 | 3.12E-09 | 5.25E-01 | 1.44E-10 | 5.08E-01 | 2.59E-21 |
| ENSG00000054803 | CBLN4 | 4.64E-01 | 2.42E-01 | 7.28E-08 | 1.08E-06 | 4.57E-01 | 2.85E-07 | 4.19E-01 | 2.71E-13 |
| ENSG00000112214 | FHL5 | 5.08E-01 | 2.08E-01 | 2.30E-09 | 6.04E-08 | 4.41E-01 | 1.43E-06 | 3.93E-01 | 2.30E-11 |
| ENSG00000131876 | SNRPA1 | 5.37E-01 | 3.11E-01 | 1.79E-10 | 7.23E-09 | 3.25E-01 | 1.34E-02 | 4.31E-01 | 3.09E-14 |
| ENSG00000178295 | GEN1 | 6.07E-01 | 3.47E-01 | 1.19E-13 | 1.87E-11 | 5.09E-01 | 1.02E-09 | 4.92E-01 | 1.05E-19 |
| ENSG00000157680 | DGKI | 5.11E-01 | 2.02E-01 | 1.83E-09 | 4.97E-08 | 4.93E-01 | 6.52E-09 | 4.14E-01 | 6.58E-13 |
| ENSG00000135914 | HTR2B | 5.12E-01 | 2.72E-01 | 1.67E-09 | 4.63E-08 | 4.46E-01 | 8.49E-07 | 2.75E-01 | 1.76E-04 |
| ENSG00000141012 | GALNS | 4.60E-01 | 2.09E-01 | 9.80E-08 | 1.39E-06 | 2.52E-01 | 8.43E-01 | 2.72E-01 | 2.37E-04 |
| ENSG00000125454 | SLC25A19 | 5.44E-01 | 2.43E-01 | 9.31E-11 | 4.11E-09 | 4.75E-01 | 4.72E-08 | 4.09E-01 | 1.51E-12 |
| ENSG00000196839 | ADA | 5.34E-01 | 2.17E-01 | 2.40E-10 | 9.27E-09 | 3.28E-01 | 1.15E-02 | 3.36E-01 | 1.13E-07 |
| ENSG00000173258 | ZNF483 | 6.24E-01 | 2.78E-01 | 1.54E-14 | 3.49E-12 | 2.27E-01 | 1.00E+00 | 2.50E-01 | 2.27E-03 |
| ENSG00000049768 | FOXP3 | 4.98E-01 | 2.04E-01 | 5.17E-09 | 1.20E-07 | 4.02E-01 | 4.84E-05 | 3.84E-01 | 1.01E-10 |
| ENSG00000205045 | SLFN12L | 5.44E-01 | 2.13E-01 | 9.04E-11 | 4.04E-09 | 4.02E-01 | 4.85E-05 | 4.07E-01 | 1.96E-12 |

**Table S3. Prediction performance for the 1011 genes on validation and test sets.**

| Ensembl_gene_id | Hgnc symbol | Coef (valid) | Rsq (valid) | P-val (valid) | adj.pval (valid) | Coef (internal) | adj.pval (internal) | Coef (external) | adj.pval (external) |
|---|---|---|---|---|---|---|---|---|---|
| ENSG00000185480 | PARPBP | 5.69E-01 | 3.18E-01 | 7.65E-12 | 5.39E-10 | 4.90E-01 | 9.35E-09 | 4.80E-01 | 1.40E-18 |
| ENSG00000206052 | DOK6 | 5.44E-01 | 2.02E-01 | 9.30E-11 | 4.11E-09 | 4.32E-01 | 3.22E-06 | 3.18E-01 | 1.09E-06 |
| ENSG00000142599 | RERE | 5.91E-01 | 2.51E-01 | 7.33E-13 | 7.68E-11 | 1.65E-01 | 1.00E+00 | 2.51E-01 | 1.90E-03 |
| ENSG00000221818 | EBF2 | 4.88E-01 | 2.30E-01 | 1.22E-08 | 2.44E-07 | 3.71E-01 | 5.61E-04 | 3.41E-01 | 5.75E-08 |
| ENSG00000186496 | ZNF396 | 5.96E-01 | 2.88E-01 | 4.31E-13 | 5.03E-11 | 4.42E-01 | 1.35E-06 | 5.01E-01 | 1.29E-20 |
| ENSG00000180660 | MAB21L1 | 5.33E-01 | 3.22E-01 | 2.66E-10 | 1.00E-08 | 4.50E-01 | 5.85E-07 | 4.51E-01 | 5.94E-16 |
| ENSG00000086475 | SEPHS1 | 3.86E-01 | 2.18E-01 | 1.12E-05 | 7.50E-05 | 4.17E-01 | 1.25E-05 | 2.92E-01 | 2.54E-05 |
| ENSG00000119772 | DNMT3A | 5.52E-01 | 2.66E-01 | 4.49E-11 | 2.25E-09 | 4.13E-01 | 1.84E-05 | 3.12E-01 | 2.51E-06 |
| ENSG00000185338 | SOCS1 | 5.48E-01 | 2.54E-01 | 6.17E-11 | 2.93E-09 | 3.55E-01 | 1.74E-03 | 2.80E-01 | 9.99E-05 |
| ENSG00000163655 | GMPS | 4.25E-01 | 2.06E-01 | 1.08E-06 | 1.05E-05 | 4.46E-01 | 8.53E-07 | 3.40E-01 | 6.72E-08 |
| ENSG00000165695 | AK8 | 4.02E-01 | 2.13E-01 | 4.45E-06 | 3.46E-05 | 4.58E-01 | 2.57E-07 | 3.30E-01 | 2.34E-07 |
| ENSG00000012822 | CALCOCO1 | 4.29E-01 | 2.85E-01 | 8.09E-07 | 8.24E-06 | 4.91E-01 | 8.18E-09 | 5.73E-01 | 5.79E-29 |
| ENSG00000140463 | BBS4 | 4.87E-01 | 2.07E-01 | 1.26E-08 | 2.50E-07 | 4.07E-01 | 3.15E-05 | 3.16E-01 | 1.56E-06 |
| ENSG00000035141 | FAM136A | 5.10E-01 | 3.18E-01 | 2.03E-09 | 5.41E-08 | 4.51E-01 | 5.45E-07 | 5.03E-01 | 7.03E-21 |
| ENSG00000167522 | ANKRD11 | 4.83E-01 | 2.00E-01 | 1.80E-08 | 3.41E-07 | 2.35E-01 | 1.00E+00 | 9.37E-02 | 1.00E+00 |
| ENSG00000090520 | DNAJB11 | 5.98E-01 | 3.54E-01 | 3.61E-13 | 4.26E-11 | 3.79E-01 | 2.88E-04 | 3.46E-01 | 2.61E-08 |
| ENSG00000004864 | SLC25A13 | 5.35E-01 | 2.75E-01 | 2.08E-10 | 8.28E-09 | 3.87E-01 | 1.63E-04 | 3.96E-01 | 1.32E-11 |
| ENSG00000174282 | ZBTB4 | 5.63E-01 | 3.15E-01 | 1.46E-11 | 9.00E-10 | 4.09E-01 | 2.48E-05 | 4.67E-01 | 2.50E-17 |
| ENSG00000134684 | YARS | 4.77E-01 | 2.08E-01 | 2.76E-08 | 4.81E-07 | 4.60E-01 | 2.15E-07 | 2.93E-01 | 2.28E-05 |
| ENSG00000177602 | GSG2 | 6.83E-01 | 3.22E-01 | 4.71E-18 | 9.26E-15 | 5.77E-01 | 1.13E-13 | 5.64E-01 | 9.01E-28 |
| ENSG00000149136 | SSRP1 | 5.05E-01 | 2.30E-01 | 3.07E-09 | 7.72E-08 | 4.15E-01 | 1.48E-05 | 1.23E-01 | 1.00E+00 |
| ENSG00000187951 | ARHGAP11B | 5.13E-01 | 2.76E-01 | 1.58E-09 | 4.41E-08 | 4.81E-01 | 2.50E-08 | 3.21E-01 | 8.31E-07 |
| ENSG00000065427 | KARS | 4.81E-01 | 2.62E-01 | 2.05E-08 | 3.83E-07 | 4.40E-01 | 1.55E-06 | 2.82E-01 | 8.37E-05 |
| ENSG00000126067 | PSMB2 | 5.92E-01 | 3.03E-01 | 6.87E-13 | 7.33E-11 | 4.78E-01 | 3.26E-08 | 3.70E-01 | 7.92E-10 |
| ENSG00000111639 | MRPL51 | 4.50E-01 | 2.25E-01 | 2.02E-07 | 2.56E-06 | 1.76E-01 | 1.00E+00 | 8.39E-02 | 1.00E+00 |
| ENSG00000141076 | CIRH1A | 4.77E-01 | 2.83E-01 | 2.75E-08 | 4.79E-07 | 4.68E-01 | 9.24E-08 | 3.80E-01 | 1.69E-10 |
| ENSG00000166598 | HSP90B1 | 4.98E-01 | 2.07E-01 | 5.38E-09 | 1.24E-07 | 4.25E-01 | 6.01E-06 | 3.05E-01 | 5.58E-06 |
| ENSG00000166272 | WBP1L | 6.67E-01 | 3.73E-01 | 5.06E-17 | 4.26E-14 | 5.21E-01 | 2.38E-10 | 5.62E-01 | 1.56E-27 |
| ENSG00000131828 | PDHA1 | 5.63E-01 | 2.12E-01 | 1.50E-11 | 9.17E-10 | 4.17E-01 | 1.28E-05 | 1.52E-01 | 1.00E+00 |
| ENSG00000146731 | CCT6A | 4.54E-01 | 2.21E-01 | 1.54E-07 | 2.04E-06 | 4.24E-01 | 7.18E-06 | 4.14E-01 | 5.84E-13 |
| ENSG00000116221 | MRPL37 | 4.63E-01 | 2.23E-01 | 7.78E-08 | 1.14E-06 | 4.21E-01 | 9.00E-06 | 8.72E-02 | 1.00E+00 |
| ENSG00000111196 | MAGOHB | 4.13E-01 | 2.17E-01 | 2.22E-06 | 1.95E-05 | 2.11E-01 | 1.00E+00 | 3.59E-01 | 4.07E-09 |
| ENSG00000166246 | C16orf71 | 4.51E-01 | 2.14E-01 | 1.86E-07 | 2.40E-06 | 4.28E-01 | 4.74E-06 | 2.98E-01 | 1.33E-05 |
| ENSG00000186310 | NAP1L3 | 4.57E-01 | 2.04E-01 | 1.20E-07 | 1.65E-06 | 4.87E-01 | 1.21E-08 | 3.79E-01 | 2.27E-10 |
| ENSG00000139180 | NDUFA9 | 3.49E-01 | 2.18E-01 | 7.97E-05 | 4.00E-04 | 2.92E-01 | 1.05E-01 | 3.51E-01 | 1.44E-08 |
| ENSG00000153879 | CEBPG | 4.76E-01 | 2.33E-01 | 2.91E-08 | 5.02E-07 | 4.00E-01 | 5.61E-05 | 3.89E-01 | 4.41E-11 |
| ENSG00000117305 | HMGCL | 4.29E-01 | 2.01E-01 | 8.11E-07 | 8.26E-06 | 3.03E-01 | 5.29E-02 | 3.89E-01 | 4.40E-11 |
| ENSG00000166965 | RCCD1 | 4.87E-01 | 2.03E-01 | 1.30E-08 | 2.56E-07 | 3.83E-01 | 2.19E-04 | 2.15E-01 | 4.78E-02 |
| ENSG00000166889 | PATL1 | 4.90E-01 | 2.75E-01 | 1.04E-08 | 2.12E-07 | 4.19E-01 | 1.06E-05 | 2.55E-01 | 1.32E-03 |
| ENSG00000168038 | ULK4 | 4.79E-01 | 2.10E-01 | 2.41E-08 | 4.34E-07 | 3.18E-01 | 2.09E-02 | 2.52E-01 | 1.85E-03 |
| ENSG00000149575 | SCN2B | 6.02E-01 | 2.39E-01 | 2.26E-13 | 3.10E-11 | 5.46E-01 | 9.63E-12 | 5.85E-01 | 1.41E-30 |
| ENSG00000138035 | PNPT1 | 4.87E-01 | 2.53E-01 | 1.29E-08 | 2.54E-07 | 3.96E-01 | 7.71E-05 | 4.16E-01 | 4.61E-13 |
| ENSG00000115091 | ACTR3 | 5.50E-01 | 2.38E-01 | 5.52E-11 | 2.67E-09 | 3.95E-01 | 7.97E-05 | 4.38E-01 | 7.45E-15 |
| ENSG00000179934 | CCR8 | 5.66E-01 | 2.53E-01 | 1.08E-11 | 7.06E-10 | 3.51E-01 | 2.31E-03 | 3.07E-01 | 4.44E-06 |
| ENSG00000132964 | CDK8 | 5.45E-01 | 2.68E-01 | 8.87E-11 | 3.98E-09 | 3.95E-01 | 8.49E-05 | 4.81E-01 | 1.16E-18 |
| ENSG00000125977 | EIF2S2 | 6.23E-01 | 3.38E-01 | 1.84E-14 | 3.92E-12 | 4.25E-01 | 6.48E-06 | 3.25E-01 | 4.52E-07 |
| ENSG00000159593 | NAE1 | 5.62E-01 | 3.32E-01 | 1.59E-11 | 9.61E-10 | 4.37E-01 | 2.02E-06 | 3.55E-01 | 7.77E-09 |
| ENSG00000188177 | ZC3H6 | 5.40E-01 | 2.01E-01 | 1.38E-10 | 5.83E-09 | 3.44E-01 | 3.78E-03 | 3.01E-01 | 9.27E-06 |
| ENSG00000165118 | C9orf64 | 4.56E-01 | 2.44E-01 | 1.32E-07 | 1.78E-06 | 4.13E-01 | 1.85E-05 | 4.35E-01 | 1.48E-14 |
| ENSG00000183765 | CHEK2 | 5.18E-01 | 2.12E-01 | 9.59E-10 | 2.93E-08 | 3.84E-01 | 2.06E-04 | 3.94E-01 | 1.87E-11 |
| ENSG00000255833 | TIFAB | 5.62E-01 | 2.20E-01 | 1.65E-11 | 9.89E-10 | 4.43E-01 | 1.16E-06 | 3.04E-01 | 6.30E-06 |
| ENSG00000197753 | LHFPL5 | 4.16E-01 | 2.39E-01 | 1.85E-06 | 1.67E-05 | 3.37E-01 | 6.13E-03 | 1.05E-01 | 1.00E+00 |
| ENSG00000132603 | NIP7 | 4.72E-01 | 2.65E-01 | 4.03E-08 | 6.60E-07 | 4.80E-01 | 2.64E-08 | 3.49E-01 | 1.91E-08 |
| ENSG00000145979 | TBC1D7 | 4.34E-01 | 2.09E-01 | 5.99E-07 | 6.37E-06 | 3.62E-01 | 1.10E-03 | 3.69E-01 | 1.03E-09 |
| ENSG00000100416 | TRMU | 4.74E-01 | 2.17E-01 | 3.60E-08 | 6.01E-07 | 4.16E-01 | 1.45E-05 | 1.59E-01 | 1.00E+00 |
| ENSG00000115241 | PPM1G | 5.31E-01 | 2.46E-01 | 3.17E-10 | 1.17E-08 | 5.03E-01 | 2.06E-09 | 2.52E-01 | 1.82E-03 |
| ENSG00000028528 | SNX1 | 4.09E-01 | 2.07E-01 | 2.95E-06 | 2.46E-05 | 3.92E-01 | 1.03E-04 | 2.02E-01 | 1.39E-01 |
| ENSG00000113391 | FAM172A | 5.81E-01 | 2.34E-01 | 2.34E-12 | 2.02E-10 | 3.03E-01 | 5.37E-02 | 3.60E-01 | 3.77E-09 |
| ENSG00000138092 | CENPO | 5.27E-01 | 2.43E-01 | 4.37E-10 | 1.53E-08 | 4.82E-01 | 2.26E-08 | 5.07E-01 | 2.80E-21 |
| ENSG00000137727 | ARHGAP20 | 5.26E-01 | 2.61E-01 | 4.83E-10 | 1.66E-08 | 4.68E-01 | 1.01E-07 | 4.33E-01 | 1.84E-14 |
| ENSG00000065600 | TMEM206 | 4.13E-01 | 2.07E-01 | 2.34E-06 | 2.04E-05 | 3.75E-01 | 3.91E-04 | 4.41E-01 | 4.67E-15 |
| ENSG00000159055 | MIS18A | 4.86E-01 | 2.26E-01 | 1.35E-08 | 2.64E-07 | 4.39E-01 | 1.64E-06 | 1.59E-01 | 1.00E+00 |
| ENSG00000159131 | GART | 4.60E-01 | 2.14E-01 | 9.80E-08 | 1.39E-06 | 5.47E-01 | 7.96E-12 | 3.11E-01 | 2.72E-06 |
| ENSG00000006283 | CACNA1G | 5.85E-01 | 2.75E-01 | 1.49E-12 | 1.38E-10 | 4.59E-01 | 2.39E-07 | 4.32E-01 | 2.19E-14 |
| ENSG00000111087 | GLI1 | 5.58E-01 | 2.42E-01 | 2.39E-11 | 1.34E-09 | 5.12E-01 | 6.96E-10 | 4.73E-01 | 6.80E-18 |
| ENSG00000000419 | DPM1 | 5.75E-01 | 2.40E-01 | 4.15E-12 | 3.25E-10 | 4.17E-01 | 1.34E-05 | 2.98E-01 | 1.33E-05 |
| ENSG00000023734 | STRAP | 4.31E-01 | 2.14E-01 | 7.17E-07 | 7.42E-06 | 2.09E-01 | 1.00E+00 | 1.90E-01 | 3.56E-01 |
| ENSG00000104147 | OIP5 | 6.12E-01 | 3.32E-01 | 6.95E-14 | 1.14E-11 | 5.73E-01 | 2.21E-13 | 3.91E-01 | 2.99E-11 |
| ENSG00000124508 | BTN2A2 | 5.25E-01 | 2.26E-01 | 5.17E-10 | 1.75E-08 | 2.58E-01 | 6.42E-01 | 2.77E-01 | 1.33E-04 |

**Table S3. Prediction performance for the 1011 genes on validation and test sets.**

| Ensembl_gene_id | Hgnc symbol | Coef (valid) | Rsq (valid) | P-val (valid) | adj.pval (valid) | Coef (internal) | adj.pval (internal) | Coef (external) | adj.pval (external) |
|---|---|---|---|---|---|---|---|---|---|
| ENSG00000155393 | HEATR3 | 6.32E-01 | 2.92E-01 | 5.91E-15 | 1.66E-12 | 4.17E-01 | 1.26E-05 | 4.63E-01 | 5.02E-17 |
| ENSG00000198055 | GRK6 | 4.51E-01 | 2.08E-01 | 1.81E-07 | 2.34E-06 | 3.39E-01 | 5.42E-03 | 2.17E-01 | 4.28E-02 |
| ENSG00000156261 | CCT8 | 4.93E-01 | 2.12E-01 | 8.22E-09 | 1.76E-07 | 4.34E-01 | 2.76E-06 | 3.39E-01 | 7.46E-08 |
| ENSG00000185261 | KIAA0825 | 5.77E-01 | 2.89E-01 | 3.68E-12 | 2.96E-10 | 3.52E-01 | 2.16E-03 | 2.97E-01 | 1.54E-05 |
| ENSG00000106615 | RHEB | 4.45E-01 | 2.06E-01 | 2.83E-07 | 3.42E-06 | 2.43E-01 | 1.00E+00 | 3.04E-01 | 6.40E-06 |
| ENSG00000055044 | NOP58 | 4.35E-01 | 3.02E-01 | 5.59E-07 | 6.00E-06 | 3.79E-01 | 3.08E-04 | 2.71E-01 | 2.68E-04 |
| ENSG00000120647 | CCDC77 | 4.07E-01 | 2.97E-01 | 3.28E-06 | 2.69E-05 | 3.84E-01 | 1.98E-04 | 2.58E-01 | 1.01E-03 |
| ENSG00000041357 | PSMA4 | 4.72E-01 | 2.18E-01 | 4.06E-08 | 6.64E-07 | 2.68E-01 | 3.81E-01 | 2.45E-01 | 3.43E-03 |
| ENSG00000160208 | RRP1B | 4.88E-01 | 2.38E-01 | 1.18E-08 | 2.36E-07 | 5.61E-01 | 1.17E-12 | 2.07E-01 | 9.83E-02 |
| ENSG00000131015 | ULBP2 | 4.66E-01 | 2.61E-01 | 6.35E-08 | 9.64E-07 | 3.75E-01 | 3.98E-04 | 4.08E-01 | 1.66E-12 |
| ENSG00000151461 | UPF2 | 4.72E-01 | 2.70E-01 | 4.15E-08 | 6.74E-07 | 4.00E-01 | 5.45E-05 | 1.83E-01 | 5.90E-01 |
| ENSG00000177807 | KCNJ10 | 5.77E-01 | 3.22E-01 | 3.34E-12 | 2.75E-10 | 4.54E-01 | 3.91E-07 | 3.72E-01 | 5.98E-10 |
| ENSG00000197044 | ZNF441 | 5.22E-01 | 2.81E-01 | 7.10E-10 | 2.30E-08 | 3.57E-01 | 1.57E-03 | 2.55E-01 | 1.38E-03 |
| ENSG00000134453 | RBM17 | 5.53E-01 | 3.39E-01 | 4.09E-11 | 2.10E-09 | 2.88E-01 | 1.32E-01 | 2.31E-01 | 1.23E-02 |
| ENSG00000115233 | PSMD14 | 5.07E-01 | 2.34E-01 | 2.49E-09 | 6.43E-08 | 4.62E-01 | 1.86E-07 | 4.82E-01 | 8.55E-19 |
| ENSG00000174442 | ZWILCH | 5.27E-01 | 2.00E-01 | 4.41E-10 | 1.54E-08 | 2.89E-01 | 1.22E-01 | 3.95E-01 | 1.51E-11 |
| ENSG00000142864 | SERBP1 | 3.78E-01 | 2.53E-01 | 1.75E-05 | 1.10E-04 | 3.99E-01 | 5.97E-05 | 1.79E-01 | 7.45E-01 |
| ENSG00000135931 | ARMC9 | 4.10E-01 | 2.26E-01 | 2.79E-06 | 2.34E-05 | 4.27E-01 | 5.04E-06 | 2.97E-01 | 1.53E-05 |
| ENSG00000187522 | HSPA14 | 5.25E-01 | 3.22E-01 | 5.33E-10 | 1.80E-08 | 4.93E-01 | 6.39E-09 | 3.41E-01 | 5.81E-08 |
| ENSG00000197275 | RAD54B | 5.01E-01 | 2.22E-01 | 4.33E-09 | 1.03E-07 | 5.47E-01 | 8.53E-12 | 3.63E-01 | 2.60E-09 |
| ENSG00000171792 | RHNO1 | 4.27E-01 | 2.12E-01 | 9.54E-07 | 9.49E-06 | 3.28E-01 | 1.15E-02 | 1.99E-01 | 1.73E-01 |
| ENSG00000183150 | GPR19 | 4.91E-01 | 2.09E-01 | 9.45E-09 | 1.97E-07 | 3.46E-01 | 3.34E-03 | 3.02E-01 | 8.51E-06 |
| ENSG00000182700 | IGIP | 5.40E-01 | 2.91E-01 | 1.43E-09 | 5.99E-08 | 4.41E-01 | 1.45E-06 | 3.39E-01 | 6.97E-08 |
| ENSG00000124787 | RPP40 | 4.70E-01 | 2.07E-01 | 4.81E-08 | 7.65E-07 | 3.67E-01 | 7.41E-04 | 2.82E-01 | 7.69E-05 |
| ENSG00000073861 | TBX21 | 5.21E-01 | 2.30E-01 | 7.85E-10 | 2.51E-08 | 4.15E-01 | 1.48E-05 | 4.11E-01 | 1.10E-12 |
| ENSG00000162676 | GFI1 | 5.48E-01 | 2.08E-01 | 6.26E-11 | 2.97E-09 | 4.65E-01 | 1.31E-07 | 4.24E-01 | 1.07E-13 |
| ENSG00000181722 | ZBTB20 | 4.97E-01 | 3.02E-01 | 5.73E-09 | 1.30E-07 | 2.98E-01 | 7.27E-02 | 4.64E-01 | 4.08E-17 |
| ENSG00000186952 | TMEM232 | 5.17E-01 | 2.24E-01 | 1.09E-09 | 3.28E-08 | 2.52E-01 | 8.51E-01 | 2.21E-01 | 3.11E-02 |
| ENSG00000050730 | TNIP3 | 5.60E-01 | 2.18E-01 | 2.02E-11 | 1.17E-09 | 4.96E-01 | 4.82E-09 | 4.25E-01 | 9.15E-14 |
| ENSG00000048991 | R3HDM1 | 4.61E-01 | 2.37E-01 | 9.08E-08 | 1.30E-06 | 3.30E-01 | 9.88E-03 | 4.23E-01 | 1.16E-13 |
| ENSG00000005302 | MSL3 | 4.78E-01 | 2.25E-01 | 2.55E-08 | 4.56E-07 | 5.42E-01 | 1.64E-11 | 4.52E-01 | 4.70E-16 |
| ENSG00000163029 | SMC6 | 4.86E-01 | 2.65E-01 | 1.43E-08 | 2.78E-07 | 3.92E-01 | 1.05E-04 | 2.11E-01 | 6.92E-02 |
| ENSG00000005194 | CIAPIN1 | 5.60E-01 | 2.87E-01 | 2.06E-11 | 1.19E-09 | 3.49E-01 | 2.76E-03 | 3.65E-01 | 1.88E-09 |
| ENSG00000148459 | PDSS1 | 5.98E-01 | 4.35E-01 | 3.54E-13 | 4.20E-11 | 5.17E-01 | 3.83E-10 | 4.84E-01 | 5.75E-19 |
| ENSG00000170275 | CRTAP | 5.45E-01 | 2.59E-01 | 8.52E-11 | 3.86E-09 | 3.14E-01 | 2.69E-02 | 2.74E-01 | 1.86E-04 |
| ENSG00000127423 | AUNIP | 4.79E-01 | 2.06E-01 | 2.35E-08 | 4.26E-07 | 4.33E-01 | 3.08E-06 | 3.93E-01 | 2.11E-11 |
| ENSG00000040275 | SPDL1 | 5.55E-01 | 2.44E-01 | 3.13E-11 | 1.68E-09 | 5.05E-01 | 1.65E-09 | 3.66E-01 | 1.64E-09 |
| ENSG00000164331 | ANKRA2 | 4.41E-01 | 2.13E-01 | 3.65E-07 | 4.22E-06 | 4.90E-01 | 9.30E-09 | 3.69E-01 | 1.02E-09 |
| ENSG00000166164 | BRD7 | 4.99E-01 | 2.23E-01 | 5.09E-09 | 1.18E-07 | 3.51E-01 | 2.42E-03 | 2.12E-01 | 6.14E-02 |
| ENSG00000109618 | SEPSECS | 5.11E-01 | 2.25E-01 | 1.84E-09 | 4.99E-08 | 2.17E-01 | 1.00E+00 | 2.17E-01 | 4.29E-02 |
| ENSG00000163811 | WDR43 | 4.03E-01 | 2.31E-01 | 4.06E-06 | 3.21E-05 | 3.30E-01 | 9.70E-03 | 3.14E-01 | 1.96E-06 |
| ENSG00000103647 | CORO2B | 5.23E-01 | 2.24E-01 | 6.58E-10 | 2.14E-08 | 5.00E-01 | 2.97E-09 | 4.12E-01 | 8.83E-13 |
| ENSG00000135045 | C9orf40 | 4.09E-01 | 2.04E-01 | 2.97E-06 | 2.47E-05 | 3.27E-01 | 1.21E-02 | 2.08E-01 | 9.10E-02 |
| ENSG00000100749 | VRK1 | 4.41E-01 | 2.62E-01 | 3.74E-07 | 4.31E-06 | 5.29E-01 | 8.97E-11 | 3.05E-01 | 5.80E-06 |
| ENSG00000155330 | C16orf87 | 4.80E-01 | 2.26E-01 | 2.26E-08 | 4.15E-07 | 3.73E-01 | 4.55E-04 | 4.11E-01 | 1.13E-12 |
| ENSG00000136122 | BORA | 4.92E-01 | 2.90E-01 | 8.56E-09 | 1.81E-07 | 3.92E-01 | 1.06E-04 | 5.01E-01 | 1.25E-20 |
| ENSG00000181938 | GINS3 | 4.98E-01 | 2.65E-01 | 5.45E-09 | 1.25E-07 | 4.26E-01 | 5.99E-06 | 3.26E-01 | 3.96E-07 |
| ENSG00000105821 | DNAJC2 | 4.54E-01 | 2.39E-01 | 1.47E-07 | 1.95E-06 | 2.50E-01 | 9.60E-01 | 1.81E-01 | 6.88E-01 |
| ENSG00000189376 | C8orf76 | 5.20E-01 | 2.15E-01 | 8.56E-10 | 2.67E-08 | 3.84E-01 | 2.07E-04 | 4.51E-01 | 5.56E-16 |
| ENSG00000159147 | DONSON | 4.56E-01 | 2.68E-01 | 1.30E-07 | 1.75E-06 | 3.65E-01 | 8.84E-04 | 2.59E-01 | 8.90E-04 |
| ENSG00000145220 | LYAR | 5.50E-01 | 2.68E-01 | 5.13E-11 | 2.52E-09 | 4.50E-01 | 5.76E-07 | 2.32E-01 | 1.13E-02 |
| ENSG00000158092 | NCK1 | 4.01E-01 | 2.17E-01 | 4.84E-06 | 3.70E-05 | 3.44E-01 | 3.76E-03 | 2.46E-01 | 3.22E-03 |
| ENSG00000119812 | FAM98A | 5.14E-01 | 2.53E-01 | 1.44E-09 | 4.11E-08 | 4.12E-01 | 2.06E-05 | 3.44E-01 | 3.89E-08 |
| ENSG00000183598 | HIST2H3D | 3.60E-01 | 2.85E-01 | 4.73E-05 | 2.57E-04 | -1.43E-03 | 1.00E+00 | 3.79E-02 | 1.00E+00 |
| ENSG00000136114 | THSD1 | 5.44E-01 | 2.15E-01 | 9.09E-11 | 4.04E-09 | 4.71E-01 | 7.39E-08 | 3.85E-01 | 8.66E-11 |
| ENSG00000152154 | TMEM178A | 4.96E-01 | 2.28E-01 | 6.08E-09 | 1.37E-07 | 3.44E-01 | 3.77E-03 | 3.97E-01 | 1.22E-11 |
| ENSG00000114354 | TFG | 5.43E-01 | 2.45E-01 | 1.07E-10 | 4.65E-09 | 2.71E-01 | 3.20E-01 | 2.98E-01 | 1.26E-05 |
| ENSG00000144395 | CCDC150 | 5.84E-01 | 2.42E-01 | 1.67E-12 | 1.53E-10 | 4.80E-01 | 2.59E-08 | 3.28E-01 | 3.28E-07 |
| ENSG00000100629 | CEP128 | 4.07E-01 | 2.03E-01 | 3.21E-06 | 2.64E-05 | 4.73E-01 | 5.67E-08 | 3.00E-01 | 9.99E-06 |
| ENSG00000115042 | FAHD2A | 5.10E-01 | 2.98E-01 | 1.90E-09 | 5.10E-08 | 3.08E-01 | 3.97E-02 | 1.31E-01 | 1.00E+00 |
| ENSG00000152457 | DCLRE1C | 4.33E-01 | 2.89E-01 | 6.51E-07 | 6.84E-06 | 4.90E-01 | 9.19E-09 | 3.73E-01 | 5.10E-10 |
| ENSG00000112130 | RNF8 | 5.00E-01 | 2.03E-01 | 4.66E-09 | 1.09E-07 | 3.98E-01 | 6.44E-05 | 1.66E-01 | 1.00E+00 |
| ENSG00000258818 | RNASE4 | 5.23E-01 | 2.10E-01 | 6.48E-10 | 2.11E-08 | 4.16E-01 | 1.37E-05 | 2.89E-01 | 3.82E-05 |
| ENSG00000226979 | LTA | 5.04E-01 | 2.11E-01 | 3.18E-09 | 7.94E-08 | 4.11E-01 | 2.09E-05 | 3.54E-01 | 8.71E-09 |
| ENSG00000216490 | IFI30 | 6.01E-01 | 2.77E-01 | 2.46E-13 | 3.27E-11 | 2.66E-01 | 4.32E-01 | 3.38E-01 | 8.14E-08 |
| ENSG00000205250 | E2F4 | 4.47E-01 | 2.14E-01 | 2.45E-07 | 3.00E-06 | 3.52E-01 | 2.28E-03 | 2.04E-01 | 1.21E-01 |
| ENSG00000166986 | MARS | 5.91E-01 | 2.25E-01 | 8.11E-13 | 8.34E-11 | 4.72E-01 | 6.39E-08 | 4.25E-01 | 8.00E-14 |
| ENSG00000182771 | GRID1 | 4.41E-01 | 2.12E-01 | 3.69E-07 | 4.26E-06 | 4.00E-01 | 5.37E-05 | 3.83E-01 | 1.16E-10 |
| ENSG00000144029 | MRPS5 | 4.70E-01 | 2.64E-01 | 4.63E-08 | 7.41E-07 | 3.64E-01 | 9.48E-04 | 2.83E-01 | 7.29E-05 |

**Table S3. Prediction performance for the 1011 genes on validation and test sets.**

| Ensembl_gene_id | Hgnc symbol | Coef (valid) | Rsq (valid) | P-val (valid) | adj.pval (valid) | Coef (internal) | adj.pval (internal) | Coef (external) | adj.pval (external) |
|---|---|---|---|---|---|---|---|---|---|
| ENSG00000131148 | EMC8 | 4.74E-01 | 2.56E-01 | 3.40E-08 | 5.72E-07 | 3.81E-01 | 2.61E-04 | 2.79E-01 | 1.06E-04 |
| ENSG00000154767 | XPC | 3.42E-01 | 2.18E-01 | 1.15E-07 | 5.47E-04 | 3.86E-01 | 1.78E-04 | 3.08E-01 | 3.76E-06 |
| ENSG00000124541 | RRP36 | 4.82E-01 | 2.28E-01 | 1.85E-08 | 3.50E-07 | 3.71E-01 | 5.53E-04 | 3.46E-01 | 2.96E-08 |
| ENSG00000171817 | ZNF540 | 5.20E-01 | 2.13E-01 | 8.47E-10 | 2.66E-08 | 3.79E-01 | 2.87E-04 | 3.26E-01 | 4.01E-07 |
| ENSG00000176407 | KCMF1 | 5.04E-01 | 3.69E-01 | 3.23E-09 | 8.00E-08 | 3.87E-01 | 1.64E-04 | 3.77E-01 | 2.93E-10 |
| ENSG00000070761 | CFAP20 | 4.74E-01 | 2.51E-01 | 3.53E-08 | 5.93E-07 | 3.56E-01 | 1.61E-03 | 4.21E-01 | 1.71E-13 |
| ENSG00000115761 | NOL10 | 4.52E-01 | 2.67E-01 | 1.68E-07 | 2.20E-06 | 4.20E-01 | 9.64E-06 | 2.97E-01 | 1.50E-05 |
| ENSG00000229474 | PATL2 | 4.94E-01 | 2.22E-01 | 7.64E-09 | 1.65E-07 | 2.37E-01 | 1.00E+00 | 3.49E-01 | 1.79E-08 |
| ENSG00000078246 | TULP3 | 3.36E-01 | 2.19E-01 | 1.51E-04 | 6.89E-04 | 9.89E-02 | 1.00E+00 | 3.74E-03 | 1.00E+00 |
| ENSG00000160801 | PTH1R | 5.08E-01 | 2.30E-01 | 2.42E-09 | 6.30E-08 | 5.20E-01 | 2.59E-10 | 4.88E-01 | 2.27E-19 |
| ENSG00000142065 | ZFP14 | 5.61E-01 | 2.44E-01 | 1.86E-11 | 1.10E-09 | 3.52E-01 | 2.16E-03 | 2.32E-01 | 1.18E-02 |
| ENSG00000215784 | FAM72D | 4.71E-01 | 2.47E-01 | 4.28E-08 | 6.93E-07 | 3.26E-01 | 1.30E-02 | 4.21E-01 | 1.73E-13 |
| ENSG00000181418 | DDN | 5.31E-01 | 2.87E-01 | 3.02E-10 | 1.12E-08 | 3.99E-01 | 6.03E-05 | 2.90E-01 | 3.23E-05 |
| ENSG00000141052 | MYOCD | 5.20E-01 | 2.12E-01 | 8.65E-10 | 2.69E-08 | 4.46E-01 | 9.14E-07 | 3.40E-01 | 6.56E-08 |
| ENSG00000197776 | KLHDC1 | 6.75E-01 | 3.40E-01 | 1.55E-17 | 1.72E-14 | 4.34E-01 | 2.84E-06 | 5.03E-01 | 8.49E-21 |
| ENSG00000196550 | FAM72A | 5.56E-01 | 3.34E-01 | 2.92E-11 | 1.58E-09 | 5.17E-01 | 3.96E-10 | 4.63E-01 | 5.43E-17 |
| ENSG00000173237 | C11orf86 | 3.91E-01 | 2.52E-01 | 8.23E-06 | 5.76E-05 | 2.73E-01 | 2.99E-01 | 2.58E-01 | 9.76E-04 |
| ENSG00000171049 | FPR2 | 6.13E-01 | 2.47E-01 | 6.03E-14 | 1.06E-11 | 3.84E-01 | 2.05E-04 | 4.02E-01 | 5.07E-12 |
| ENSG00000153574 | RPIA | 4.06E-01 | 2.82E-01 | 3.53E-06 | 2.85E-05 | 3.88E-01 | 1.47E-04 | 2.56E-01 | 1.18E-03 |
| ENSG00000152464 | RPP38 | 3.96E-01 | 2.05E-01 | 6.40E-06 | 4.66E-05 | 2.53E-01 | 8.38E-01 | 2.91E-01 | 2.89E-05 |
| ENSG00000171865 | RNASEH1 | 4.69E-01 | 2.04E-01 | 5.18E-08 | 8.16E-07 | 5.08E-01 | 1.15E-09 | 2.90E-01 | 3.35E-05 |
| ENSG00000143942 | CHAC2 | 4.54E-01 | 2.41E-01 | 1.55E-07 | 2.04E-06 | 3.89E-01 | 1.38E-04 | 1.93E-01 | 2.84E-01 |
| ENSG00000168490 | PHYHIP | 5.35E-01 | 2.28E-01 | 2.12E-10 | 8.38E-09 | 3.21E-01 | 1.73E-02 | 4.50E-01 | 8.15E-16 |
| ENSG00000169857 | AVEN | 3.80E-01 | 2.03E-01 | 1.57E-05 | 1.00E-04 | 2.96E-01 | 8.15E-02 | 1.76E-01 | 9.15E-01 |
| ENSG00000181544 | FANCB | 5.49E-01 | 2.68E-01 | 5.80E-11 | 2.79E-09 | 5.23E-01 | 1.77E-10 | 3.76E-01 | 3.20E-10 |
| ENSG00000103995 | CEP152 | 4.70E-01 | 2.71E-01 | 4.65E-08 | 7.43E-07 | 3.91E-01 | 1.19E-04 | 3.16E-01 | 1.48E-06 |
| ENSG00000107951 | MTPAP | 3.62E-01 | 2.09E-01 | 4.13E-05 | 2.29E-04 | 2.83E-01 | 1.67E-01 | 2.17E-01 | 4.39E-02 |
| ENSG00000165733 | BMS1 | 4.63E-01 | 2.13E-01 | 7.65E-08 | 1.12E-06 | 4.25E-01 | 6.38E-06 | 2.33E-01 | 1.02E-02 |
| ENSG00000126895 | AVPR2 | 4.57E-01 | 2.17E-01 | 1.22E-07 | 1.68E-06 | 4.58E-01 | 2.70E-07 | 3.14E-01 | 1.92E-06 |
| ENSG00000148572 | NRBF2 | 4.23E-01 | 2.09E-01 | 1.21E-06 | 1.16E-05 | 4.10E-01 | 2.46E-05 | 3.60E-01 | 3.77E-09 |
| ENSG00000138395 | CDK15 | 5.22E-01 | 2.34E-01 | 6.91E-10 | 2.24E-08 | 4.78E-01 | 3.23E-08 | 2.48E-01 | 2.56E-03 |
| ENSG00000168872 | DDX19A | 4.58E-01 | 2.16E-01 | 1.11E-07 | 1.54E-06 | 3.95E-01 | 8.59E-05 | 1.86E-01 | 4.61E-01 |
| ENSG00000198522 | GPN1 | 4.45E-01 | 2.10E-01 | 2.77E-07 | 3.36E-06 | 2.79E-01 | 2.13E-01 | 2.44E-01 | 3.99E-03 |
| ENSG00000087274 | ADD1 | 4.93E-01 | 2.68E-01 | 7.88E-09 | 1.70E-07 | 2.57E-01 | 6.60E-01 | 3.53E-01 | 1.05E-08 |
| ENSG00000143184 | XCL1 | 5.03E-01 | 2.47E-01 | 3.63E-09 | 8.88E-08 | 3.76E-01 | 3.85E-04 | 3.13E-01 | 2.01E-06 |
| ENSG00000188610 | FAM72B | 6.34E-01 | 3.99E-01 | 4.39E-15 | 1.30E-12 | 5.57E-01 | 2.26E-12 | 4.47E-01 | 1.31E-15 |
| ENSG00000111537 | IFNG | 5.03E-01 | 2.41E-01 | 3.58E-09 | 8.77E-08 | 4.92E-01 | 6.97E-09 | 4.03E-01 | 3.94E-12 |
| ENSG00000138073 | PREB | 5.16E-01 | 2.08E-01 | 1.20E-09 | 3.55E-08 | 3.27E-01 | 1.23E-02 | 3.42E-01 | 4.65E-08 |
| ENSG00000011021 | CLCN6 | 5.80E-01 | 2.68E-01 | 2.67E-12 | 2.26E-10 | 2.43E-01 | 1.00E+00 | 2.99E-01 | 1.19E-05 |
| ENSG00000144231 | POLR2D | 5.39E-01 | 2.68E-01 | 1.51E-10 | 6.25E-09 | 4.67E-01 | 1.07E-07 | 5.18E-01 | 1.91E-22 |
| ENSG00000196275 | GTF2IRD2 | 5.85E-01 | 2.52E-01 | 1.54E-12 | 1.43E-10 | 4.11E-01 | 2.24E-05 | 1.03E-01 | 1.00E+00 |
| ENSG00000198939 | ZFP2 | 4.78E-01 | 2.17E-01 | 2.68E-08 | 4.72E-07 | 3.01E-01 | 5.99E-02 | 3.21E-01 | 7.89E-07 |
| ENSG00000115607 | IL18RAP | 5.93E-01 | 2.14E-01 | 6.02E-13 | 6.61E-11 | 3.82E-01 | 2.40E-04 | 3.44E-01 | 3.74E-08 |
| ENSG00000055130 | CUL1 | 4.32E-01 | 2.05E-01 | 6.69E-07 | 7.00E-06 | 1.63E-01 | 1.00E+00 | 2.09E-01 | 7.92E-02 |
| ENSG00000126953 | TIMM8A | 4.77E-01 | 2.11E-01 | 2.75E-08 | 4.79E-07 | 4.31E-01 | 3.58E-06 | 3.54E-01 | 9.03E-09 |
| ENSG00000168883 | USP39 | 5.28E-01 | 3.23E-01 | 4.07E-10 | 1.44E-08 | 3.48E-01 | 2.89E-03 | 4.17E-01 | 3.72E-13 |
| ENSG00000165861 | ZFYVE1 | 5.02E-01 | 2.66E-01 | 3.82E-09 | 9.28E-08 | 3.65E-01 | 8.93E-04 | 2.34E-01 | 1.00E-02 |
| ENSG00000167178 | ISLR2 | 5.83E-01 | 2.77E-01 | 1.93E-12 | 1.72E-10 | 5.18E-01 | 3.54E-10 | 4.14E-01 | 5.87E-13 |
| ENSG00000144580 | RQCD1 | 3.97E-01 | 2.31E-01 | 5.89E-06 | 4.36E-05 | 4.28E-01 | 4.80E-06 | 3.25E-01 | 4.56E-07 |
| ENSG00000144130 | NT5DC4 | 5.11E-01 | 2.61E-01 | 1.87E-09 | 5.05E-08 | 5.40E-01 | 2.03E-11 | | |
| ENSG00000132837 | DMGDH | 5.02E-01 | 2.29E-01 | 3.94E-09 | 9.56E-08 | 2.38E-01 | 1.00E+00 | 2.82E-01 | 8.09E-05 |
| ENSG00000135298 | BAI3 | 5.25E-01 | 3.26E-01 | 5.34E-10 | 1.80E-08 | 3.79E-01 | 2.88E-04 | 2.94E-01 | 2.15E-05 |
| ENSG00000120328 | PCDHB12 | 5.71E-01 | 2.43E-01 | 6.75E-12 | 4.84E-10 | 4.43E-01 | 1.19E-06 | 2.94E-01 | 2.14E-05 |
| ENSG00000150628 | SPATA4 | 4.89E-01 | 2.12E-01 | 1.06E-08 | 2.16E-07 | 3.58E-01 | 1.49E-03 | 4.17E-01 | 3.73E-13 |
| ENSG00000183395 | PMCH | 4.73E-01 | 2.66E-01 | 3.66E-08 | 6.10E-07 | 4.93E-01 | 6.16E-09 | 3.17E-01 | 1.36E-06 |
| ENSG00000132300 | PTCD3 | 4.12E-01 | 2.11E-01 | 2.38E-06 | 2.07E-05 | 3.01E-01 | 5.93E-02 | 2.71E-01 | 2.70E-04 |
| ENSG00000180061 | TMEM150B | 5.03E-01 | 2.08E-01 | 3.58E-09 | 8.77E-08 | 2.47E-01 | 1.00E+00 | 4.40E-01 | 5.04E-15 |
| ENSG00000266964 | FXYD1 | 4.47E-01 | 2.07E-01 | 2.39E-07 | 2.95E-06 | 4.53E-01 | 4.47E-07 | 3.53E-01 | 1.04E-08 |
| ENSG00000042317 | SPATA7 | 4.93E-01 | 2.40E-01 | 8.17E-09 | 1.75E-07 | 3.19E-01 | 1.97E-02 | 3.30E-01 | 2.44E-07 |
| ENSG00000131019 | ULBP3 | 4.79E-01 | 2.32E-01 | 2.31E-08 | 4.21E-07 | 4.53E-01 | 4.51E-07 | 3.76E-01 | 3.34E-10 |
| ENSG00000269067 | ZNF728 | 5.03E-01 | 2.24E-01 | 3.64E-09 | 8.89E-08 | 4.10E-01 | 2.42E-05 | | |
| ENSG00000166527 | CLEC4D | 5.18E-01 | 2.03E-01 | 9.62E-10 | 2.93E-08 | 2.61E-01 | 5.39E-01 | 2.69E-01 | 3.19E-04 |
| ENSG00000109680 | TBC1D19 | 4.71E-01 | 2.27E-01 | 4.51E-08 | 7.24E-07 | 3.10E-01 | 3.63E-02 | 2.23E-01 | 2.46E-02 |
| ENSG00000197054 | ZNF763 | 5.33E-01 | 2.16E-01 | 2.61E-10 | 9.88E-09 | 3.60E-01 | 1.21E-03 | 2.26E-01 | 1.99E-02 |
| ENSG00000117598 | | 5.77E-01 | 2.36E-01 | 3.67E-12 | 2.96E-10 | 4.34E-01 | 2.82E-06 | | |
| ENSG00000198785 | GRIN3A | 4.56E-01 | 2.45E-01 | 1.28E-07 | 1.74E-06 | 2.37E-01 | 1.00E+00 | 3.51E-01 | 1.38E-08 |
| ENSG00000205846 | CLEC6A | 4.38E-01 | 2.60E-01 | 4.67E-07 | 5.18E-06 | 3.91E-01 | 1.12E-04 | 3.43E-01 | 4.26E-08 |
| ENSG00000151687 | ANKAR | 4.34E-01 | 2.04E-01 | 5.75E-07 | 6.14E-06 | 3.01E-01 | 6.15E-02 | 1.71E-01 | 1.00E+00 |
| ENSG00000183775 | KCTD16 | 3.80E-01 | 2.14E-01 | 1.55E-05 | 9.97E-05 | 2.99E-01 | 6.73E-02 | 1.51E-01 | 1.00E+00 |

**Table S3. Prediction performance for the 1011 genes on validation and test sets.**

| Ensembl_gene_id | Hgnc symbol | Coef (valid) | Rsq (valid) | P-val (valid) | adj.pval (valid) | Coef (internal) | adj.pval (internal) | Coef (external) | adj.pval (external) |
|---|---|---|---|---|---|---|---|---|---|
| ENSG00000260456 | C16orf95 | 3.90E-01 | 2.09E-01 | 8.99E-06 | 6.21E-05 | 2.34E-01 | 1.00E+00 | 1.49E-02 | 1.00E+00 |
| ENSG00000261594 | TPBGL | 3.48E-01 | 2.15E-01 | 8.69E-05 | 4.31E-04 | 1.46E-01 | 1.00E+00 | | |
| ENSG00000186105 | LRRC70 | 5.30E-01 | 2.13E-01 | 3.49E-10 | 1.26E-08 | 4.16E-01 | 1.41E-05 | 2.93E-01 | 2.44E-05 |
| ENSG00000160808 | MYL3 | 5.59E-01 | 2.58E-01 | 2.27E-11 | 1.28E-09 | 5.56E-01 | 2.55E-12 | 4.96E-01 | 4.35E-20 |
| ENSG00000070729 | CNGB1 | 5.05E-01 | 2.75E-01 | 3.07E-09 | 7.72E-08 | 3.53E-01 | 2.08E-03 | 2.43E-01 | 4.40E-03 |
| ENSG00000180772 | AGTR2 | 3.47E-01 | 2.39E-01 | 9.10E-05 | 4.48E-04 | 1.85E-01 | 1.00E+00 | 2.69E-01 | 3.26E-04 |
| ENSG00000206559 | ZCWPW2 | 5.13E-01 | 2.31E-01 | 1.59E-09 | 4.43E-08 | 2.74E-01 | 2.79E-01 | 2.63E-01 | 6.27E-04 |
| ENSG00000189430 | NCR1 | 5.83E-01 | 2.10E-01 | 1.84E-12 | 1.66E-10 | 3.51E-01 | 2.35E-03 | 3.00E-01 | 1.00E-05 |
| ENSG00000137757 | CASP5 | 5.47E-01 | 2.75E-01 | 7.43E-11 | 3.45E-09 | 4.29E-01 | 4.36E-06 | 3.83E-01 | 1.18E-10 |
| ENSG00000187758 | ADH1A | 4.10E-01 | 2.25E-01 | 2.74E-06 | 2.32E-05 | 1.96E-01 | 1.00E+00 | 3.33E-01 | 1.70E-07 |
| ENSG00000141371 | C17orf64 | 4.40E-01 | 2.45E-01 | 4.05E-07 | 4.60E-06 | 1.09E-01 | 1.00E+00 | 1.02E-01 | 1.00E+00 |
| ENSG00000131142 | CCL25 | 6.13E-01 | 2.84E-01 | 6.33E-14 | 1.09E-11 | 2.32E-01 | 1.00E+00 | 1.68E-01 | 1.00E+00 |
| ENSG00000158578 | ALAS2 | 3.89E-01 | 2.15E-01 | 9.77E-06 | 6.64E-05 | 5.08E-02 | 1.00E+00 | 6.12E-02 | 1.00E+00 |
| ENSG00000048545 | GUCA1A | 4.72E-01 | 2.29E-01 | 4.04E-08 | 6.61E-07 | 1.86E-01 | 1.00E+00 | 2.42E-01 | 4.47E-03 |
| ENSG00000138684 | IL21 | 5.54E-01 | 3.35E-01 | 3.78E-11 | 1.95E-09 | 4.12E-01 | 1.92E-05 | 3.31E-01 | 2.12E-07 |
| ENSG00000140481 | CCDC33 | 5.15E-01 | 2.20E-01 | 1.29E-09 | 3.77E-08 | 8.70E-02 | 1.00E+00 | 2.41E-02 | 1.00E+00 |
| ENSG00000102794 | IRG1 | 5.66E-01 | 2.46E-01 | 1.10E-11 | 7.19E-10 | 3.98E-01 | 6.65E-05 | 1.29E-01 | 1.00E+00 |
| ENSG00000178997 | EXD1 | 5.63E-01 | 2.70E-01 | 1.43E-11 | 8.90E-10 | 1.39E-01 | 1.00E+00 | 1.03E-01 | 1.00E+00 |
| ENSG00000175065 | DSG4 | 4.52E-01 | 2.69E-01 | 1.76E-07 | 2.28E-06 | -3.86E-02 | 1.00E+00 | 2.67E-02 | 1.00E+00 |
| ENSG00000166856 | GPR182 | 4.61E-01 | 2.66E-01 | 8.83E-08 | 1.27E-06 | 1.17E-01 | 1.00E+00 | 1.81E-01 | 6.57E-01 |
| ENSG00000205929 | C21orf62 | 4.17E-01 | 2.13E-01 | 1.73E-06 | 1.57E-05 | 1.40E-01 | 1.00E+00 | -5.18E-02 | 1.00E+00 |
| ENSG00000179774 | ATOH7 | 4.81E-01 | 2.01E-01 | 2.13E-08 | 3.94E-07 | 2.84E-01 | 1.66E-01 | 2.09E-01 | 8.10E-02 |
| ENSG00000251692 | PTX4 | 4.09E-01 | 2.32E-01 | 2.90E-06 | 2.42E-05 | 1.89E-01 | 1.00E+00 | -7.29E-03 | 1.00E+00 |
| ENSG00000266733 | TBC1D29 | 5.46E-01 | 3.21E-01 | 8.17E-11 | 3.78E-09 | -6.44E-02 | 1.00E+00 | -6.39E-03 | 1.00E+00 |
| ENSG00000267368 | UPK3BL | 3.45E-01 | 2.07E-01 | 9.80E-05 | 4.78E-04 | 2.07E-01 | 1.00E+00 | 5.75E-02 | 1.00E+00 |
| ENSG00000164400 | CSF2 | 4.73E-01 | 2.03E-01 | 3.63E-08 | 6.06E-07 | 2.18E-01 | 1.00E+00 | 1.93E-01 | 2.72E-01 |
| ENSG00000231274 | SBK3 | 3.97E-01 | 2.01E-01 | 5.88E-06 | 4.35E-05 | 2.17E-01 | 1.00E+00 | 1.22E-01 | 1.00E+00 |
| ENSG00000177669 | MBOAT4 | 3.69E-01 | 2.15E-01 | 2.91E-05 | 1.68E-04 | 3.50E-02 | 1.00E+00 | -7.17E-02 | 1.00E+00 |
| ENSG00000234469 | | 3.90E-01 | 2.36E-01 | 8.87E-06 | 6.13E-05 | 2.59E-02 | 1.00E+00 | | |
| ENSG00000117400 | MPL | 4.57E-01 | 2.14E-01 | 1.21E-07 | 1.66E-06 | 1.08E-01 | 1.00E+00 | 8.36E-02 | 1.00E+00 |
| ENSG00000221986 | MYBPHL | 3.42E-01 | 2.29E-01 | 1.14E-04 | 5.44E-04 | 2.03E-01 | 1.00E+00 | -1.41E-04 | 1.00E+00 |
| ENSG00000181625 | SLX1B | 4.02E-01 | 2.66E-01 | 4.44E-06 | 3.46E-05 | 2.21E-02 | 1.00E+00 | -1.95E-01 | 2.38E-01 |
| ENSG00000204071 | TCEAL6 | 3.97E-01 | 2.10E-01 | 6.00E-06 | 4.42E-05 | 2.27E-01 | 1.00E+00 | 1.58E-01 | 1.00E+00 |
| ENSG00000141437 | SLC25A52 | 3.60E-01 | 2.42E-01 | 4.60E-05 | 2.51E-04 | 2.26E-01 | 1.00E+00 | | |
| ENSG00000064489 | MEF2BNB-MEF | 3.79E-01 | 2.10E-01 | 1.72E-05 | 1.09E-04 | 2.46E-01 | 1.00E+00 | -1.62E-02 | 1.00E+00 |
| ENSG00000161973 | CCDC42 | 3.96E-01 | 2.47E-01 | 6.42E-06 | 4.67E-05 | 1.66E-01 | 1.00E+00 | 8.63E-02 | 1.00E+00 |
| ENSG00000157211 | CDCP2 | 3.92E-01 | 2.52E-01 | 7.78E-06 | 5.51E-05 | 2.32E-01 | 1.00E+00 | -3.18E-02 | 1.00E+00 |
| ENSG00000260272 | | 3.60E-01 | 2.37E-01 | 4.69E-05 | 2.55E-04 | 2.51E-01 | 9.21E-01 | | |
| ENSG00000272896 | | 4.12E-01 | 2.05E-01 | 2.46E-06 | 2.12E-05 | 1.10E-01 | 1.00E+00 | | |
| ENSG00000236334 | PPIAL4G | 3.69E-01 | 2.12E-01 | 2.92E-05 | 1.68E-04 | 1.43E-01 | 1.00E+00 | 8.92E-02 | 1.00E+00 |
| ENSG00000255292 | | 3.65E-01 | 3.05E-01 | 3.53E-05 | 1.99E-04 | 3.08E-01 | 4.03E-02 | | |
| ENSG00000258644 | SYNJ2BP-COX | 3.86E-01 | 2.12E-01 | 1.12E-05 | 7.48E-05 | 1.55E-01 | 1.00E+00 | 2.02E-02 | 1.00E+00 |
| ENSG00000214866 | DCDC2C | 3.77E-01 | 2.02E-01 | 1.91E-05 | 1.18E-04 | 2.08E-01 | 1.00E+00 | | |
| ENSG00000189152 | GRAPL | 3.46E-01 | 2.16E-01 | 9.63E-05 | 4.70E-04 | 2.94E-01 | 9.34E-02 | -4.95E-02 | 1.00E+00 |
| ENSG00000255641 | | 5.20E-01 | 2.19E-01 | 8.53E-10 | 2.67E-08 | 1.90E-01 | 1.00E+00 | | |

**Table S4. ST gene panel.**

| Probe Name | Probe Name | Probe Name | Probe Name |
|---|---|---|---|
| AKT1 | CMKLR1 | IFNG | POLR2A |
| ARG1 | CSF1R | IFNGR1 | PSMB10 |
| B2M | CTLA4 | IL12B | PTEN |
| BATF3 | CTNNB1 | IL15 | PTPRC |
| BCL2 | CXCL10 | IL6 | RAB7A |
| C10orf54 | CXCL9 | ITGAM | SDHA |
| CCL5 | CXCR6 | ITGAV | STAT1 |
| CCND1 | DKK2 | ITGAX | STAT2 |
| CD27 | EPCAM | ITGB2 | STAT3 |
| CD274 | FAS | ITGB8 | TBX21 |
| CD276 | FOXP3 | LAG3 | TIGIT |
| CD3E | GZMB | LY6E | TNF |
| CD4 | HAVCR2 | MKI67 | TNFRSF9 |
| CD40 | HIF1A | MS4A1 | UBB |
| CD40LG | HLA-DQ1 | multi-KRT | VEGFA |
| CD44 | HLA-DRB1 | NKG7 | NegPrb1 |
| CD47 | HLA-E | OAZ1 | NegPrb2 |
| CD68 | ICAM1 | pan-melanocyte | NegPrb3 |
| CD74 | ICOSLG | PDCD1 | NegPrb4 |
| CD86 | IDO1 | PDCD1LG2 | NegPrb5 |
| CD8A | IFNAR1 | PECAM1 | NegPrb6 |

# Supplementary Figures

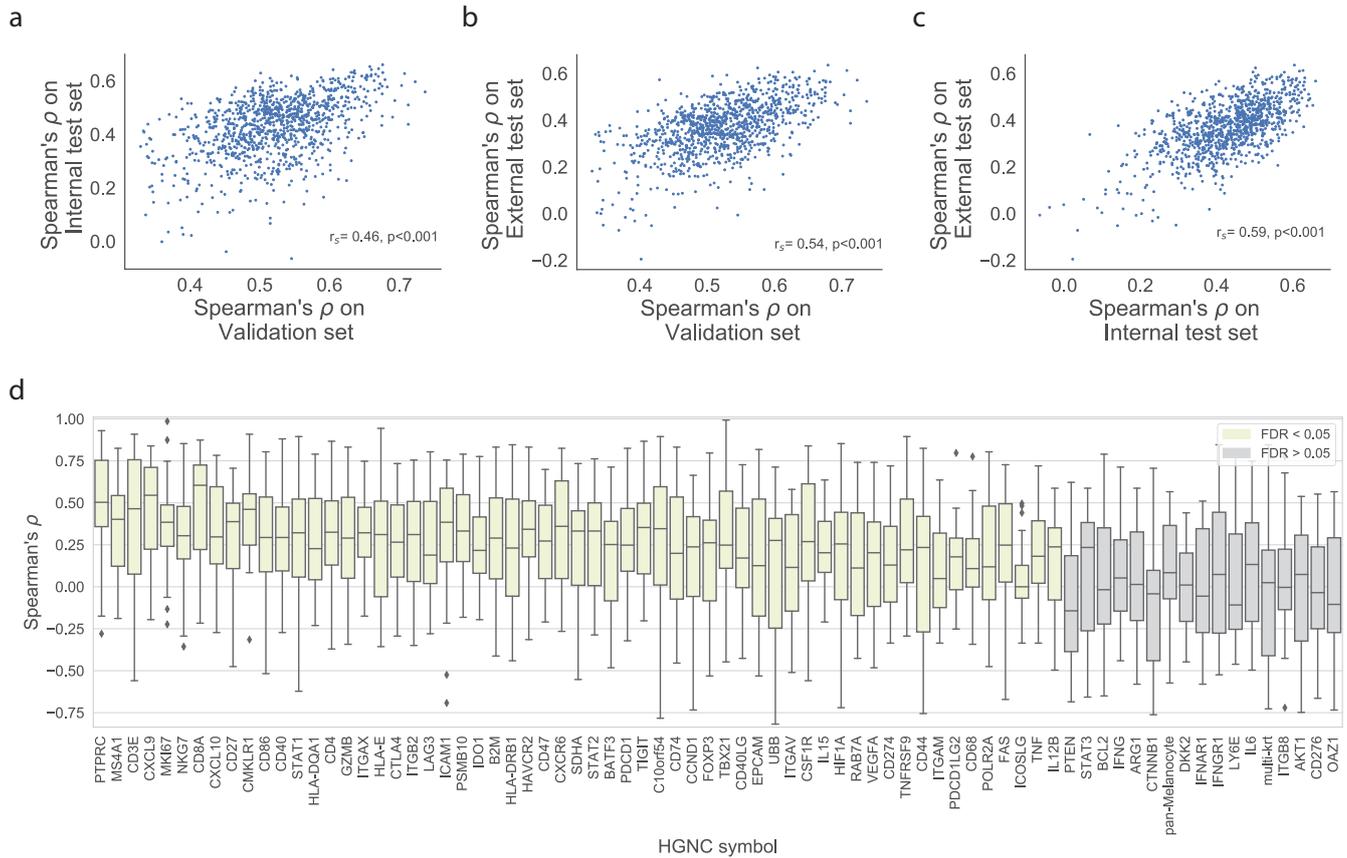

**Figure S1 a**, Concordance of Spearman correlation coefficients, describing the relationship between RNA-seq expression and predicted expression for each gene, in the validation and internal test set. **b,** Concordance of Spearman correlation coefficients between validation and external test set. **c,** Concordance of Spearman correlation coefficients between internal and external test sets. **d.** Distribution of Spearman correlation coefficients, derived from within-slide correlations (across ROIs) for each slide and each gene, between the EMO-spatial predictions and ST expression profiling estimates.

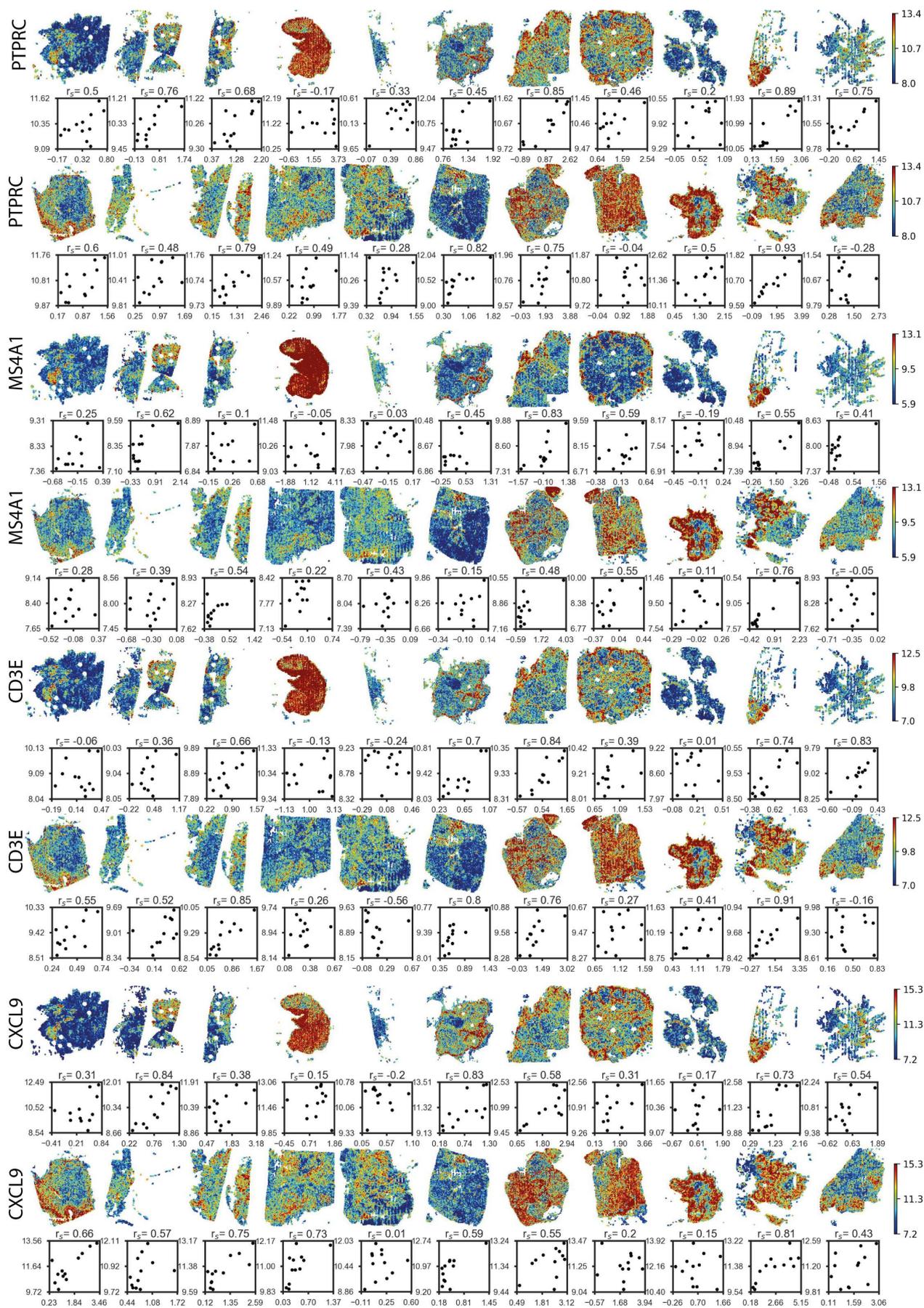

**Figure S2.** Prediction results across 22 WSIs for top performing genes (PTPRC,MS4A1,CD3E,CXCL9). For each gene, the first and third line shows prediction heatmaps per WSI, and the scatter plots in the second and fourth line shows corresponding measured EMO-spatial predictions and ST measurements, each dot represents a value within one ROI.

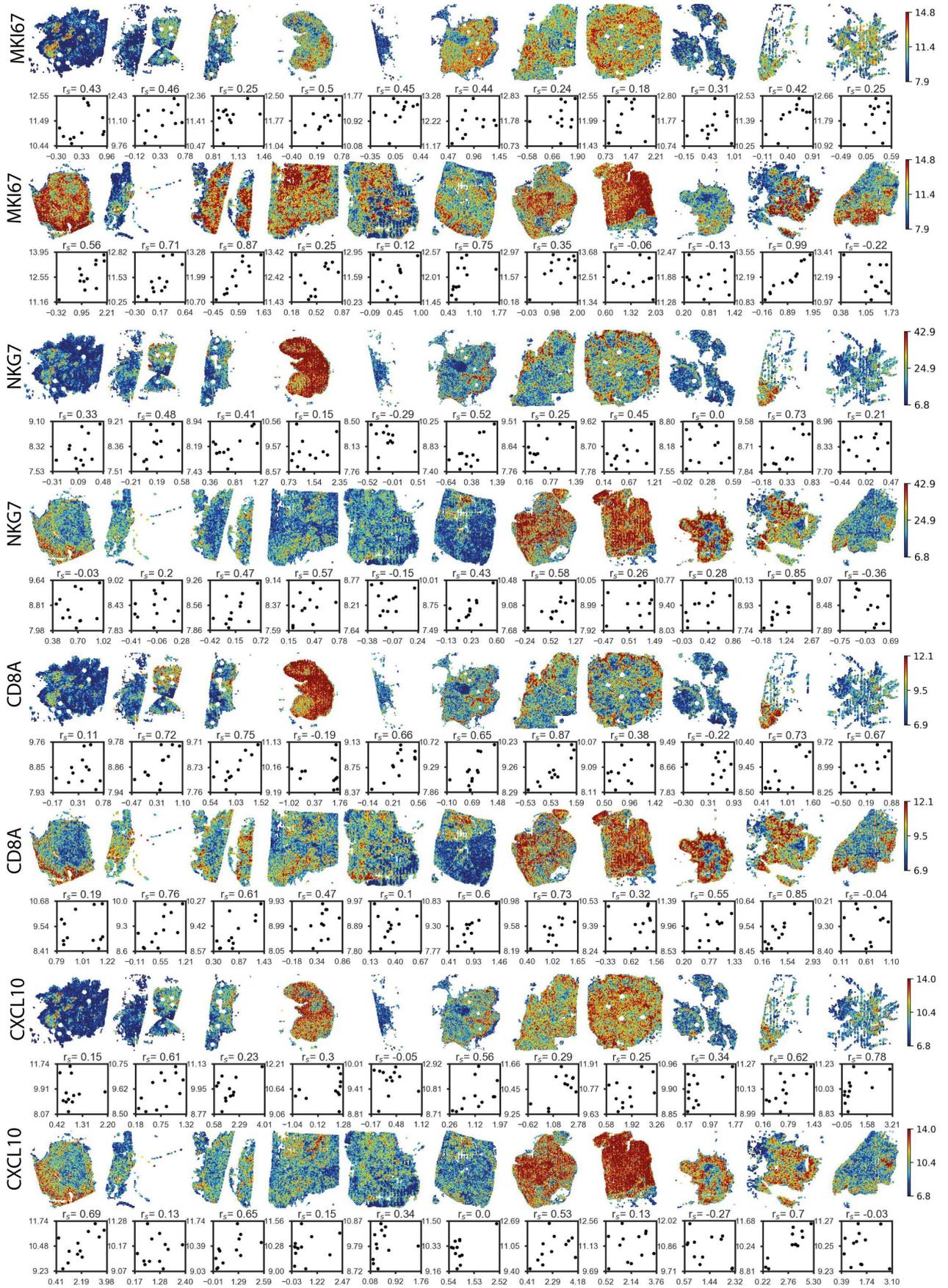

**Figure S3.** Prediction results across 22 WSIs for the next 4 top performing genes (MKI67,NKG7,CD8A,CXCL10). For each gene, the first and third line shows prediction heatmaps per WSI, and the scatter plots in the second and fourth line shows corresponding measured EMO-spatial predictions and ST measurements, each dot represents a value within one ROI.